\documentclass[apj]{emulateapj}
\slugcomment{{\sc Accepted to ApJ:} June 3, 2010}
\def\gtorder{\mathrel{\raise.3ex\hbox{$>$}\mkern-14mu
                \lower0.6ex\hbox{$\sim$}}}
\def\ltorder{\mathrel{\raise.3ex\hbox{$<$}\mkern-14mu
                \lower0.6ex\hbox{$\sim$}}}

\newcommand{\noprint}[1]{}

\shorttitle{Metal-Absorption in Evaporating Clouds}
\shortauthors{Gnat, Sternberg, \& McKee}

\begin{document}
\title{Metal-Ion Absorption in Conductively Evaporating Clouds}

\vspace{1cm}
\author{Orly Gnat\altaffilmark{1,2}  Amiel Sternberg\altaffilmark{3} and Christopher F. McKee\altaffilmark{4}}

\altaffiltext{1}{Theoretical Astrophysics, California Institute of Technology, 
        MC 350-17, Pasadena, CA 91125.}
\altaffiltext{2}{Chandra Fellow}
\altaffiltext{3}{School of Physics and Astronomy and the Wise Observatory,
        The Beverly and Raymond Sackler Faculty of Exact Sciences,
        Tel Aviv University, Tel Aviv 69978, Israel}
\altaffiltext{4}{Departments of Physics and Astronomy, 
        University of California, Berkeley, 
        Campbell Hall, Berkeley, CA 94720-7304, USA}
\email{orlyg@tapir.caltech.edu}

\begin{abstract}
We present computations of the ionization structure and metal-absorption
properties of thermally conductive interface layers that surround evaporating
warm spherical clouds, embedded in a hot medium. We rely on the analytical
steady-state formalism of Dalton and Balbus to calculate the temperature profile in
the evaporating gas, and we explicitly solve the time-dependent ionization 
equations for H, He, C, N, O, Si, and S in the conductive interface.
We include photoionization by an external field. We estimate how departures
from equilibrium ionization affect the resonance-line cooling efficiencies
in the evaporating gas, and determine the conditions for which radiative
losses may be neglected in the solution for the evaporation dynamics and
temperature profile. Our results indicate that non-equilibrium cooling
significantly increases the value of the saturation parameter $\sigma_0$
at which radiative losses begin to affect the flow dynamics. 
As applications we calculate the ion fractions and projected column
densities arising in the evaporating layers surrounding dwarf-galaxy-scale
objects
that are also photoionized by metagalactic radiation. We compare our results
to the UV metal-absorption column densities observed in local highly-ionized
metal-absorbers, located in the Galactic corona or intergalactic medium.
Conductive interfaces significantly enhance the formation of high-ions such
as C$^{3+}$, N$^{4+}$, and O$^{5+}$ relative to purely photoionized clouds,
especially for clouds embedded in a high-pressure corona. However, the
enhanced columns are still too low to account for the \ion{O}{6} columns
($\sim10^{14}$~cm$^{-2}$) observed in the local high-velocity metal-ion
absorbers. We find that column densities larger than $\sim10^{13}$~cm$^{-2}$
cannot be produced in evaporating clouds. 
Our results do support the
conclusion of Savage and Lehner, that absorption due to evaporating
\ion{O}{6} likely occurs in the local interstellar medium, with characteristic
columns of $\sim10^{13}$~cm$^{-2}$.
\end{abstract}


\section{Introduction}
\label{introduction}

The physics of conductively evaporating clouds has applications
to a variety of astrophysical environments. This includes the study of the
Local Cloud and Local Bubble (Slavin~1989; Smith \& Cox~2001, Slavin
\& Frisch~2002; Savage \& Lehner~2006; Jenkins~2009), cloud evaporation 
in the interstellar medium
(e.g.~B$\ddot{\rm o}$hringer \& Hartquist~1987; Dalton \& Balbus~1993;
Nagashima et al.~2007; Slavin~2007; Vieser \& Hensler~2007),
supernovae remnants (e.g.~Shelton~1998; Slavin \& Cox~1992; 
Balsara et al.~2008), active galactic nuclei (e.g.~McKee \& Begelman~1990),
and the galaxy formation process (Nipoti \& Binney~2007).

In this paper, we present new computations of the non-equilibrium ionization
states and metal absorption line signatures of thermally conductive interfaces
that surround evaporating warm  ($\sim10^4$~K) spherical gas clouds embedded in 
a hot ionized medium (HIM, $\gtrsim10^6$~K). In such interfaces, the warm
clouds undergo steady evaporation, while heat from the hot ambient medium
flows into the clouds (Cowie \& McKee~1977; McKee \& Cowie~1977). 
We compute the integrated metal-ion column densities
through the conductive interfaces, for comparison to absorption line 
observations. We consider the effect of photoionization by the metagalactic 
radiation field on the non-equilibrium ionization states in the conduction 
fronts.

As matter flows from the warm cloud toward the hot ambient medium, 
its ionization state changes continuously.
Non-equilibrium effects become significant when 
the ionization time-scale is long compared to the
rate of temperature change. The gas then tends to remain
``underionized'' compared to gas in ionization equilibrium
at the same temperature.
Because the energy losses in the evaporating gas are dominated
by atomic and ionic line emissions from many species, the
cooling efficiencies are affected by the non-equilibrium
abundances. For underionized gas, cooling is enhanced.

Our theoretical work is motivated by recent ultraviolet and X-ray 
absorption line spectroscopy of hot gas in the Galactic halo, 
around higher redshift galaxies, and in intergalactic environments
(e.g.~Sembach et al.~2003; Collins et al.~2005; Savage et al.~2005;
Tumlinson et al.~2005; Fang et al.~2006; Stocke et al.~2006;
Narayanan et al.~2010).
We are also motivated by observations of discrete 
ionized ``high-velocity clouds'' (Sembach et al.~1999, 2000, 2002, 
2003; Murphy et al.~2000; Wakker et al.~2003; 
Collins et al.~2003, 2004, 2007; Fox et al.~2005, 2006).
These observations indicate that the ionized gas is not in
equilibrium ionization (Sembach et al.~2002; Fox et al.~2004,
2005; Gnat \& Sternberg~2004), and non-equilibrium processes such as 
time-dependent radiative cooling (e.g.~Gnat \& Sternberg~2007;
Sutherland \& Dopita~1993), shock ionization (e.g.~Allen et al.~2008;
Gnat \& Sternberg~2009), or ionization in conductive interfaces
(this paper) may be at play in these objects.

Theoretical models of cloud evaporation in a hot medium have been 
studied extensively.
Cowie \& McKee~(1977) developed an analytical solution for the mass loss rate
and temperature profile in static conductive interfaces around spherical 
clouds. In their solution, they included the possible effects of heat flow 
saturation that occurs when the electron mean free path is 
not small compared to the temperature scale height (see \S2).
The saturation in the flow can be measured by one global parameter,
the ``saturation parameter'' $\sigma_0$.
McKee \& Cowie~(1977) studied the effects of radiative losses on the
interface. In particular, they demarcated the range of saturation parameters
for which radiative losses may be neglected.
The situation becomes much more complicated if the dynamics of the evaporation
are affected by magnetic fields (e.g.~Slavin~1989; Borkowski, Balbus \& 
Fristrom~1990) or if the evaporation is time-dependent 
(e.g.~Shelton~1998; Smith \& Cox~2001; Vieser \& Hensler~2007).

Non-equilibrium ionization in conductive interfaces was
considered previously 
in several studies
(Ballet, Arnaud \& Rothenflug~1986; 
B$\ddot{\rm o}$hringer \& Hartquist~1987; Dalton \& Balbus~1993;
Slavin~1989; Slavin \& Cox~1992; Shelton ~1998; Smith \& Cox~2001).
However, most of these works explored only a limited range of parameters
(with $P_{\rm HIM}/k_{\rm B}\sim10^4$~cm$^{-3}$~K and $R_{\rm cloud}\sim1$~pc),
appropriate for interstellar clouds.
None of the previous computations included photoionization by an external
radiation field.

Here we calculate the ion fractions created in the conductive
interfaces surrounding evaporating gas clouds.
We consider pressures between $0.1$ and
$10^4$~cm$^{-3}$~K, cloud radii 
between $1$~pc and $100$~kpc, and ambient
temperatures between $5\times10^5$~K
and $10^7$~K.
We study how the metal absorption line properties
depend on the interface parameters. We consider the effects
of non-equilibrium cooling and reexamine the criterion derived
by McKee \& Cowie~(1977) for the neglect of radiative losses
in the evaporative flows.

As applications, we then focus on the example of the local
``ionized high-velocity clouds''. Over the past decade, UV
absorption line observations have revealed a population of
local high-velocity metal-ion absorbers 
(Sembach et al.~1999, 2000, 2002, 2003; Murphy et al.~2000; 
Wakker et al.~2003; Collins et al.~2004; Fox et al.~2005, 2006).
The observations, carried out with the {\it Goddard High Resolution 
Spectrometer} (GHRS) and {\it Space Telescope Imaging Spectrograph} (STIS)
on board HST, and more recently with the {\it Far Ultraviolet Spectroscopic
Explorer} (FUSE), indicate the presence of many 
high-ionization metal absorption lines, including 
\ion{Si}{3} $\lambda~1206.5$, 
\ion{Si}{4} $\lambda\lambda~1393.8,~1402.8$, and 
\ion{C}{4}  $\lambda\lambda~1548.2,~1550.8$
with little or no associated \ion{H}{1} 
(N$_{\rm{HI}}\lesssim10^{15}-10^{16}$~cm$^{-2}$). 
Of particular interest is \ion{O}{6} $\lambda~1031.9$, 
for which absorbing column densities of order
$10^{14}$~cm$^{-2}$ have been measured (Collins et al.~2004; 
Fox et al.~2006; Sembach et al.~2000, 2003).

Many of these ionized absorbers can be identified with the
diffuse nearby \ion{H}{1} high-velocity cloud complexes, 
such as the Magellanic Stream or Complex C (Sembach et 
al.~2003). However, some of these absorbers do not appear 
directly linked to such structures, and could be clouds at
much larger distances, perhaps pervading the Local Group. 
The absorbers towards Mrk~509 and PKS~2155-304 are 
examples of such isolated high-velocity metal-ion
absorbers (Sembach et al.~1999; Collins et al.~2004; Fox et al.~2005).
As a guide to these observations, we list in Table~\ref{collins}
the data presented by Collins et al.~(2004) for the metal-ion
absorbers towards Mrk~509.

In Gnat \& Sternberg~(2004, hereafter GS04), we modeled the high-velocity
metal ion absorbers as photoionized gas clouds associated 
with low-mass dark matter ``minihalos'' (see also Kepner et al.~1999). 
For this purpose, we studied the metal photoionization properties of
hydrostatic gas clouds embedded in gravitationally dominant
dark-matter halos, that are photoionized by the present-day 
cosmological metagalactic radiation field, and are 
pressure-confined
by an external hot medium.
We considered minihalo models for dwarf-galaxy-scale objects
and for the lower mass compact \ion{H}{1} high velocity
clouds (CHVCs), based on the properties derived by 
Sternberg, McKee, \& Wolfire\footnote{Sternberg, McKee, \& 
Wolfire~(2002) constructed explicit minihalo models for Local Group
dwarf galaxies, based on the observed H~I properties of Leo~A
and Sag~DIG. They then searched for minihalo models for the H~I
CHVCs, that would simultaneously account for the range of observed
H~I column densities, the existence of multiphased (cold/warm) cores, 
and the total number of objects.} (2002; see also Giovanelli et al. 2010). 

For low mass objects ($M_{\rm vir}\sim10^8$~M$_{\odot}$)
embedded in the relatively high-pressure environment
of the Galactic corona ($P\sim50$~cm$^{-3}$~K), we found that the
ionization parameter is too low to efficiently produce \ion{C}{4}
and other high ions.
However, for the more massive dwarf-galaxy-scale halos 
($M_{\rm vir}\sim2\times10^9$~M$_\odot$, $P\lesssim1$~cm$^{-3}$~K) embedded
in the low-pressure intergalactic medium (IGM), we found that their
ionized envelopes are natural sites for the formation of high-ions.
Photoionized envelopes of dwarf-galaxy-scale halos could be detectable 
as UV metal-line absorbers, with ionization states similar to
those observed in the ionized high-velocity absorbers.
\ion{O}{6} remains an important exception, as it is
inefficiently produced by photoionization.

The strong \ion{O}{6} absorption in the high-velocity metal-ion
absorbers strongly suggests that photoionization is not the
only ionization mechanism at play in these objects 
(Sembach et al.~1999; Fox et al.~2004, 2005; GS04). 
The contribution of additional ionization processes may 
produce high-ions that are suppressed in the purely photoionized clouds.

The fact that minihalo clouds may be embedded in a hot plasma,
such as the Galactic corona or the intergalactic medium
at various redshifts, suggests 
the possibility of significant 
collisional interactions between the hot gas and the warm clouds.
Here we use our evaporating cloud models to study
the interface layers surrounding
dwarf-galaxy-scale minihalos
($0.1<P/k_{\rm B}<50$~cm$^{-3}$~K, $R_{\rm cloud}\sim$~kpc,
$T_{\rm HIM}\gtrsim10^6$~K, $Z=0.1$~solar)
for comparison with the observations of the high-velocity metal
ion absorbers. 
As we describe below, 
we include the effect of photoionization by the
metagalactic external radiation on the non-equilibrium ionization
properties of the gas in the interface.

In~\S2 we describe the basic equations 
and numerical method.
In~\S3 we present a set of computations of the ion fractions in
conductive interfaces, and we discuss the resulting metal absorption column
densities for the dwarf-galaxy-scale models in~\S4. 
We find that conductive interfaces significantly
enhance the formation of high ions such as C$^{3+}$, N$^{4+}$ and O$^{5+}$
relative to purely photoionized clouds, especially for clouds embedded in
a high-pressure medium. However, the enhanced columns are still too low to
account for the \ion{O}{6} columns ($\sim10^{14}$~cm$^{-2}$) observed in the
high-velocity metal-ion absorbers. We find that \ion{O}{6} column densities
larger than $\sim10^{13}$~cm$^{-2}$ cannot be produced in evaporating clouds.
Our models do support the conclusion by Savage \& Lehner~(2006) that 
evaporating \ion{O}{6} absorption occurs in the local ISM, with 
characteristic columns of $\sim10^{13}$~cm$^{-2}$.
We summarize in~\S5.

\begin{deluxetable}{lll}
\tablewidth{0pt}
\tablecaption{Column-Density Measurements in MrK~509\label{collins}}
\tablehead{
\colhead{Ion}&
\colhead{-300 km s$^{-1}$}&
\colhead{-240 km s$^{-1}$}
}
\startdata
\ion{C}{2} &   $4.79^{+0.97}_{-0.90}\times10^{13}$  &  $<2.09\times10^{13}$                 \\
\ion{N}{1} &   $<3.39\times10^{13}$                 &  $<2.40\times10^{13}$                 \\
\ion{Si}{2}&   $<1.23\times10^{12}$                 &  $8.51^{+2.05}_{-1.33}\times10^{12}$  \\
\ion{S}{2} &   $<1.00\times10^{14}$                 &  $<7.08\times10^{13}$                 \\
\\
\ion{C}{4} &   $1.41^{+0.37}_{-0.21}\times10^{14}$  &  $3.39^{+0.41}_{-0.44}\times10^{13}$  \\
\ion{N}{5} &   $<1.74\times10^{13}$                 &  $<1.20\times10^{13}$                 \\
\ion{O}{6} &   $8.51^{+0.82}_{-0.93}\times10^{13}$  &  $7.76^{+0.75}_{-0.68}\times10^{13}$  \\
\ion{Si}{3}&   $2.04^{+0.15}_{-0.18}\times10^{13}$  &  $2.75^{+0.71}_{-0.66}\times10^{12}$  \\
\ion{Si}{4}&   $3.39^{+1.86}_{-0.88}\times10^{13}$  &  $<3.09\times10^{12}$                 \\
\ion{S}{3} &   $<8.51\times10^{13}$                 &  $<6.03\times10^{13}$                 \\
\enddata
\tablecomments{Column densities for two velocity
components toward Mrk~509  by Collins et al. (2004).
The error estimates are the $1\sigma$ estimates,
and the upper limits are $3\sigma$ limits.}
\end{deluxetable}

\section{Basic Equations and Processes}
\label{physics}

\subsection{Evaporation}

We are interested in studying the evolving non-equilibrium ionization states
in thermally conductive gas that evaporates from warm purely 
photoionized clouds into a hot ambient medium.
In our treatment of the ionization in the evaporating layers
we use an analytic solution for the temperature profiles in the conductive
interfaces surrounding steady-state, spherical, non-magnetic clouds.
The temperature profiles are obtained by solving the equation of energy
conservation in the flow, balancing the outward energy flux
with the inward heat flux. Within the interface, the heat flux 
changes from a classical diffusive form, 
\begin{equation}
\mathbf{q}_{\rm cl}=-\kappa \nabla T
\end{equation} 
(Spitzer~1962) that applies when the electron mean free path
is small compared to the temperature scale height ($T/\nabla T$), 
to a saturated form, 
\begin{equation}
|\mathbf{q}_{\rm sat}|=5\phi_s\rho c^3
\end{equation}
(Cowie \& McKee~1977)
when the mean free path is comparable to, or greater than,
the temperature scale height.
In the above expressions, $\kappa$ is the thermal
conductivity\footnote{$\kappa = 1.84\times10^{-5}
T^{5/2} (\ln\Lambda)^{-1}$~erg~s$^{-1}$~cm$^{-1}$~K$^{-1}$
where $\ln\Lambda=29.7+\ln(n_e^{-1/2}T_6)$,
$n_e$ is the electron density, and $T_6 = T/(10^6~K)$ (Spitzer~1962).},
$\rho$ is the mass density, $c$ is the sound speed, and $\phi_s$ is a factor
of order unity
that allows for various uncertainties in the estimation of
$q_{\rm sat}$. 

The local saturation in the flow is defined by the
ratio (Cowie \& McKee~1977),
\begin{equation}
\sigma = \frac{q_{\rm cl}}{q_{\rm sat}}\;,
\end{equation}
and is proportional to the ratio of the mean free path
to the temperature scale height.

For a classical diffusive interface, the rate of mass loss
\begin{equation}
\dot{m}_{\rm cl} = \frac{16 \pi \mu R \kappa_{\rm HIM}}{25 k_{\rm B}}
\end{equation}
(Cowie \& McKee~1977), where $R$ is the cloud radius, $\mu$ is the mean
mass per particle, $k_B$ is the Boltzmann constant, and  subscript ``HIM''
refers to the surrounding hot ionized medium. As the flow saturates,
the heat flux is limited to a maximal
rate given by equation (2), and the classical mass loss
rate is reduced by a factor $\omega$ such
that $\dot{m}=\omega\dot{m}_{\rm cl}$.
The global level of saturation in the flow is parametrized
by the saturation parameter,
\begin{equation}
\sigma_0 
\equiv \frac{2\kappa_{\rm HIM} T_{\rm HIM}}{25 \phi_s \rho_{\rm HIM} c_{\rm HIM}^3 R}
\propto \frac{T_{\rm HIM}^2}{\rho_{\rm HIM} R} \propto \frac{T_{\rm HIM}^3}{P R}\;
\label{sigma0}
\end{equation}
(Cowie \& McKee~1977). In this expressions $T$ is the temperature, and 
$P=\rho k_{\rm B} T / \mu$ is the gas pressure.

To obtain the steady state temperature profile in the 
conductive interface, the equation of energy conservation,
\begin{equation}
\label{fullE}
\nabla\cdot\left[ \frac{\rho \mathbf{v}}{2} (\mathbf{v}^2+5c^2) \right]
  + \left( n^2\Lambda-n\Upsilon \right) = 
-\nabla\cdot \mathbf{q}
\end{equation}
must be solved, where $\Lambda$ is the cooling efficiency
(erg cm$^3$ s$^{-1}$), and $\Upsilon$ is the heating rate (erg s$^{-1}$). 
The first term on the left hand side of
equation~(\ref{fullE}) is a bulk kinetic energy term; 
the second is due to internal energy and $PdV$ work;
and the next two terms account for radiative losses
(cooling) and gains (heating). 
This outward energy flux is balanced by an inward heat
flux represented by the right hand side of equation~(\ref{fullE}).

We rely on the formalism developed by Dalton \& Balbus~(1993, hereafter DB93)
for determining the temperature profile in the interface.
The DB93 formalism allows the thermal conductivity to change continuously
from the classical diffusive form to a saturated form, by using
an effective heat flux $q$, defined as the harmonic mean of
the local classical and saturated fluxes,
\begin{equation}
\frac{1}{q} 
\equiv  \frac{1}{q_{\rm cl}} + \frac{1}{q_{\rm sat}}\;.
\end{equation}
DB93 then solve for the steady state temperature profile that develops
in the interface layer between the warm cloud and the hot medium
surrounding it.

\begin{figure}[!h]
\epsscale{1.0}
\plotone{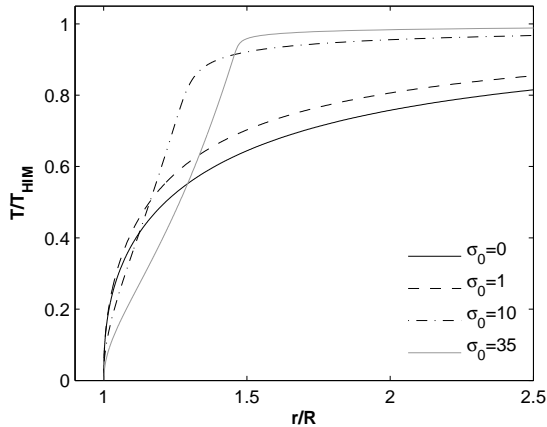}
\caption{Scaled temperature, $T/T_{\rm HIM}$ versus scaled distance
$r/R$, computed using the DB93 analytical solutions, 
for a range of saturation parameters.}
\label{tau-y}
\end{figure}

When solving the energy equation, DB93 neglect the kinetic energy term,
as well as the radiative losses (cooling) term.
We discuss both assumptions in detail below.
With these simplifying assumptions, the equation of energy
conservation becomes,
\begin{equation}
\nabla\cdot\left(\frac{5}{2}\rho \mathbf{v} c^2\right) = - \nabla\cdot \mathbf{q}\;.
\end{equation}
The temperature profile in the conductive interface\footnote{
DB93 write the solution for the scaled radius, $y=r/R$, as
a function of the scaled temperature, $\tau=T/T_{\rm HIM}$, as
$y=\frac{11}{10}\omega\chi\tau^{1/4}$
$\frac{K_{5/11}(\chi)I_{-(6/11)}(\chi\tau^{11/4})+
K_{6/11}(\chi\tau^{11/4})I_{-(6/11)}(\chi)}
{K_{5/11}(\chi\tau^{11/4})I_{5/11}(\chi)-
K_{5/11}(\chi)I_{5/11}(\chi\tau^{11/4})}$.
In this expression $\chi=(\frac{10}{11})(\frac{\sigma_0}{\omega})^{1/2}$,
$I$ are the modified Bessel functions of the 
first kind, and $K$ are the modified Bessel functions of the
second kind.
}
then depends only
on the global saturation parameter, $\sigma_0$, which is a function 
of the cloud radius, and the ambient pressure and temperature,
as given by equation~(\ref{sigma0}).
Temperature profiles for various values of $\sigma_0$ are 
shown in Figure~\ref{tau-y}.

\subsubsection{On the Validity of the Subsonic Approximation}
\label{mach}

The kinetic energy term may be neglected when 
$v^2 \ll 5c^2$ (see equation \ref{fullE}),
or equivalently when $M^2\ll 5$.
If the kinetic energy is small relative to the internal energy,
the flow is isobaric,  with a constant gas pressure 
$P = \rho_{\rm HIM} k_{\rm B} T_{\rm HIM} / \mu$.

In classical flows, $M$ is always less than unity, and 
the subsonic approximation holds. However, as the flow saturates,
the maximal Mach number increases.
In fully saturated flows, the heat flux is
given by equation~(2), and equation~(6) reads
\begin{equation}
5/2\;\;\rho v r^2 c^2 (1 + M^2/5) - 5 \phi_s \rho c^3 r^2 = 0,
\end{equation}
This implies that in fully saturated flows
\begin{equation}
M(1+M^2/5) = 2 \phi_s
\end{equation}
(e.g.~Cowie \& McKee~1977).
Therefore, for $\phi_s<0.6$, the subsonic
approximation holds throughout the flow. 

As a consistency check, we also consider the local Mach number
in the analytic profiles derived by Dalton \& Balbus neglecting
the kinetic energy term. In this case
\begin{equation}
M = T^{1/2} \frac{k_{\rm B}}{P} \frac{\omega}{4\pi r^2}
\left( \frac{16\pi \mu^{1/2} \kappa_{\rm HIM} R}{25 k_{\rm B}^3} \right)
\label{DB-B3}
\end{equation}
(DB93 equation B3).
Given a set of temperature profiles appropriate for 
a range of saturation parameters (see Figure~\ref{tau-y}), 
this equation shows that the maximal Mach number is an
{\it increasing}$\,$\footnote
{DB93 argued, incorrectly, that the maximal Mach number is a 
decreasing function of $\sigma_0$.
They used a specific set
 of physical variables 
($P/k_{\rm B}=10^4$~cm$^{-3}$~K, 
$T_{\rm HIM}=7\times10^5$~K, $R=1$~pc) to derive the numerical
expression $M = 0.66 \omega\sqrt{T / T_{\rm HIM}} / (r/R)^2$
(equation B5 in DB93).
They then used this expression to evaluate the Mach number
for a range of saturation parameters, and concluded that
the maximal Mach number is a decreasing function of $\sigma_0$.
However, the numerical coefficient in their expression was computed for
interface variables implying $\sigma_0=0.32$ and is therefore
not valid for other saturation parameters.} function of $\sigma_0$.
With the kinetic energy term neglected, the maximal 
Mach number approaches $2\phi_s$ in highly saturated flows\footnote{
As opposed to max$(M)=1.42$ $(1.0,\; 0.56)$ resulting from equation
(10) which includes the kinetic energy term, for $\phi_s=1.0$ $(0.6,\;0.3)$.},
implying that for $\phi_s<0.5$ the subsonic
approximation holds even for large $\sigma_0$.

Experimental evidence suggests that $\phi_s\sim0.3$
(see Balbus \& McKee~1982).
Throughout this paper we therefore assume $\phi_s=0.3$, 
consistent
with the use of subsonic approximation.

\subsubsection{On the Validity of the Non-Radiative Approximation}
\label{nonrad}

The fate of warm clouds
that are embedded in a hot ambient medium
depends of the delicate balance between the conductive heating 
and the radiative losses 
in the interface. Evaporation 
takes place if the radiative losses are small compared to
conductive heating. If, on the other hand, cooling overcomes the 
conductive heating, the evaporation stops, and instead the ambient 
medium  condenses onto the warm cloud.

McKee \& Cowie~(1977) used an approximate analytical method
to 
determine the critical cloud radii at which radiative losses
balance the conductive heating. They used fits to the
equilibrium cooling functions of solar metallicity gas to 
derive these critical radii as a function of the external
temperature and density.
McKee \& Cowie found that radiative losses are important 
only at low saturations, $\sigma_0\lesssim0.03/\phi_s$. 
Such 
small
values of $\sigma_0$ are generally out of the regime
``occupied'' by dwarf-galaxy scale halos (for which $\sigma_0>0.2$).

Our models differ from those 
presented by McKee \& Cowie.
First, we explicitly follow the
non-equilibrium ion fractions in the interface.
As mentioned above, the outflowing gas may be
underionized relative to gas in equilibrium at the same
temperature. Underionized gas tends to radiate more
efficiently (McCray~1987; Gnat \& Sternberg 2007), thus increasing the 
contributions
of radiative losses. Second, we include photoionization 
by an external background radiation field. This both
increases the ionization level in the gas and provides
heating, both effects acting to reduce the importance
of radiative losses. Finally, we 
assume gas metallicity of $0.1$~solar, thus reducing
the cooling efficiency.
The relative importance of these three effects must
be estimated numerically.

Using our numerical results we check, for each model,
whether the emissivity associated with the heat flux
is indeed everywhere larger than the radiative cooling.
Given our non-equilibrium ion fractions (including photoionization,
see section~\ref{ionization}), we use the cooling
functions included in Cloudy (version 07.02, Ferland et al.~1998)
to compute the local cooling efficiencies. 
We then compare these values with $\nabla\cdot \mathbf{q}$.
\begin{figure}[!h]
\epsscale{1.0}
\plotone{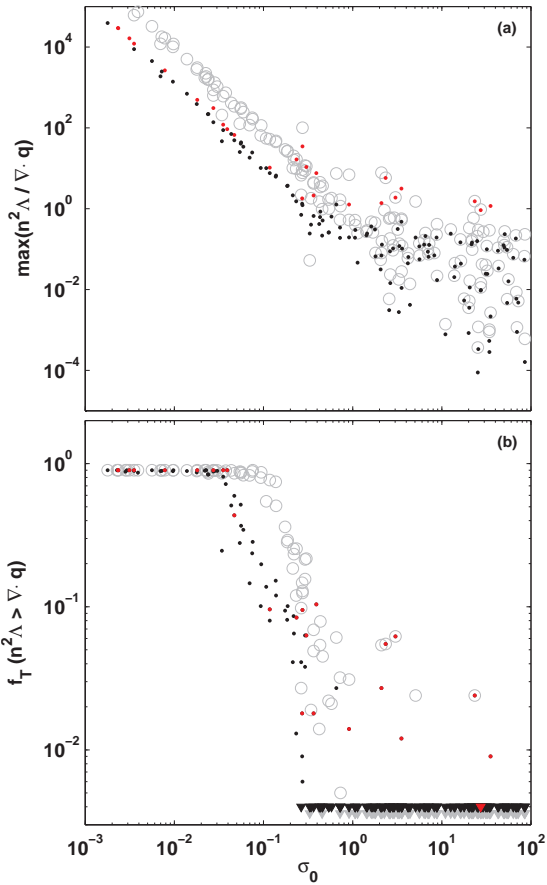}
\caption{(a) Maximum of $n^2\Lambda / \nabla\cdot \mathbf{q}$ versus
saturation parameter $\sigma_0$.
The black ($P/k_{\rm B}\le10^3$~cm$^{-3}$~K) 
and red ($P/k_{\rm B}=10^4$~cm$^{-3}$~K) points
are for $Z=0.1$~solar metallicity gas.
For $P/k_{\rm B}=10^4$~cm$^{-3}$~K the 
neutral fraction on the cloud surface is considerable ($x_{\rm HI}=0.574$).
The gray circles are for solar metallicity gas (all pressures).
(b) Fraction of interface temperature range over which the cooling
is locally larger than $\nabla\cdot \mathbf{q}$. Triangles indicate
models for which this fraction is smaller than $0.003$.}
\label{no-cool?}
\end{figure}
Figure~\ref{no-cool?} displays our results.
Panel~(a) shows the maximal ratio of the radiative to
conductive emissivity in each 
model\footnote{We only display the maximum for 
$T<0.9T_{\rm HIM}$. As the temperature approaches the
ambient temperature, the heat flux approaches zero but
the cooling does not.} as a function of the saturation.
Panel~(b) shows the fraction of the interface temperature 
over which the cooling is larger than the conductive
heating as a function of saturation.
The points (black and red) are for our 
assumed heavy element abundances of $Z=0.1$~times solar.
The gray circles show results for solar metallicity gas,
for a direct comparison with the critical $\sigma_0$ derived
by McKee \& Cowie (1977).

For $Z=0.1$~times solar, both panels show that for 
$\sigma_0\lesssim 0.2$ (for $\phi_s=0.3$) radiative losses
{\it are} significant and may not be neglected. 
For larger values of $\sigma_0$ the importance of radiative
cooling is generally smaller, and depends on the specific
parameters of the evaporating cloud. 

For $\sigma_0\gtrsim 0.2$, enhanced cooling and larger
values of $n^2\Lambda/\nabla\cdot \mathbf{q}$ are associated with
interfaces of higher gas densities, with larger neutral 
fractions at the cloud surface. We find that for our assumed
$10^4$~K-photoionized clouds, the neutral
fraction on the surface is significant when 
$P\gtrsim10^4$~cm$^{-3}$~K. For example, for pressures of  
$P=10^4$, $10^3$, and $100$~cm$^{-3}$~K, the neutral fractions
are $0.57$, $0.20$, and $0.03$, respectively.
Neutral gas cools very efficiently, and in underionized
flows the lingering and enhanced contribution of \ion{H}{1}
Ly$\alpha$ cooling significantly affects the energy balance. 

To study the impact of the surface ionization on the flow
dynamics, we distinguish between ``neutral cloud'' models for
which $P/k_{\rm B}\gtrsim10^4$~cm$^{-3}$~K (red points),
and ``ionized cloud'' models with $P/k_{\rm B}\le10^3$~cm$^{-3}$~K (black 
points)\footnote{$P/k_{\rm B}=10^3$~cm$^{-3}$~K corresponds to an ionization 
parameter of $10^{-5}$ on the cloud surface (see Section~3, equation~\ref{U}).}.
Figure~\ref{no-cool?} shows that for $\sigma_0\gtrsim 0.2$, radiative
losses are negligible for bounding pressures $\le10^3$~cm$^{-3}$~K.
However, in more neutral clouds 
cooling may be significant even at high saturations.

The fact that the limiting saturation parameter
that we find is similar to the one derived by McKee \& Cowie
(1977, $\sigma_0 = 0.03/\phi_s$)~-- despite the $10$ times
lower metallicity that we use, and the inclusion
of photoionization~-- implies that the cooling efficiencies
in our non-equilibrium interfaces are considerably larger than
their equilibrium counterparts. In fact, we find that Ly$\alpha$
from underionized He$^+$ dramatically increases the cooling efficiency in the
interface. We discuss this point in detail in section~\ref{results}.

For $Z=1$~solar, we find that radiative losses are important 
up to a critical saturation parameter of $\sigma_0\sim0.5$
(for $\phi_s=0.3$), and become less
significant for larger values of $\sigma_0$.
Since underionized He$^+$ is one of the dominant coolants in 
the interface, increasing the metallicity from a tenth-solar
to solar changes the critical saturation parameter by a factor 
smaller than $10$. Figure~\ref{no-cool?} shows that
this factor is $\sim3$.

For the tenth solar metallicity dwarf-galaxy-scale
models that we use for comparison with the observations 
of the high-velocity metal-ion absorbers, the radiative 
losses are, in fact, everywhere smaller than the conductive
heating.
\begin{figure*}
\epsscale{1.0}
\plotone{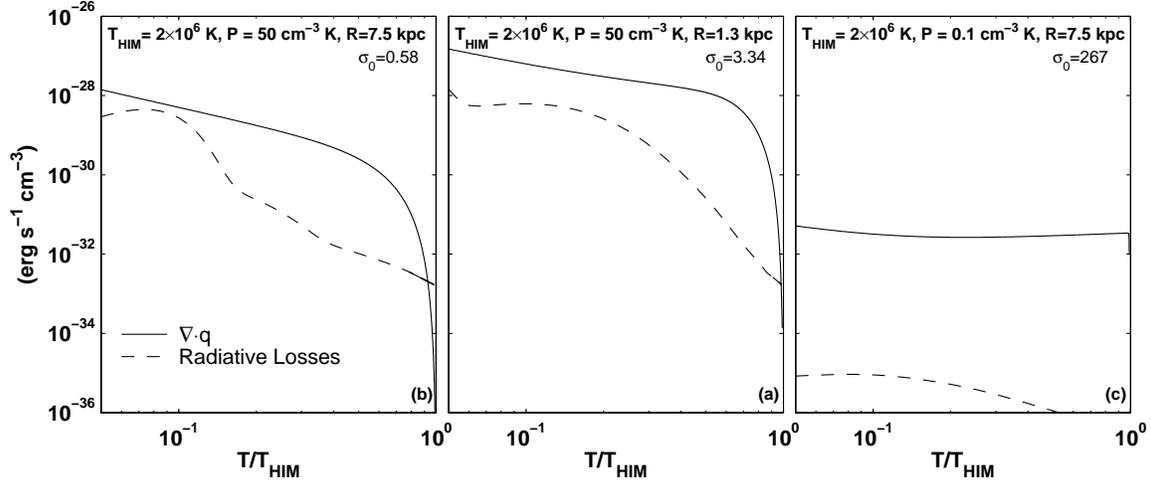}
\caption{Comparison of the radiative losses and $\nabla\cdot \mathbf{q}$ 
(erg~s$^{-1}$~cm$^{-3}$), for three interface models,
with $\sigma_0$ of $3.34$ (a), $0.58$ (b), and $267$ (c).}
\label{num-rad}
\end{figure*}
Some examples are shown in Figure~\ref{num-rad},
in which we compare $\nabla\cdot \mathbf{q}$ (solid curves) with
the radiative losses (dashed curves) in three interface
models.
The left hand panels are for interfaces embedded in a
high pressure corona, with $P/k_{\rm B} = 50$~cm$^{-3}$~K.
We show results for a small, marginally saturated cloud
($R=1.3$~kpc, $\sigma_0\sim3$) in panel (a), and
for a classical interface surrounding 
a larger cloud ($R=7.5$~kpc, $\sigma_0\sim0.6$) in panel (b).
In panel (c) we present a highly saturated interface, 
with $\sigma_0\sim270$, surrounding a $7.5$~kpc
cloud, embedded in a low pressure intergalactic medium
($P/k_{\rm B} = 0.1$~cm$^{-3}$~K).
In all cases, the radiative losses are everywhere smaller
than the emissivity associated with the heat flux.
These plots confirm that the significance of radiative losses
is larger for lower saturation parameters, and becomes
small at larger $\sigma_0$.

\subsection{Ionization}
\label{ionization}

Given the DB93 temperature profile, we solve
for the non-equilibrium abundances of the various
species in the interface. 
We follow the time-dependent 
ionization in the evaporating gas, given the initial photoionization
equilibrium state in the warm cloud.
We consider all ionization stages of the elements 
H, He, C, N, O, Si, and S.
The temperature-dependent
ionization and recombination processes that we include are
collisional ionization by thermal electrons (Voronov 1997),
radiative recombination (Aldrovandi \& Pequignot~1973; 
Shull \& van Steenberg~1982;
Landini \& Monsignori Fossi~1990; Landini \& Fossi~1991;
Pequignot, Petitjean, \& Boisson~1991; Arnaud \& Raymond~1992;
Verner et al.~1996),
dielectronic recombination (Aldrovandi \& Pequignot~1973;
Arnaud \& Raymond~1992;
Badnell et al.~2003, Badnell~2006;
Colgan et al.~2003, 2004, 2005;
Zatsarinny et al.~2003, 2004a, 2004b, 2005a, 2005b, 2006;
Altun et al.~2004, 2005, 2006;
Mitnik \& Badnell~2004), and
neutralization and ionization by charge transfer reactions with
hydrogen and helium atoms and ions
(Kingdon \& Ferland fits\footnote{See:
http://www-cfadc.phy.ornl.gov/astro/ps/data/cx
/hydrogen/rates/ct.html},
based on Kingdon \& Ferland~1996,
Ferland et al.~1997, Clarke et al.~1998, Stancil et al.~1998;
Arnaud \& Rothenflug~1985). 

The time-dependent equations for the ion abundance
fractions, $x_i$, of element $m$ in ionization stage $i$ are,
\begin{equation}
\label{ion-eq}
\begin{array}{l}
v\;\;d{x_i}/dr = x_{i-1}~~[q_{i-1}n_{\rm e} + \Gamma_{i-1}
+ k^{\rm H}_{\uparrow i-1}n_{\rm H^+}\\
\;\;\;\;\;\;\;\;\;\;\;\;\;\;\;\;\;\;\;\;\;\;\;\;\;\;\;\;\;\;
+ k^{\rm He}_{\uparrow i-1}n_{\rm He^+}]\\
\;\;\;\;\;\;\; + x_{i+1}~~[\alpha_{i+1}n_{\rm e} +
k^{\rm H}_{\downarrow i+1}n_{\rm H^0}
+ k^{\rm He}_{\downarrow i+1}n_{\rm He^0}] \\
\;\;\;\;\;\;\; - x_{i}~~[(q_{i} + \alpha_{i})n_{\rm e} + \Gamma_i +
k^{\rm H}_{\downarrow i}n_{\rm H^0}
+ k^{\rm He}_{\downarrow i}n_{\rm He^0}\\
\;\;\;\;\;\;\;\;\;\;\;\;\;\;\;\;\;
+ k^{\rm H}_{\uparrow i}n_{\rm H^+}
+ k^{\rm He}_{\uparrow i}n_{\rm He^+}] \ \ \ .
\end{array}
\end{equation}
In this expression, $v$ is the gas velocity at radius $r$
in the evaporating flow. The parameters $q_i$ and $\alpha_i$ 
are the rate coefficients
for collisional ionization and recombination (radiative plus
dielectronic), and
$k^{\rm H}_{\downarrow i}$, $k^{\rm H}_{\uparrow i}$,
$k^{\rm He}_{\downarrow i}$, and $k^{\rm He}_{\uparrow i}$
are the rate coefficients for charge transfer reactions
with hydrogen and helium that lead to ionization or neutralization.
The quantities
$n_{\rm H^0}$, $n_{\rm H^+}$, $n_{\rm He^0}$, $n_{\rm He^+}$,
and $n_{\rm e}$ are the particle densities (cm$^{-3}$)
for neutral hydrogen, ionized hydrogen, neutral helium, singly ionized
helium, and electrons, respectively. $\Gamma_{i}$ are the photoionization
rates of ions $i$, due to externally incident radiation. 

\begin{figure}[!h]
\epsscale{1.0}
\plotone{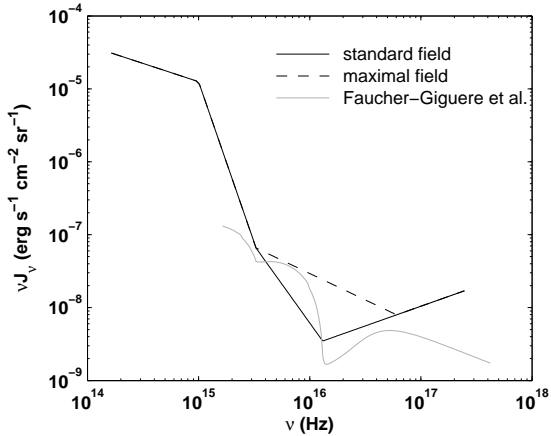}
\caption{
The present-day metagalactic radiation field.
The solid dark curve is for our standard field as represented by 
Sternberg et al.~(2002; see text and Equation~\ref{radn}),
the gray curve is for the Faucher-Giguere et al. (2009) field,
and the dashed curve is for the shallower "maximal field" considered
in Appendix~A (Gnat \& Sternberg~2004).
}
\label{radfield}
\end{figure}

For the external radiation, we use the Sternberg, McKee \& Wolfire~(2002) 
fit for the present-day metagalactic field,
\begin{equation}
\label{radn}
J_{\nu} = \left\{ \begin{array}{lcl}
   1.051\times10^2J_{\nu0}(\frac{\nu}{\nu_0})^{-1.5}&,& \frac{\nu}{\nu_0}~<~0.3\\
   J_{\nu0}(\frac{\nu}{\nu_0})^{-5.41}&,& 0.3~<~\frac{\nu}{\nu_0}~<~1\\
   J_{\nu0}(\frac{\nu}{\nu_0})^{-3.13}&,& 1~<\frac{\nu}{\nu_0}~<~4\\
   2.512\times10^{-2}J_{\nu0}(\frac{\nu}{\nu_0})^{-0.46}&,& 4~<~\frac{\nu}{\nu_0} \\
                  \end{array}\right.
\end{equation}
to compute $\Gamma_i$. In this expression
$J_{\nu0} = 2\times10^{-23}$ erg s$^{-1}$ cm$^{-2}$ Hz$^{-1}$ sr$^{-1}$,
and $\nu_0$ is the Lyman limit frequency. This fit, which we
plot in Figure~\ref{radfield}, is based on
observational constraints for the optical/UV and X-rays,
(Bernstein et al. 2002; Martin et al. 1991; Chen et al. 1997) 
and on theoretical models (Haardt \& Madau 1996) for the unobservable
radiation from the Lyman limit to $\sim0.25$~keV.

Recent observations and modeling have suggested that the spectral
slope of the metagalactic radiation may be shallower near the
Lyman limit, with $J_\nu$ varying as $\nu^{-1.8}$, and steepen gradually with
increasing energy (e.g.~Shull et al.~1999; Telfer et al.~2002;
Scott et al.~2004; Shull et al.~2004; Zheng et al.~2004; 
Faucher-Giguere et al.~2009). In Appendix~A we consider a radiation
field (also plotted in Figure~\ref{radfield}) that varies as $\nu^{-1.725}$ 
from the Lyman limit to $0.25$~keV, and investigate the sensitivity of our results to
the spectral slope. For reference, we also plot in Figure 4 the 
Faucher-Giguere et al.~(2009) model computation of the present-day 
metagalactic field.  In Appendix~A we show that the column densities
of the high ions (\ion{O}{6} and \ion{N}{5}) are insensitive to the
spectral slope at the Lyman limit, while the column densities
of lower ions (e.g.~\ion{Si}{3}) may be affected at a level of 
$<0.5$~dex, mostly for {\it low} external bounding pressures.

The present-day metagalactic radiation field.
The solid dark line is for the standard field as represented by 
Sternberg et al.~(2002; see text and Equation~\ref{radn}),
the gray line is for the Faucher-Giguere et al. (2009) field,
and the dashed line is for the shallower "maximal field" considered
in Appendix~A (Gnat \& Sternberg~2004).

For each element $m$, the ion fractions 
$x_i\equiv n_{i,m}/(n_{\rm H} A_m)$
satisfy
\begin{equation}
\sum x_i = 1
\end{equation}
where $n_{i,m}$ is the density (cm$^{-3}$) of ions in ionization stage $i$
of element $m$,
$n_{\rm H}$ is total hydrogen density, and $A_m$ is the
abundance of element $m$ relative to hydrogen. 
The sum is over all ionization stages of the element.

For the atomic elements that we include,
equations~(12) are a set of $54$ coupled
ordinary differential equations (ODEs).
For the initial conditions, we assume that the cloud surface
is in photoionization equilibrium with the metagalactic filed,
and has a temperature of $10^4$~K. We advance the solutions 
in small spatial steps $\Delta r$, where $r$ is the current
distance from the cloud center ($r>R$), which is associated with
a temperature $T(r)$ through the analytical solutions 
outlined by DB93.
We integrate equations~(12) over the interval $\Delta r$
using the same Livermore ODE solver\footnote{
See: http://www.netlib.org/odepack/} 
(Hindmarsh~1983) we used in Gnat \& Sternberg (2007).
We verified that our code converges to the equilibrium abundances 
given a constant temperature, by comparing our solutions to those
found by solving the set of algebraic equations for $x_i$ obtained
by setting $dx_i/dr=0$ in equation~(12), as appropriate for steady 
state.
We verified our time-dependent computations 
by comparing with previous non-equilibrium models 
(e.g.~Sutherland \& Dopita~1993), as discussed in detail in 
Gnat \& Sternberg~(2007).


\section{Metal Ions Fractions and Cooling in Evaporating Clouds}
\label{results}

We computed the ionization states of the 
elements H, He, C, N, O, Si, and S in thermally conductive interfaces
surrounding evaporating warm gas clouds.
We calculate the non-equilibrium ionization states as a function of
distance from the warm cloud surface.
We consider a grid of models with pressures between 
$0.1$ and $10^4$~cm$^{-3}$~K, cloud radii 
between $1$~pc and $100$~kpc, and ambient
(HIM) temperatures between $5\times10^5$ and $10^7$~K.
We assume heavy element abundance of $0.1$ times solar, as listed
in Table~\ref{met-table}.
We begin with a brief discussion of our
non-equilibrium conductive interfaces,
and later focus on models for dwarf-galaxy-scale objects.

\begin{deluxetable}{lr}
\tablewidth{0pt}
\tablecaption{$0.1$~solar metallicity abundances\label{Holmet}}
\tablehead{
\colhead{Element} & 
\colhead{Abundance}\\
\colhead{} &
\colhead{(X/H)$_{\odot}$} }
\startdata
Carbon   & $2.45\times10^{-5}$  \\
Nitrogen & $6.03\times10^{-6}$  \\
Oxygen   & $4.57\times10^{-5}$  \\
Silicon  & $3.24\times10^{-6}$  \\
Sulfur   & $1.38\times10^{-6}$  \\
\enddata
\label{met-table}
\end{deluxetable}

As the evaporating gas flows into the external, hotter, parts
of the interface layer, its overall ionization state gradually
increases. If the gas is heated faster than it is ionized, nonequilibrium
effects become significant and the gas remains ``underionized''
throughout the flow. This happens when the heating time-scale,
which is set by the flow time through the interface,
\begin{equation}
t_{\rm flow} \equiv \int{\frac{dr}{v}} \propto \frac{R^3 \rho_{\rm HIM}}
{\dot{m}},
\end{equation}
becomes short compared to the ionization time scale for ion $i$,
$t_{{\rm ion}, i}$.

In the absence of external photoionization, the ionization time scale
is given by 
\begin{equation}
t_{{\rm ion}, i}^{\rm no-rad}\simeq\frac{1}{n_e C_i(T)},
\end{equation}
where $C_i(T)$ is the collisional ionization coefficient for ion $i$ at
temperature $T$. The ionization state remains close to collisional
ionization equilibrium when $t_{\rm flow}/t_{{\rm ion}, i}\gg1$.
Nonequilibrium effects become significant when 
$t_{\rm flow}/t_{{\rm ion}, i}\le1$. With no photoionization, the
ratio of the flow to ionization time scale,
\begin{equation}
\label{timesratio}
\frac{t_{\rm flow}}{t_{{\rm ion}, i}^{\rm no-rad}} 
\propto \frac{T^{3/2}C_i(T)}{\sigma_0^2 \omega(\sigma_0)},
\end{equation}
is a function just of the saturation parameter and ambient
temperature. The ion fractions and integrated metal-ion column
densities then depend on just the two parameters $\sigma_0$ and $T_{\rm HIM}$.

Here we also include photoionization by the
external metagalactic radiation field, which increases the
level of ionization in the gas, and provides heating.
The ionization parameter at the surface of a photoionized,
$10^4$~K cloud is given by,
\begin{equation}
\label{U}
U = \frac{4\pi}{n_{\rm H} c} \int_{\nu_0}^{\infty}\frac{J_\nu}{h\nu}d\nu=
4.3\times10^{-7}~n_{\rm H}^{-1},
\end{equation}
where $\nu_0$ is the Lyman limit frequency, $c$ is the speed of light,
and $n_{\rm H}$ is the hydrogen density (cm$^{-3}$) at the cloud surface.
Photoionization is more efficient in lower density gas.
With photoionization included, the ionization states and integrated
column densities depend on all three interface parameters,
$T_{\rm HIM}$, $P/k_{\rm B}$, and $R$. 
In general, we find that the ionization lag is so
great that despite the inclusion of photoionization, the 
gas remains highly underionized, not only compared to
photoionization equilibrium, but also compared to collisional
ionization equilibrium (CIE).

Because the energy losses are dominated by
atomic and ionic emission lines, the nonequilibrium cooling rates
also differ (are enhanced for underionized gas) compared to
equilibrium cooling. The lower ionization species that remain present
at each temperature, most significantly He$^+$, 
are more efficiently excited
by the thermal electrons 

\begin{figure}[!h]
\epsscale{1.0}
\plotone{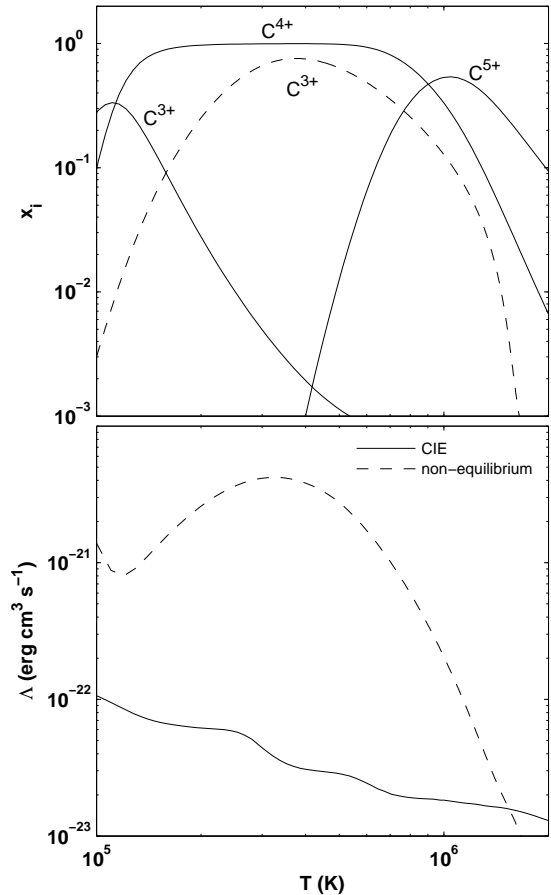}
\caption{Underionized gas in a model with $R=1.3$~kpc, 
$P_{\rm HIM}/k=50$~cm$^{-3}$~K, $T_{\rm HIM}=2\times10^6$~K.
(a) Non-equilibrium C$^{3+}$ ion fraction (dashed curve). The 
carbon CIE ion fractions are shown by the solid curves for comparison.
(b) Enhanced cooling in the underionized interface. The dashed 
curve shows the non-equilibrium cooling efficiency, and the solid 
curve shows the CIE cooling efficiency (for $Z=0.1$~solar).
}
\label{noneq}
\end{figure}

We demonstrate the effect of ``under-ionization''
in Figure~\ref{noneq}, for a model with
$R=1.3$~kpc, $P_{\rm HIM}/k=50$~cm$^{-3}$~K,  and
$T_{\rm HIM}=2\times10^6$~K ($\sigma_0=3.34$).
In the upper panel we show the nonequilibrium C$^{3+}$ ion fraction
versus temperature (dashed curve), and the CIE ion fractions 
of C$^{3+}$-C$^{5+}$ (solid curves, Gnat \& Sternberg~2007). 
The nonequilibrium C$^{3+}$ abundance peaks at a temperature
of $\sim4\times10^5$~K, whereas the CIE C$^{3+}$ abundance
peaks at $\sim10^5$~K .
For CIE, the dominant ion at 
$4\times 10^5$~K is C$^{4+}$.

The modified ionization states enhance the cooling efficiency in this
interface. We computed the local cooling efficiencies using
Cloudy (version 07.02, Ferland et al.~1998) given our nonequilibrium
ion fractions. This is shown in panel (b).
The CIE cooling efficiency is shown by the solid curve (Gnat \& Sternberg~2007).
The non-equilibrium cooling efficiency (dashed curve)
is an order of magnitude larger in this model,
over a significant temperature range.
This is mostly due to He$^+$ which persists to much higher temperatures
in the nonequilibrium gas compared to CIE.

\begin{figure}[!h]
\epsscale{1.0}
\plotone{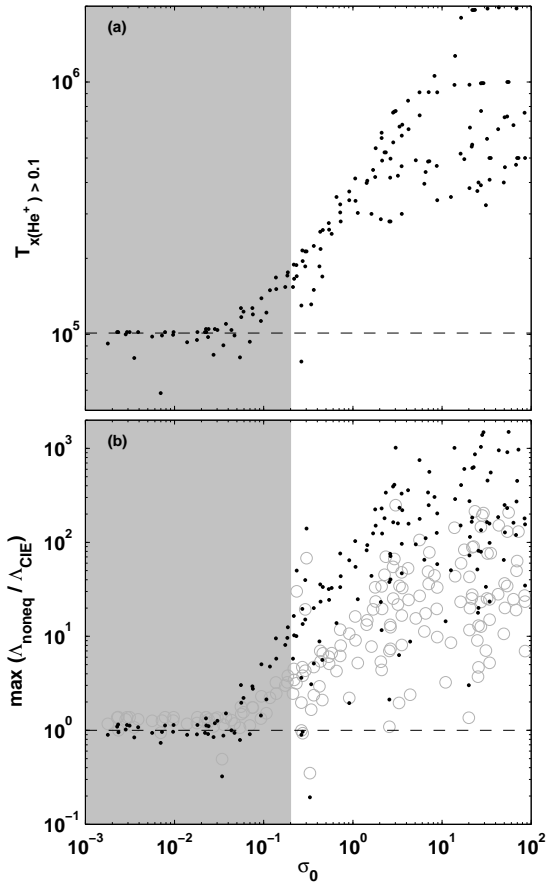}
\caption{Departures from equilibrium in conductive flows.
(a) The temperature to which He$^+$ persists, with a fractional
abundance larger than $10\%$, in the non-equilibrium gas. The CIE value
is shown by the dashed line. These results are independent of $Z$.
(b) Maximal ratio of non-equilibrium to equilibrium 
cooling versus saturation for $Z=0.1$~solar (black points),
and for $Z=1$~solar (gray circles).
Models inside the shaded areas are not self-consistent, as cooling
may not be neglected (see section~\ref{nonrad}).
}
\label{noneq-sigma}
\end{figure}
In Figure~\ref{noneq-sigma} we demonstrate the
effect of non-equilibrium ionization for our complete set of models.
In panel (a) we show the ionization lag for He$^+$.
For each model, we plot the temperature to which He$^+$
persists, with a fractional abundance larger than $10\%$,
versus its saturation parameter. These values are
independent of the gas metallicity.
The corresponding
CIE value ($\sim10^5$~K)
is shown by the dashed line.
The shaded region indicates
the range of saturation parameters for which no self-consistent 
evaporating solution exists (for $Z=0.1$~solar).
In this region, radiative cooling may not be neglected
(see section~\ref{nonrad}), and 
results in condensation.
For the evaporating models, departures from equilibrium
ionization can clearly be seen. He$^+$ persists to higher
temperatures than in CIE, despite the 
contribution of external photoionization by the metagalactic radiation field.
For highly saturated models, in which the temperature profile 
is extremely steep, He$^+$ may persist to $T\sim10^6$~K.

The impact of these extended ion distributions on the cooling
efficiencies is shown in panel (b).
For each model we plot the maximal ratio of the
non-equilibrium- to CIE-cooling efficiencies
through the interface. The black points are for $Z=0.1$~times 
solar, and the gray circles are for $Z=1$.
For the evaporating models,
cooling is indeed enhanced by factors $\sim5$ to $10^3$ 
relative to the CIE cooling efficiency.

These enhanced cooling efficiencies explain the results presented in
section~\ref{nonrad} (Figure~\ref{no-cool?}). 
We found that for solar and $0.1$~solar metallicity gas, 
radiative cooling has a small impact on the dynamics of
the flow for saturation parameters $\sigma_0>0.5$ and 
$\sigma_0>0.2$, respectively.
Our results indicate that non-equilibrium effects significantly
increase the value of $\sigma_0$ at which radiative effects set in.
For comparison, in CIE radiative losses in solar metallicity gas become
significant for $\sigma_0<0.1$ (for $\phi_s=0.3$, McKee and Cowie~1977),
a factor of $\sim5$ lower than for nonequilibrium cooling.
The reason for this is that the cooling efficiencies in
the underionized gas are larger, due to the 
enhanced contributions of He$^+$ and metal-ion 
resonance-line cooling.

\begin{figure*}
\epsscale{1.0}
\plotone{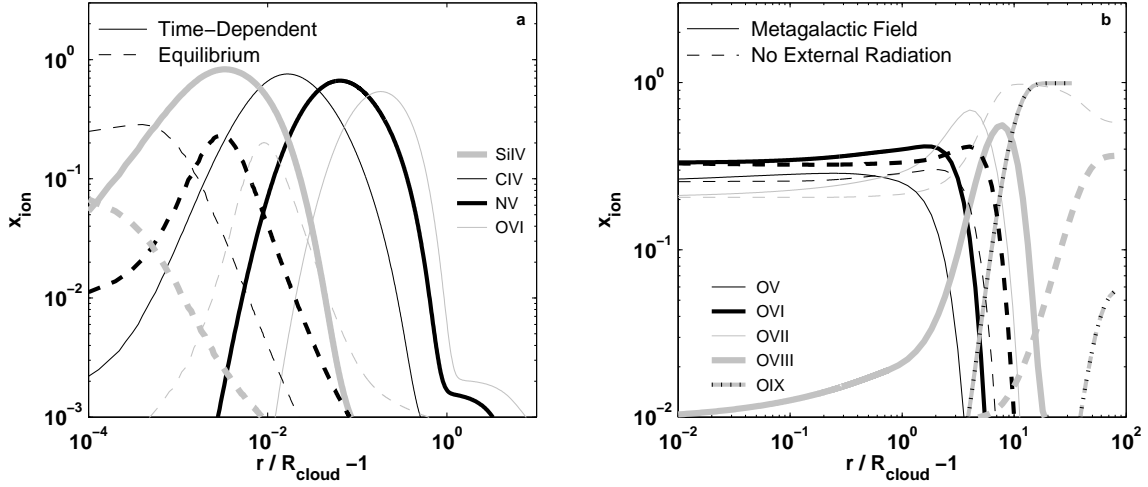}
\caption{
(a) Non-equilibrium ionization in a high-pressure model with $R=1.3$~kpc, 
$P_{\rm HIM}/k=50$~cm$^{-3}$~K, and $T_{\rm HIM}=2\times10^6$~K.
The solid and dashed lines show the non-equilibrium and equilibrium
abundances, respectively. The non-equilibrium abundances reach their 
peaks further away from the cloud center and span larger path-lengths.
(b) The effect of photoionization by an external
radiation field, for low-pressure model with $R=7.5$~kpc, 
$P_{\rm HIM}/k=0.1$~cm$^{-3}$~K, and $T_{\rm HIM}=2\times10^6$~K.
The solid and dashed lines show the abundances including photoionization
by the metagalactic radiation file, and in the absence of external radiation,
respectively. Photoionization increases the level of ionization in the interface.
}
\label{CIchecks}
\end{figure*}

Does non-equilibrium ionization modify the observational
signatures of evaporating clouds?
In Figure~\ref{CIchecks}(a) we compare the equilibrium (dashed curves)
and non-equilibrium (solid curves) ion distributions versus
{\it position} in the evaporating layers,
for a model with $R=1.3$~kpc, $P_{\rm HIM}/k=50$~cm$^{-3}$~K,
and $T_{\rm HIM}=2\times10^6$~K.
We show the abundances of C$^{3+}$, N$^{4+}$, O$^{5+}$, 
and Si$^{3+}$ including the effect of photoionization.
The difference between the equilibrium and non-equilibrium
curves is apparent: the non-equilibrium abundances reach their 
peaks further away from the cloud center (at higher temperatures)
as the gas remains underionized, and they span considerably
larger path-lengths than the equilibrium abundances.
These larger path lengths result in enhanced column densities.

Finally, in panel (b) of Figure~\ref{CIchecks} we demonstrate
the impact of photoionization by the metagalactic radiation field 
on the ionization states.
Here we show a low-pressure model with 
$P_{\rm HIM}/k=0.1$~cm$^{-3}$~K, $T_{\rm HIM}=2\times10^6$~K,
and $R=7.5$~kpc.
We compare the ionization state including photoionization
by the metagalactic radiation field to the ionization in the
absence of any external radiation. The presence of the external 
field increases the level of ionization both in the interface
and in the surrounding HIM.

\begin{deluxetable*}{lllllllll}
\tablewidth{0pt}
\tablecaption{Model Parameters}
\tablehead{
\colhead{$T_{\rm HIM}$}&
\colhead{$R$}&
\colhead{$P/k_{\rm B}$}&
\colhead{$U^{\ast}$}&
\colhead{}&
\colhead{$\dot{m}_{\rm cl}$}&
\colhead{}&
\colhead{}&
\colhead{$m_7/\dot{m}^{\ddagger}$} \\
\colhead{(K)} &
\colhead{(kpc)}&
\colhead{(cm$^{-3}$ K)}&
\colhead{}&
\colhead{$\sigma_0$}&
\colhead{(g s$^{-1}$)}&
\colhead{$\omega$}&
\colhead{max($M$)$^{\dagger}$}&
\colhead{Gyr}
}
\startdata
$10^6$              & $3$         & $0.1$       & $9.7\times10^{-2}$ & $85.62$        & $6.45\times10^{22}$       & $0.03$       & $0.60$ & $330$  \\
$10^6$              & $3$         & $1$         & $9.7\times10^{-3}$ & $8.83$         & $6.64\times10^{22}$       & $0.20$       & $0.58$ & $47.8$ \\
$10^6$              & $3$         & $5$         & $1.9\times10^{-3}$ & $1.81$         & $6.80\times10^{22}$       & $0.63$       & $0.38$ & $14.8$ \\
$10^6$              & $3$         & $10$        & $9.7\times10^{-4}$ & $0.91$         & $6.86\times10^{22}$       & $0.80$       & $0.23$ & $11.6$ \\
$10^6$              & $3$         & $20$        & $4.8\times10^{-4}$ & $0.46$         & $6.93\times10^{22}$       & $0.89$       & $0.13$ & $10.3$ \\
$10^6$              & $3$         & $50$        & $1.9\times10^{-4}$ & $0.19$         & $7.02\times10^{22}$       & $0.96$       & $0.06$ & $9.4$  \\
${\bf 2\times10^6}$ & ${\bf 1.3}$ & ${\bf 50}$  & $1.9\times10^{-4}$ & ${\bf 3.34}$   & ${\bf 1.67\times10^{23}}$ & ${\bf 0.43}$ & ${\bf 0.50}$ & ${\bf 8.8}$\\
$2\times10^6$       & $3$         & $0.1$       & $9.7\times10^{-2}$ & $666.8$        & $3.56\times10^{23}$       & $0.01$       & $0.60$ & $178$  \\
$2\times10^6$       & $3$         & $1$         & $9.7\times10^{-3}$ & $68.70$        & $3.66\times10^{23}$       & $0.04$       & $0.60$ & $43.4$ \\
$2\times10^6$       & $3$         & $5$         & $1.9\times10^{-3}$ & $14.04$        & $3.74\times10^{23}$       & $0.13$       & $0.59$ & $13.1$ \\
$2\times10^6$       & $3$         & $10$        & $9.7\times10^{-4}$ & $7.08$         & $3.77\times10^{23}$       & $0.24$       & $0.57$ & $7.0$  \\
$2\times10^6$       & $3$         & $20$        & $4.8\times10^{-4}$ & $3.58$         & $3.81\times10^{23}$       & $0.41$       & $0.51$ & $4.1$  \\
$2\times10^6$       & $3$         & $50$        & $1.9\times10^{-4}$ & $1.45$         & $3.86\times10^{23}$       & $0.69$       & $0.33$ & $2.4$  \\
${\bf 2\times10^6}$ & ${\bf 4.2}$ & ${\bf 1}$   & $9.7\times10^{-3}$ & ${\bf 49.40}$  & ${\bf 5.09\times10^{23}}$ & ${\bf 0.05}$ & ${\bf 0.60}$ & ${\bf 24.9}$ \\
$2\times10^6$       & $7.5$       & $0.1$       & $9.7\times10^{-2}$ & $266.7$        & $8.87\times10^{23}$       & $0.01$       & $0.60$ & $71.6$ \\
$2\times10^6$       & $7.5$       & $1$         & $9.7\times10^{-3}$ & $27.48$        & $9.14\times10^{23}$       & $0.08$       & $0.60$ & $8.7$  \\
$2\times10^6$       & $7.5$       & $5$         & $1.9\times10^{-3}$ & $5.62$         & $9.34\times10^{23}$       & $0.29$       & $0.56$ & $2.3$  \\
$2\times10^6$       & $7.5$       & $10$        & $9.7\times10^{-4}$ & $2.83$         & $9.43\times10^{23}$       & $0.49$       & $0.47$ & $1.4$ \\
$2\times10^6$       & $7.5$       & $20$        & $4.8\times10^{-4}$ & $1.43$         & $9.52\times10^{23}$       & $0.70$       & $0.33$ & $0.95$ \\
$2\times10^6$       & $7.5$       & $50$        & $1.9\times10^{-4}$ & $0.58$         & $9.64\times10^{23}$       & $0.87$       & $0.16$ & $0.77$ \\
$10^7$              & $3$         & $10$        & $9.7\times10^{-4}$ & $831.1$        & $1.98\times10^{25}$       & $0.007$      & $0.60$ & $4.7$  \\
$10^7$              & $3$         & $20$        & $4.8\times10^{-4}$ & $419.3$        & $2.00\times10^{25}$       & $0.008$      & $0.60$ & $4.0$  \\
$10^7$              & $3$         & $50$        & $1.9\times10^{-4}$ & $169.7$        & $2.02\times10^{25}$       & $0.02$       & $0.60$ & $1.9$  \\
$10^7$              & $7.5$       & $10$        & $9.7\times10^{-4}$ & $332.4$        & $4.95\times10^{25}$       & $0.01$       & $0.60$ & $1.3$  \\
$10^7$              & $7.5$       & $20$        & $4.8\times10^{-4}$ & $167.7$        & $4.99\times10^{25}$       & $0.02$       & $0.60$ & $0.75$ \\
$10^7$              & $7.5$       & $50$        & $1.9\times10^{-4}$ & $67.9$         & $5.05\times10^{25}$       & $0.04$       & $0.60$ & $0.35$ \\
\enddata 
\tablecomments{\\Results are shown for $\phi_s=0.3$.\\
$\ast$ -- The ionization parameter $U=9.7\times10^{-3}~(P/k_{\rm B})^{-1}$.\\
$\dagger$ -- The maximal Mach number in the interface, max$(M)$. \\
$\ddagger$ -- The evaporation time, $t_{\rm evap} \equiv m_7/\dot{m}$~Gyr,
where $m_7$ is the warm cloud gas mass in units of $10^7$~M$_\odot$,
and $\dot{m} = \omega \dot{m_{\rm cl}}$.\\
The highlighted lines correspond to the models considered in 
sections~4.1 and 4.2 below.
}
\label{models-table}
\end{deluxetable*}

\subsection{Metal Ion Fractions in Dwarf-Galaxy-Scale Objects}
We now focus on the ion fractions in conductive interfaces
surrounding dwarf galaxy-scale halos for comparison with
the observed metal-ion absorbers.
As a guide to the physical properties of dwarf-scale
halos, we rely on the results presented in GS04, 
where we considered median halos with a
virial mass $\sim2\times10^9$~M$_\odot$.
For the dwarf-scale halos, we assume ambient temperatures of $10^6$~K,
$2\times10^6$~K, and $10^7$~K,
gas cloud radii of $3$~kpc, $4.2$, and $7.5$~kpc, and bounding pressures
in the range $0.1-50$~cm$^{-3}$~K. 
For bounding pressures in this range, the hydrogen in the $10^4$~K 
photoionized clouds is fully ionized at the surface, and the ionization 
parameter may be written as $U=9.7\times10^{-3}~(P/k_{\rm B})^{-1}$.

In Table~\ref{models-table} we list the input parameters for the 
dwarf-scale models that we consider.
These include the ambient temperature $T_{\rm HIM}$, the cloud radius $R$,
and the gas pressure $P/k_{\rm B}$.
We also list in Table~\ref{models-table}
the ionization parameter $U$, saturation parameter $\sigma_0$,
the classical (diffusive) mass loss rate $\dot{m}_{\rm cl}$, the mass-loss
suppression factor $\omega = \dot{m} / \dot{m}_{\rm cl}$, the maximal
Mach number in the flow, max$(M)$, and the evaporation time scale,
$t_{\rm evap} \equiv m_7/\dot{m}$, for clouds with warm gas masses
of $10^7$~M$_\odot$.
All the values listed in Table~\ref{models-table} are computed 
assuming $\phi_s=0.3$.

\begin{figure}[!h]
\figurenum{8.1}
\epsscale{1.0}
\plotone{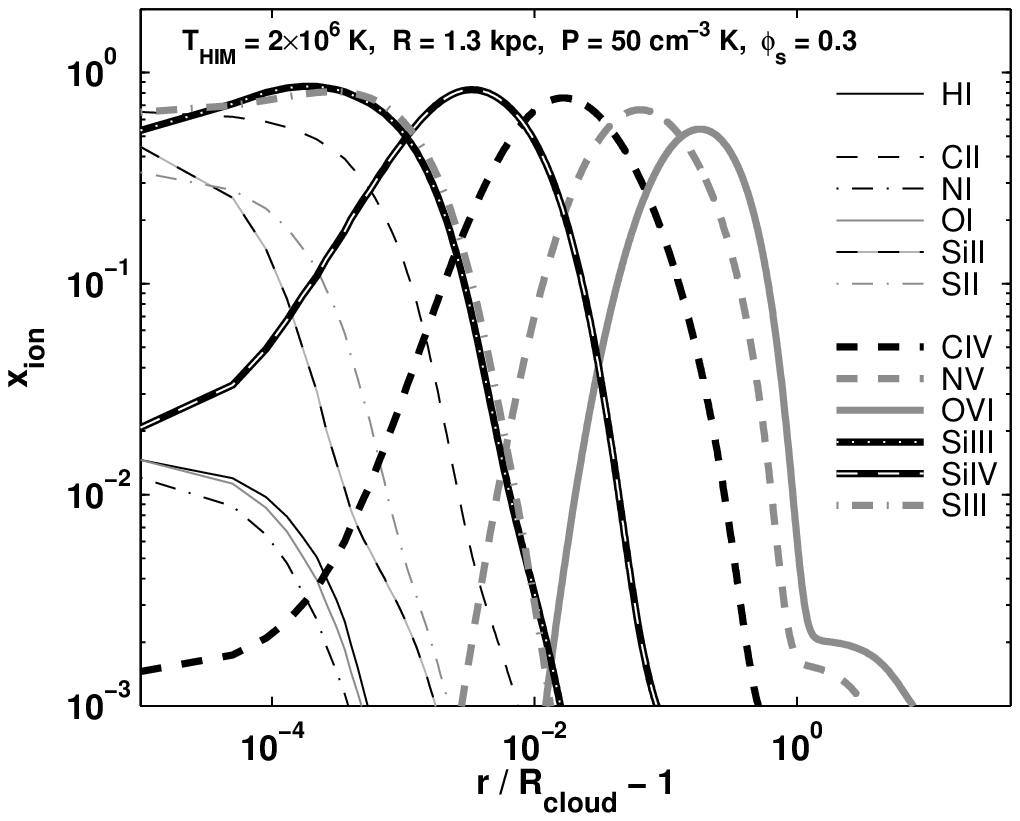}
\label{dwarf-fig}
\caption{Ion fractions as a function of scaled radius
$r/R-1$, in conductive interfaces surrounding dwarf-galaxy-halos. 
The ambient temperature, $T_{\rm HIM}$, cloud radius, $R$, 
and bounding pressure, $P$ are indicated in each panel.
In Fig~8.1 the non-equilibrium ion fractions are for 
$T_{\rm HIM}=2\times10^6$~K, $R=1.3$~kpc, and $P/k_{\rm B}=50$~cm$^{-3}$~K.}
\end{figure}

\begin{figure}[!h]
\figurenum{8.2}
\label{chvc-fig}
\epsscale{1.0}
\plotone{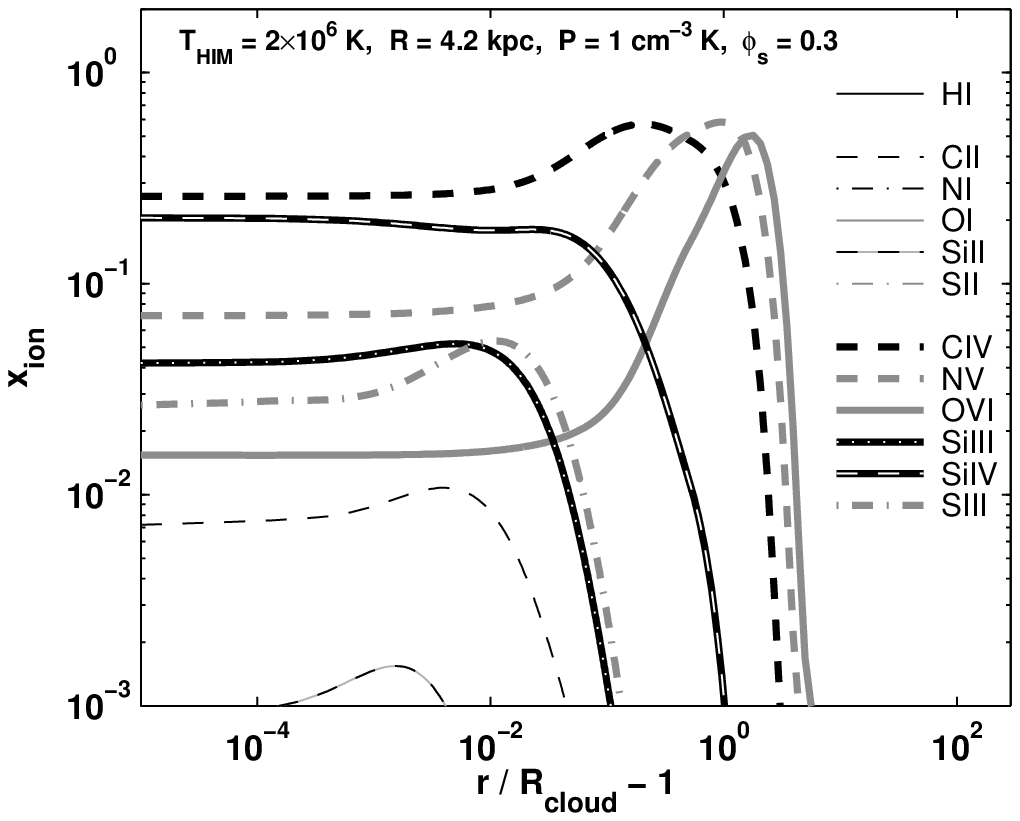}
\caption{Same as figure 8.1 but for 
$T_{\rm HIM}=2\times10^6$~K, $R=4.2$~kpc, and $P/k_{\rm B}=1$~cm$^{-3}$~K.}
\end{figure}

In Figure set~8,
we present the ion fractions as a function of
distance from the cloud surface for the grid of dwarf galaxy-scale
models. The ambient temperature, $T_{\rm HIM}$, cloud radius, $R$, 
and bounding pressure, $P/k_{\rm B}$ are indicated in each panel.
Figure~8
displays the ion fractions of the low ions
\ion{C}{2}, \ion{N}{1}, \ion{O}{1}, \ion{Si}{2}, and \ion{S}{2},
and the high-ions \ion{C}{4}, \ion{N}{5}, \ion{O}{6},
\ion{Si}{3}, \ion{Si}{4}, and \ion{S}{3}. These are the ions that
were measured
by Sembach et al. (1999, 2000), Collins et al. (2004), and
Fox et al.~(2005) in the high-velocity metal ion absorbers
(see e.g.~Table~\ref{collins}).

\setcounter{figure}{8}

\section{Metal-Absorption Column Densities}
\label{columns}

In this section we present computations of the integrated metal-ion
column densities that are produced in the conductively evaporating 
layers. 

The characteristic column density in an interface,
$N_{\rm HIM} \equiv n_{\rm HIM} R$, is given by (see equation~\ref{sigma0}),
\begin{equation}
N_{\rm HIM} = 2\times \frac{2\kappa_{\rm HIM} T_{\rm HIM}}
{25\phi_s \mu_{\rm HIM} c^3_{\rm HIM}} = 
1.7\times10^{17}\; \frac{T_{6,{\rm HIM}}^2}{\sigma_0 \phi_{0.3}}\;
\;{\rm cm}^{-2},
\end{equation}
where $\mu_{\rm HIM}$ is the mean mass per particle,
$T_{6,{\rm HIM}}$ is the HIM temperature in units of $10^6$~K,
and $\phi_{0.3}=\phi_{\rm s} / 0.3$.
This equation shows that in general, models with lower
saturation parameters
produce larger columns.
The column density of a given ionic species is related to this
characteristic column by an ``ionization correction'', that
depends on the interface parameters.

For gas in collisional ionization equilibrium, the local ion
fractions are a function of the temperature only.
The ionization correction and implied metal-ion column densities
therefore depend only on $\sigma_0$ (through its control of
the temperature profile, see Figure~\ref{tau-y}) and on the
ambient temperature, $T_{\rm HIM}$. 
However, the functional dependence on 
$\sigma_0$ and $T_{\rm HIM}$ may be complicated.

For non-equilibrium conditions, departures from equilibrium ionization
and the ion fractions at a given temperature depend on the ratio of
the heating time-scale and the ionization time-scale. 
Equation~\ref{timesratio} shows that
in the absence of photoionization,  the non-equilibrium abundances
and columns still depend only on  $\sigma_0$ and $T_{\rm HIM}$.
The inclusion of photoionization by the metagalactic radiation field
introduces an implicit dependence on the gas density (or ionization
parameter, see equation~\ref{U}). 
The non-equilibrium column densities are therefore
{\it general} functions of all three interface parameters\footnote{We find that the
column densities listed in Table~\ref{columns-table} cannot be
expressed as simple power-law fits in $T_{\rm HIM}$, $P/k_{\rm B}$,
and $R$.}.

Here we wish to find conditions that yield
metal-ion column densities comparable to those observed in the
high-velocity ionized absorbers, including the high observed
\ion{O}{6} column ($N_{\rm O\,VI}\sim10^{14}$~cm$^{-2}$).
We  focus on the column densities of the high ions \ion{C}{4},
\ion{N}{5}, \ion{O}{6}, \ion{Si}{3}, \ion{Si}{4}, and \ion{S}{3},
as a function of the impact parameter from the cloud center,~$b$.
In Table~\ref{columns-table}, we list the (two-sided) central
($b=0$) and maximal column densities for the different models
considered in section~3.
For the evaporating gas, the line-of-sight column density of ion $i$
of element $m$ observed at an impact parameter $b$ is,
\begin{equation}
\label{Ni}
N_i=2\int_{{\rm max}(b,R)}^\infty \frac{n_{\rm H}(r) A_m x_i(r)rdr}{\sqrt{r^2-b^2}},
\end{equation}
where $A_m$ is the abundance of element $m$ relative to hydrogen, 
$n_{\rm H}(r)$ is the position dependent hydrogen density, 
and $x_i(r)$ are the position-dependent ion fractions, 
as presented in section~\ref{results}.
The maximal columns are evaluated for each ion independently,
and occur at different impact parameters for each ion and model.

The results presented in Table~\ref{columns-table} are computed for
$Z=0.1$~times solar.
Since the fractional ion abundances 
are independent of metallicity, the integrated
columns are proportional to $Z$. (However,
as discussed in~\ref{nonrad}, the range
of $\sigma_0$ values for which a self-consistent evaporating
solution can be found {\it is} a function of $Z$.)

Table~\ref{columns-table} lists the central column densities
produced in the interface layers. When observing
evaporating clouds, the observed absorption line properties will
generally depend not only on the conductive interface, but also on the
ionization properties of the embedded photoionized clouds.
In the text that follows we discuss, as illustrative
examples, the total (cloud $+$ interface) column densities that
arise in two specific models. First we consider a low-ionization cloud
embedded in a high-pressure corona, with $P/k_{\rm B}=50$~cm$^{-3}$~K (\S4.1), 
and then we study a more highly-ionized cloud embedded a low-pressure medium,
with $P/k_{\rm B}=1$~cm$^{-3}$~K (in \S4.2). The lines representing these
two illustrative models are highlighted in Table~\ref{models-table}.

\subsection{A Low-Ionization Cloud Embedded in a High Pressure Medium}

As an example for a low-ionization photoionized cloud embedded in a
high-pressure corona, we rely on the physical properties of the 
``CHVC-scale'' model presented in GS04 (``model A'' in GS04). 
This model represents a $10^8$~M$_\odot$ dark-matter halo that contains 
a warm gas mass of $1.1\times10^6$~M$_\odot$, which is subjected to an
external bounding pressure, $P/k_{\rm B}=50$~cm$^{-3}$~K.
The resulting photoionized cloud has a radius $R=1.3$~kpc.
The ionization parameter on the cloud surface is low, 
$U=1.9\times10^{-4}$, and high ions are thus inefficiently
produces in this cloud.
Here we assume that the cloud is surrounded by a hot Galactic corona,
with a temperature $T_{\rm HIM}=2\times10^6$~K, and photoionized by the
metagalactic radiation field.

These parameters imply that the cloud evaporates in a 
marginally-saturated flow, with $\sigma_0\simeq3.3$.
The mass loss rate is suppressed by a factor $0.43$ relative to
that in a classical diffusive flow, and the maximal Mach number in the flow
is $0.50$.

Figure~8.1 displays the ion fractions that
arise in the conductive
interface in this model. 
In evaluating the absorption-line signatures that arise from
the evaporating cloud, we use the abundances
given in GS04 for ``model A'' (Figure 3 in GS04)
for the ion fractions within the photoionized cloud,
and the ion fractions computed here and presented in Figure~8.1
for the evaporating layers. 
We then evaluate the projected column densities as a function of 
impact parameter from the cloud center. These column densities
include the contributions
of both the photoionized cloud {\it and} the conductive interface
surrounding it.

\begin{figure}[!h]
\epsscale{1.0}
\plotone{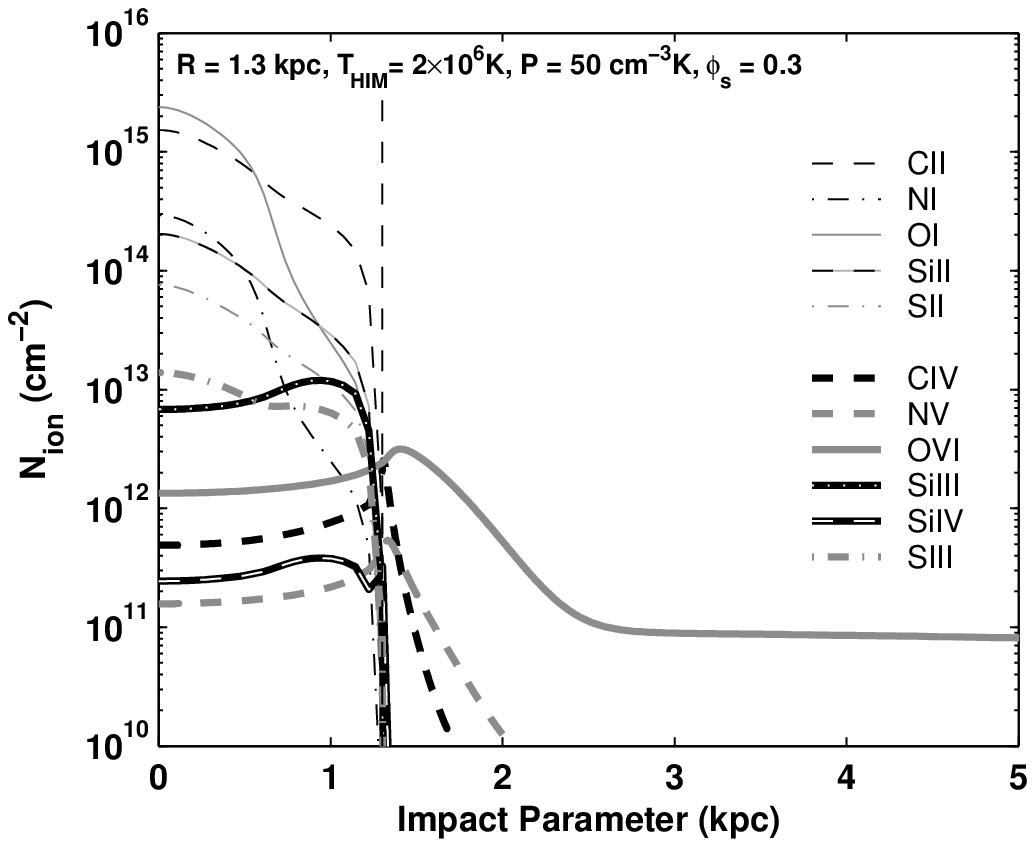}
\caption{Projected column densities as a function of impact parameter
in a high-pressure model. For this model $R=1.3$~kpc, 
$P/k_{\rm B}=50$~cm$^{-3}$~K, and $T_{\rm HIM}=2\times10^6$~K.}
\label{chvc-columns}
\end{figure}

Figure~\ref{chvc-columns} shows the projected column densities in our
evaporating high-pressure model.
The ionization in this conductive interface 
results in a maximal \ion{O}{6} column density of 
$3.2\times10^{12}$~cm$^{-2}$ (compared to 
$4.5\times10^{6}$~cm$^{-2}$ in the purely photoionized cloud),
and a \ion{C}{4} column density of $2.5\times10^{12}$~cm$^{-2}$
(compared to $2.6\times10^{11}$~cm$^{-2}$ in the purely 
photoionized cloud).
The evaporating layers increase the \ion{O}{6}
column by $\sim6$ orders of magnitude, and the \ion{C}{4}
column by $\sim1$ order of magnitude relative
to the purely photoionized cloud discussed in GS04.
However, for this model, the enhanced columns still 
do not bring the model into harmony with the observations 
of the high-velocity metal-ion absorbers (see e.g.~Table~\ref{collins}).

For pressure supported photoionized clouds
in high-pressure environments ($P\sim50$~cm$^{-3}~K$), the ionization 
parameter is generally too low to allow for efficient production of high
ions such as C$^{3+}$ and O$^{5+}$. The high-ion column densities are 
therefore dominated by ions produced in the evaporating layers, and are
not significantly affected by the details of the cloud's ionization state.

\subsection{A Highly-ionized Cloud Embedded in a Low Pressure Medium}

As an illustrative example for the physical properties of a 
more highly ionized photoionized cloud embedded in a low-pressure
medium, we rely on the standard dwarf-scale model presented in GS04
(``model C'' in GS04), based on the observed properties
of Local Group dwarf galaxies Leo~A and Sag~DIG (Sternberg et al.~2002). 
This model represents a $2\times10^9$~M$_\odot$ dark-matter halo
that contains a total warm gas mass of $1.6\times10^7$~M$_\odot$, and is
subjected to an external bounding pressure $P/k_{\rm B}=1$~cm$^{-3}$~K.
The resulting cloud has a radius $R\sim4.2$~kpc.
The ionization parameter on the surface of this cloud is higher,
$U=9.7\times10^{-3}$, and high ions such as C$^{3+}$ and N$^{4+}$
are efficiently produced in the ionized envelopes of this cloud.
Here we assume that the halo is surrounded by a hot
intergalactic medium, with a temperature $T_{\rm HIM}=2\times10^6$~K,
and is photoionized by the metagalactic radiation field.

\begin{deluxetable*}{lllllllllll}
\tablewidth{0pt}
\tablecaption{Column Densities}
\tablehead{
\colhead{T$_{\rm HIM}$}&
\colhead{R}&
\colhead{P/k$_{\rm B}$}&
\colhead{$\sigma_0$}&
\colhead{}&
\colhead{C~IV}&
\colhead{N~V}&
\colhead{O~VI} &
\colhead{Si~III} &
\colhead{Si~IV}&
\colhead{S~III}}
\startdata
$10^{6}$        &  $3$    &  $ 0.1$ & 85.62 & central &   $8.9\times10^{9}$  &  $7.1\times10^{9}$  &  $8.6\times10^{10}$ &  $0.0            $  &  $9.0\times10^{6}$  &  $1.9\times10^{5}$ \\
                &         &         &       & maximal &   $2.9\times10^{10}$ &  $1.8\times10^{10}$ &  $1.9\times10^{11}$ &  $0.0            $  &  $4.7\times10^{7}$  &  $6.2\times10^{5}$ \\
$10^{6}$        &  $3$    &  $ 1$   &  8.83 & central &   $1.6\times10^{11}$ &  $5.0\times10^{10}$ &  $3.8\times10^{11}$ &  $5.2\times10^{8}$  &  $4.0\times10^{9}$  &  $2.8\times10^{8}$ \\
                &         &         &       & maximal &   $4.8\times10^{11}$ &  $7.8\times10^{11}$ &  $5.6\times10^{11}$ &  $6.3\times10^{9}$  &  $2.9\times10^{10}$ &  $2.8\times10^{9}$ \\
$10^{6}$        &  $3$    &  $ 5$   &  1.81 & central &   $3.6\times10^{11}$ &  $1.3\times10^{11}$ &  $1.1\times10^{12}$ &  $5.4\times10^{9}$  &  $2.0\times10^{10}$ &  $2.1\times10^{9}$ \\
                &         &         &       & maximal &   $1.1\times10^{12}$ &  $2.7\times10^{11}$ &  $2.0\times10^{12}$ &  $9.9\times10^{10}$ &  $1.5\times10^{11}$ &  $3.6\times10^{10}$\\
$10^{6}$        &  $3$    &  $10$   &  0.91 & central &   $4.5\times10^{11}$ &  $1.7\times10^{11}$ &  $1.7\times10^{12}$ &  $1.2\times10^{10}$ &  $2.9\times10^{10}$ &  $5.5\times10^{9}$ \\
                &         &         &       & maximal &   $2.1\times10^{12}$ &  $4.9\times10^{11}$ &  $3.5\times10^{12}$ &  $2.9\times10^{11}$ &  $2.6\times10^{11}$ &  $1.3\times10^{11}$\\
$10^{6}$        &  $3$    &  $20$   &  0.46 & central &   $5.4\times10^{11}$ &  $2.2\times10^{11}$ &  $2.5\times10^{12}$ &  $1.3\times10^{10}$ &  $3.6\times10^{10}$ &  $6.9\times10^{9}$ \\
                &         &         &       & maximal &   $3.7\times10^{12}$ &  $8.5\times10^{11}$ &  $6.2\times10^{12}$ &  $3.7\times10^{11}$ &  $4.6\times10^{11}$ &  $1.8\times10^{11}$\\
$10^{6}$        &  $3$    &  $50$   &  0.19 & central &   $7.4\times10^{11}$ &  $3.5\times10^{11}$ &  $4.5\times10^{12}$ &  $5.2\times10^{10}$ &  $5.1\times10^{10}$ &  $2.9\times10^{10}$\\
                &         &         &       & maximal &   $7.0\times10^{12}$ &  $1.6\times10^{12}$ &  $1.2\times10^{13}$ &  $1.8\times10^{12}$ &  $8.8\times10^{11}$ &  $9.3\times10^{12}$\\
$2\times10^{6}$ &  $1.3$  &  $50$   &  3.34 & central &   $4.2\times10^{11}$ &  $1.6\times10^{11}$ &  $1.3\times10^{12}$ &  $7.6\times10^{9}$  &  $2.5\times10^{10}$ &  $3.6\times10^{9}$ \\
                &         &         &       & maximal &   $2.5\times10^{12}$ &  $5.4\times10^{11}$ &  $3.2\times10^{12}$ &  $2.3\times10^{11}$ &  $3.3\times10^{11}$ &  $1.0\times10^{11}$\\
$2\times10^{6}$ &  $3$    &  $ 0.1$ & 666.8 & central &   $8.2\times10^{9}$  &  $6.1\times10^{9}$  &  $6.9\times10^{10}$ &  $0.0            $  &  $8.8\times10^{6}$  &  $2.8\times10^{5}$ \\
                &         &         &       & maximal &   $3.1\times10^{10}$ &  $1.9\times10^{10}$ &  $1.8\times10^{11}$ &  $0.0            $  &  $5.1\times10^{7}$  &  $1.5\times10^{6}$ \\
$2\times10^{6}$ &  $3$    &  $ 1$   & 68.70 & central &   $1.5\times10^{11}$ &  $3.6\times10^{10}$ &  $2.3\times10^{11}$ &  $4.5\times10^{8}$  &  $3.8\times10^{9}$  &  $2.1\times10^{8}$ \\
                &         &         &       & maximal &   $4.2\times10^{11}$ &  $5.9\times10^{10}$ &  $3.1\times10^{11}$ &  $4.8\times10^{9}$  &  $2.5\times10^{10}$ &  $1.7\times10^{9}$ \\
$2\times10^{6}$ &  $3$    &  $ 5$   & 14.04 & central &   $3.1\times10^{11}$ &  $9.4\times10^{10}$ &  $6.1\times10^{11}$ &  $4.9\times10^{9}$ &   $1.8\times10^{10}$ &  $1.8\times10^{9}$ \\
                &         &         &       & maximal &   $9.4\times10^{11}$ &  $1.7\times10^{11}$ &  $9.5\times10^{11}$ &  $9.3\times10^{10}$ &  $1.4\times10^{11}$ &  $3.3\times10^{10}$\\
$2\times10^{6}$ &  $3$    &  $10$   &  7.08 & central &   $3.7\times10^{11}$ &  $1.2\times10^{11}$ &  $9.0\times10^{11}$ &  $6.3\times10^{9}$  &  $2.2\times10^{10}$ &  $2.8\times10^{9}$ \\
                &         &         &       & maximal &   $1.5\times10^{12}$ &  $30.\times10^{11}$ &  $1.6\times10^{12}$ &  $1.5\times10^{11}$ &  $2.0\times10^{11}$ &  $6.6\times10^{10}$\\
$2\times10^{6}$ &  $3$    &  $20$   &  3.58 & central &   $4.1\times10^{11}$ &  $1.5\times10^{11}$ &  $1.3\times10^{12}$ &  $1.7\times10^{10}$ &  $2.6\times10^{10}$ &  $8.0\times10^{9}$ \\
                &         &         &       & maximal &   $2.4\times10^{12}$ &  $5.0\times10^{11}$ &  $2.9\times10^{12}$ &  $6.7\times10^{11}$ &  $3.1\times10^{11}$ &  $3.2\times10^{11}$\\
$2\times10^{6}$ &  $3$    &  $50$   &  1.45 & central &   $4.7\times10^{11}$ &  $1.9\times10^{11}$ &  $2.0\times10^{12}$ &  $8.2\times10^{9}$  &  $2.9\times10^{10}$ &  $3.9\times10^{9}$ \\
                &         &         &       & maximal &   $4.7\times10^{12}$ &  $1.1\times10^{12}$ &  $6.8\times10^{12}$ &  $3.6\times10^{11}$ &  $6.0\times10^{11}$ &  $1.6\times10^{11}$\\
$2\times10^{6}$ &  $4.2$  &  $ 1$   & 49.40 & central &   $1.7\times10^{11}$ &	$4.5\times10^{10}$ &  $2.8\times10^{11}$ &  $5.1\times10^{8}$  &  $4.2\times10^{9}$  &  $2.4\times10^{8}$ \\
                &         &         &       & maximal &   $5.7\times10^{11}$ &  $7.9\times10^{10}$ &  $3.9\times10^{11}$ &  $6.8\times10^{9}$  &  $3.5\times10^{10}$ &  $2.4\times10^{9}$ \\
$2\times10^{6}$ &  $7.5$  &  $ 0.1$ & 226.7 & central &   $1.2\times10^{10}$ &  $9.7\times10^{9}$  &  $1.1\times10^{11}$ &  $0.0            $  &  $1.0\times10^{7}$  &  $0.0$             \\
                &         &         &       & maximal &   $5.2\times10^{10}$ &  $3.3\times10^{10}$ &  $3.2\times10^{11}$ &  $0.0            $  &  $7.9\times10^{7}$  &  $0.0$             \\
$2\times10^{6}$ &  $7.5$  &  $ 1$   & 27.48 & central &   $2.1\times10^{11}$ &  $6.2\times10^{10}$ &  $4.0\times10^{11}$ &  $6.3\times10^{8}$  &  $4.4\times10^{9}$  &  $3.7\times10^{8}$ \\
                &         &         &       & maximal &   $8.4\times10^{12}$ &  $1.2\times10^{11}$ &  $5.6\times10^{11}$ &  $1.1\times10^{10}$ &  $4.7\times10^{10}$ &  $5.5\times10^{9}$ \\
$2\times10^{6}$ &  $7.5$  &  $ 5$   &  5.62 & central &   $3.4\times10^{11}$ &  $1.3\times10^{11}$ &  $9.9\times10^{12}$ &  $8.1\times10^{9}$  &  $2.1\times10^{10}$ &  $4.0\times10^{9}$ \\
                &         &         &       & maximal &   $1.7\times10^{12}$ &  $3.5\times10^{11}$ &  $1.9\times10^{12}$ &  $3.0\times10^{11}$ &  $2.7\times10^{11}$ &  $1.4\times10^{11}$\\
$2\times10^{6}$ &  $7.5$  &  $10$   &  2.83 & central &   $4.1\times10^{11}$ &  $1.6\times10^{11}$ &  $1.4\times10^{12}$ &  $4.3\times10^{9}$  &  $2.2\times10^{10}$ &  $1.9\times10^{9}$ \\
                &         &         &       & maximal &   $2.8\times10^{12}$ &  $6.0\times10^{11}$ &  $3.6\times10^{12}$ &  $1.7\times10^{11}$ &  $3.3\times10^{11}$ &  $7.1\times10^{10}$\\
$2\times10^{6}$ &  $7.5$  &  $20$   &  1.43 & central &   $4.7\times10^{11}$ &  $1.9\times10^{11}$ &  $1.9\times10^{12}$ &  $2.0\times10^{10}$ &  $3.1\times10^{10}$ &  $9.7\times10^{0}$ \\
                &         &         &       & maximal &   $4.7\times10^{12}$ &  $1.1\times10^{12}$ &  $6.7\times10^{12}$ &  $1.2\times10^{12}$ &  $6.1\times10^{11}$ &  $5.6\times10^{11}$\\
$2\times10^{6}$ &  $7.5$  &  $50$   &  0.58 & central &   $5.7\times10^{11}$ &  $2.4\times10^{11}$ &  $2.9\times10^{12}$ &  $3.5\times10^{10}$ &  $4.2\times10^{10}$ &  $1.7\times10^{10}$\\
                &         &         &       & maximal &   $9.6\times10^{12}$ &  $2.1\times10^{12}$ &  $1.4\times10^{13}$ &  $2.3\times10^{12}$ &  $1.3\times10^{12}$ &  $1.1\times10^{12}$\\
$10^7$          &  $3$    &  $10$   &  831.1& central &   $3.9\times10^{11}$ &  $1.3\times10^{11}$ &  $9.6\times10^{11}$ &  $1.1\times10^{10}$ &  $2.2\times10^{10}$ &  $4.7\times10^{09}$\\
                &  	  &         &       & maximal &   $1.7\times10^{12}$ &  $3.4\times10^{11}$ &  $1.7\times10^{12}$ &  $3.4\times10^{11}$ &  $2.1\times10^{11}$ &  $1.4\times10^{11}$\\
$10^7$          &  $3$    &  $20$   &  419.3& central &   $4.1\times10^{11}$ &  $1.5\times10^{11}$ &  $1.2\times10^{12}$ &  $1.2\times10^{10}$ &  $2.5\times10^{10}$ &  $6.0\times10^{09}$\\
                &  	  &         &       & maximal &   $2.4\times10^{12}$ &  $4.8\times10^{11}$ &  $2.6\times10^{12}$ &  $6.1\times10^{11}$ &  $3.1\times10^{11}$ &  $3.0\times10^{11}$\\
$10^7$          &  $3$    &  $50$   & 169.7 & central &   $4.4\times10^{11}$ &  $1.7\times10^{11}$ &  $1.5\times10^{12}$ &  $7.6\times10^{09}$ &  $2.6\times10^{10}$ &  $3.4\times10^{09}$\\
                &  	  &         &       & maximal &   $4.1\times10^{12}$ &  $8.7\times10^{11}$ &  $5.0\times10^{12}$ &  $3.2\times10^{11}$ &  $5.2\times10^{11}$ &  $1.4\times10^{11}$\\
$10^7$          &  $7.5$  &  $10$   & 332.4 & central &   $4.2\times10^{11}$ &  $1.5\times10^{11}$ &  $1.3\times10^{12}$ &  $1.2\times10^{10}$ &  $2.4\times10^{10}$ &  $5.4\times10^{09}$\\
                &  	  &         &       & maximal &   $2.7\times10^{12}$ &  $5.5\times10^{11}$ &  $3.0\times10^{12}$ &  $5.1\times10^{11}$ &  $3.4\times10^{11}$ &  $2.3\times10^{11}$\\
$10^7$          &  $7.5$  &  $20$   & 167.7 & central &   $4.4\times10^{11}$ &  $1.7\times10^{11}$ &  $1.5\times10^{12}$ &  $1.6\times10^{10}$ &  $2.7\times10^{10}$ &  $7.1\times10^{09}$\\
                &  	  &         &       & maximal &   $4.1\times10^{12}$ &  $8.7\times10^{11}$ &  $5.0\times10^{12}$ &  $8.9\times10^{11}$ &  $5.3\times10^{11}$ &  $3.9\times10^{11}$\\
$10^7$          &  $7.5$  &  $50$   & 67.88 & central &   $4.6\times10^{11}$ &  $1.8\times10^{11}$ &  $1.7\times10^{12}$ &  $2.3\times10^{10}$ &  $3.0\times10^{10}$ &  $9.5\times10^{09}$\\
                &  	  &         &       & maximal &   $7.0\times10^{12}$ &  $1.5\times10^{12}$ &  $9.0\times10^{12}$ &  $1.7\times10^{12}$ &  $9.3\times10^{11}$ &  $6.7\times10^{11}$\\
\enddata 
\tablecomments{Assuming $\phi_s=0.3$. Columns were computed
for $Z=0.1$ times solar, and are proportional to $Z$.}
\label{columns-table}
\end{deluxetable*}

This cloud then evaporates in a saturated flow, with $\sigma_0\sim50$.
The mass-loss rate is suppressed by a factor $0.05$ relative to
a classical diffusive flow, and the maximal Mach number in the flow
is $0.60$.

Figure~8.2 displays the ion fractions of the low ions
\ion{C}{2}, \ion{N}{1}, \ion{O}{1}, \ion{Si}{2}, and \ion{S}{2},
and the high-ions \ion{C}{4}, \ion{N}{5}, \ion{O}{6},
\ion{Si}{3}, \ion{Si}{4}, and \ion{S}{3} in the conductive interface
surrounding this dwarf scale halo.

For the column densities produced in this model, we rely on the ion-abundances
given in GS04 (``model C'', Figure 5) for
the photoionized component, and the ion fractions computed here and presented
in Figure~8.2 for the conductive interface surrounding it. 
We compute the projected column densities as a function of impact parameter
from the cloud center,
again including both the photoionized core and the surrounding evaporating
envelope.

\begin{figure}[!h]
\epsscale{1.0}
\plotone{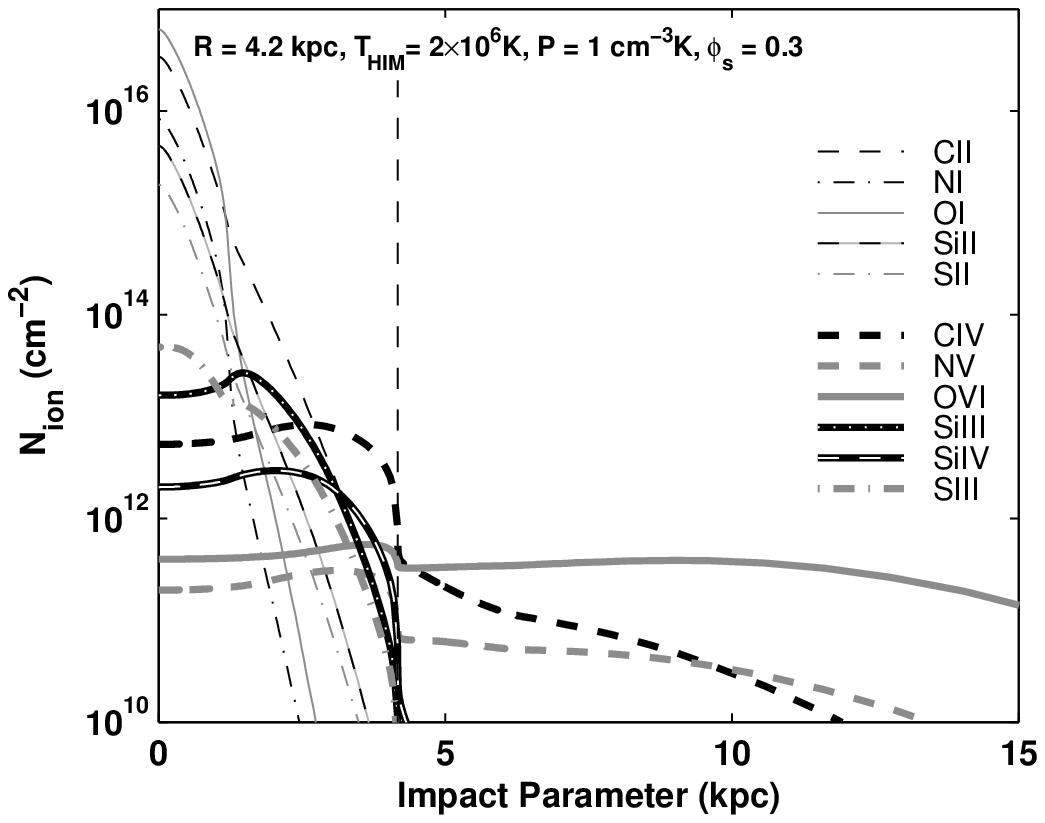}
\caption{Projected column densities as a function of impact parameter
in a dwarf galaxy-scale model. For this model $R=4.2$~kpc, 
$P/k_{\rm B}=1$~cm$^{-3}$~K, and $T_{\rm HIM}=2\times10^6$~K.}
\label{dwarf-columns}
\end{figure}

Figure~\ref{dwarf-columns} shows the projected column densities in our
dwarf-galaxy-scale model.
The ionization in this conductive interface results in a maximal
\ion{O}{6} column density of $5.5\times10^{11}$~cm$^{-2}$
(compared to $2.9\times10^{11}$~cm$^{-2}$ in the
photoionized cloud), and a maximal \ion{C}{4} column density of
$8.3\times10^{12}$~cm$^{-2}$, comparable to that produced in the
photoionized cloud). This is
only a minor enhancement relative to the purely photoionized cloud, 
and the columns in this dwarf galaxy-scale model are therefore still 
too low to account for the observed properties
of the metal-ion absorbers (e.g.~Table~\ref{collins}).

\subsection{Diagnostics}

Diagnostic diagrams for evaporating gas may be constructed using
Equation~\ref{Ni} and the computational data presented in \S~\ref{results},
and used to probe the ionization mechanisms at play in observed metal-ion 
absorbers. As an example, 
\begin{figure*}
\epsscale{0.95}
\plotone{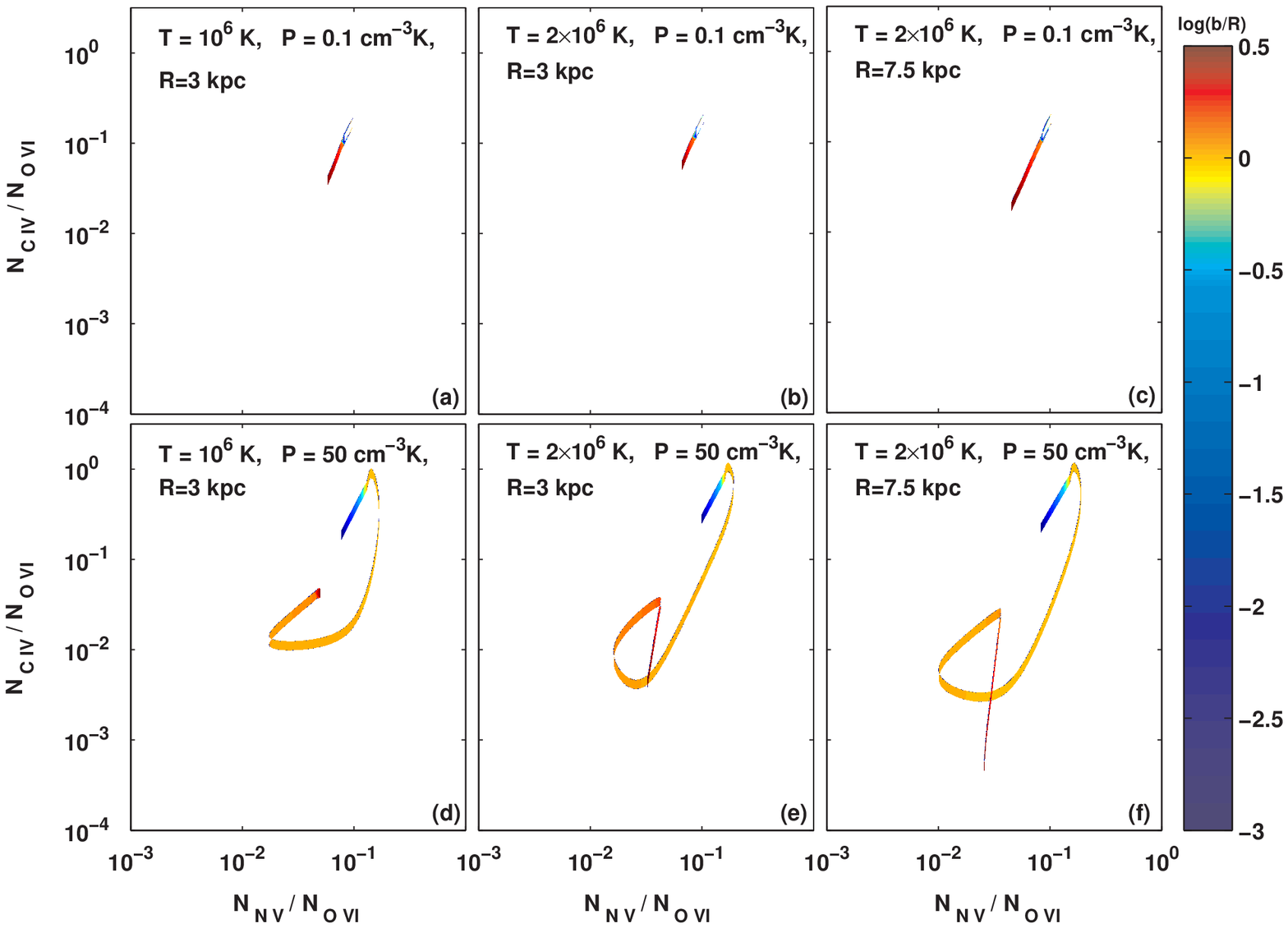}
\caption{Column density ratios $N_{\rm C~IV} / N_{\rm O~VI}$ versus 
$N_{\rm N~V} / N_{\rm O~VI}$ for evaporating gas. The impact parameter $b$
is indicated by the color along the trajectories, from the cloud center (blue)
to the surrounding HIM (red). The different panels are for different 
interface parameters, as indicated by the legends.
These ion ratios
do {\it not} include any contribution from the embedded warm cloud.
}
\label{ratios}
\end{figure*}
in Figure ~\ref{ratios} we display trajectories
for $N_{\rm C~IV} / N_{\rm O~VI}$ versus 
$N_{\rm N~V} / N_{\rm O~VI}$, as a function of impact parameter,
$b$ through the evaporative interface. 
The impact parameter is represented by the color along the 
curves, from the cloud center where $b=0$ (blue) to the
surrounding HIM where $b>R$ (red).
The different panels are for non-equilibrium evaporation for
a range of dwarf-scale objects, drawn from the range of cloud
radii, bounding pressures, and HIM temperatures 
considered in Table~\ref{columns-table}. These diagrams
show the possible range of ion ratios produced in evaporating 
dwarf-scale models. Since the column densities are proportional
to the metallicity, the ion ratios are independent of $Z$.
Note that these ion ratios do {\it not} include any
contribution from the embedded warm cloud, which depends on
the astrophysical setting.

O$^{5+}$ is the most highly ionized of the three ions that we
consider. $N_{\rm C~IV} / N_{\rm O~VI}$ and 
$N_{\rm N~V} / N_{\rm O~VI}$ are therefore large at 
low temperatures where the O$^{5+}$ fraction is small, 
and decrease at higher temperature as O$^{5+}$ becomes
more abundant.
Small impact parameters (in blue) sample the lower-$T$
parts of the interface, where $N_{\rm C~IV} / N_{\rm O~VI}$ 
and $N_{\rm N~V} / N_{\rm O~VI}$ are large. The cooling 
trajectories therefore start in the upper right corner
of the parameter space. Larger impact parameters
only sample the hotter parts of the interface, and the
trajectories evolve in the direction of decreasing
$N_{\rm C~IV} / N_{\rm O~VI}$ and 
$N_{\rm N~V} / N_{\rm O~VI}$. Finally, as the temperature
approaches $T_{\rm HIM}$ ($2\times10^6$~K in Figure~\ref{ratios}), 
O$^{5+}$ is replaced by higher ionization states, and the
evolutionary tracks turn up again.

An additional ion that is commonly observed in high-velocity 
absorbers, and that has been used as a diagnostic
for the physical conditions prevailing in these systems, is 
\ion{Si}{3} (Collins et al.~2004; Collins et al.~2009; Shull et al.~2009).
The typical observed column densities of \ion{Si}{3} in the high-velocity
absorbers are $\sim3\times10^{13}$~cm$^{-2}$ (Shull et al.~2009). 
\ion{Si}{3} is never efficiently produced in evaporating
dwarf-galaxy-scale clouds. For low bounding pressures the ionization parameter
is too high to efficiently produce \ion{Si}{3} in the evaporating layers, 
but even at high pressures ($P\sim50$~cm$^{-3}$~K), for which \ion{Si}{3}
is the dominant ionization state at the cloud surface, columns in excess
of $10^{12}$~cm$^{-2}$ are rarely produced. \ion{Si}{3} must therefore
be produced by alternative ionization mechanisms. It may be efficiently
produced by photoionization within an embedded warm cloud, as we demonstrate
in Figures~\ref{chvc-columns} and~\ref{dwarf-columns}.

\subsection{High \ion{O}{6} columns in evaporating clouds?}

\begin{figure}[!h]
\epsscale{1.0}
\plotone{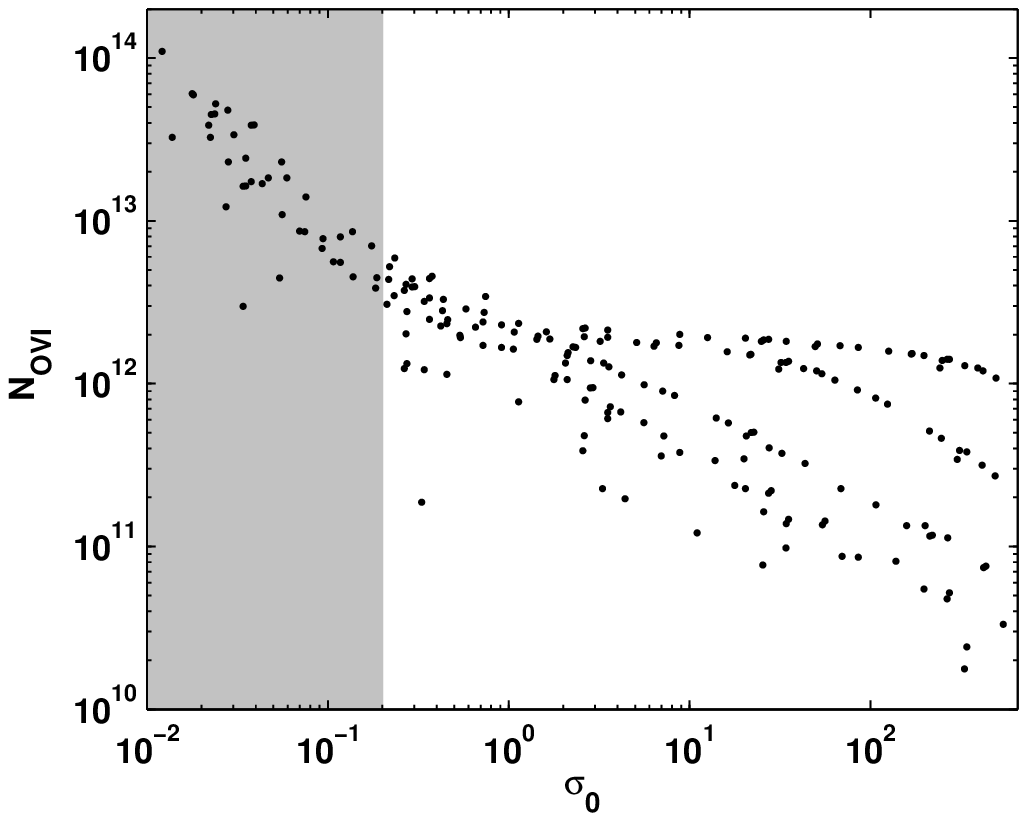}
\caption{Central (two-sided) \ion{O}{6} column density versus saturation.
Models inside the shaded areas are not self-consistent, as cooling
may not be neglected (see section~\ref{nonrad}). 
}
\label{NOVI}
\end{figure}

The results presented in Table~\ref{columns-table} 
and illustrated by the examples discussed above,
show that the typical \ion{O}{6} columns that are
produced in the different dwarf-scale models are
$\le10^{13}$~cm$^{-2}$. Can column densities of order
$10^{14}$~cm$^{-2}$ be produced in conductively evaporating
clouds?

In Figure~\ref{NOVI} we show the central (two-sided) \ion{O}{6} column
density as a function of saturation parameter for our full grid of models,
assuming $Z=0.1$.
The range of saturation parameters for which no self-consistent 
evaporating solution can be found has been shaded in the plot.
In the shaded region, cooling overcomes conductive heating, evaporation stops,
and instead the ambient medium condenses onto the clouds.
Our theory does not apply to condensing clouds.
Figure~\ref{NOVI} shows that central \ion{O}{6} column densities of
order $10^{14}$~cm$^{-2}$ cannot be obtained for any $Z=0.1$ evaporating 
model. The largest central \ion{O}{6} columns, obtained for
$\sigma_0\sim0.2$, are $\sim2\times10^{12}$~cm$^{-2}$.

We note that for solar metallicity clouds, the \ion{O}{6}
columns are a factor of $10$ larger.
However, since for $Z=1$ self-consistent evaporating solutions 
can only be found for $\sigma_0>0.5$, central \ion{O}{6} columns
of order $10^{14}$~cm$^{-2}$ can still not be obtained.

Since the observed column density is a function of the impact
parameter, we also computed the maximal \ion{O}{6} column density
obtained in each self-consistent evaporating model.
Inspecting our entire parameter space 
($0.1<P/k_{\rm B}<10^4$~cm$^{-3}$~K, $1$~pc~$<R<100$~kpc,
$5\times10^5<T<10^7$~K),
we find that for $Z=0.1$, maximal column densities of order
$10^{14}$~cm$^{-2}$ only occur in evaporating clouds
($\sigma_0>0.2$) for extreme combinations of parameters, 
with $T\sim10^7$~K, and $P\sim10^4$~cm$^{-3}$~K.
We therefore conclude that non-equilibrium ionization in
conductively evaporating clouds an unlikely explanation of
the high-ions observed in the high-velocity metal-ion absorbers.
An alternative explanation is that \ion{O}{6} is created
in a flow of cooling gas (e.g.~Edgar \& Chevalier~1986; Heckman 
et al.~2002; Gnat \& Sternberg~2007; Gnat \& Sternberg~2009).

While a conductive origin for the \ion{O}{6} absorption observed in the ionized
high-velocity clouds appears unlikely, recent observations indicate
that evaporating \ion{O}{6} may be present in  more local environments.
Savage \& Lehner~(2006) report on {\it FUSE} absorption line observations
toward a sample of local white dwarfs. \ion{O}{6} was detected along 
24 lines of sight, with column densities between $2.5\times10^{12}$
and $4.0\times10^{13}$~cm$^{-2}$. Based on the observed columns
and on the kinematics of the absorption, Savage \& Lehner suggest
that this sample of local absorbers may probe conductive interfaces
in the local interstellar medium. Their analysis relied on the 
plane-parallel conduction-front models of Borkowski et al.~(1990). 
Our spherical models for typical interstellar parameters ($T\sim10^6$~K,
$P\sim4\times10^3$~cm$^{-3}$~K, and $R\sim15$~pc) support this interpretation.
For interstellar conditions the saturation parameter $\sigma_0\sim0.5$, 
and we find that \ion{O}{6} column densities of order $10^{13}$~cm$^{-2}$ are
expected for solar metallicity gas. The predicted Doppler widths ($15-30$~km~s$^{-1}$)
and flow velocities ($5-55$~km~s$^{-1}$) are also consistent with the 
Savage \& Lehner observations.

\section{Summary}

In this paper we present computations of the non-equilibrium
ionization states of H, He, C, N, O, Si, and S in thermally
conductive interfaces, surrounding warm 
(and photoionized)
gas clouds that undergo
steady  evaporation into a hot ambient medium.

We rely on the formalism developed by 
Dalton \& Balbus (1993; DB93) for solving the temperature profile in the
evaporating gas. This formalism allows the thermal conductivity
to change continuously from the classical 
diffusive
form (Spitzer 1962) to a flux-limited saturated form (Cowie \& McKee 1977).
The temperature profile then depends only on the global level
of saturation in the flow, as described by the saturation parameter,
$\sigma_0$.

In their analytic solution for the temperature profile, DB93 neglect the kinetic energy
term as well as the radiative losses term in the energy equation.
As opposed to previous claims, we find that the neglect of
the kinetic energy term is accurate only for classical, unsaturated
flows,
and applies to 
saturated flows
only if the parameter $\phi_s$ 
(see Equation 2)
is 
less  than $0.6$.

We explicitly solve the
time-dependent ionization equations for
the evaporating gas,
including also photoionization by the present-day metagalactic field.
As the gas evaporates, heating can become
fast compared to the ionization time-scale, and the gas tends to
remain under-ionized.
We study how departures from equilibrium ionization affects the
ion distributions and resulting cooling rates in the evaporating 
layers. We show that the existence of underionized species,
most significantly He$^+$, significantly enhances the cooling rates.
We estimate numerically the range of saturation parameters
for which radiative cooling plays a significant role in determining
the dynamics of the gas.
We find that for a $0.1$ ($1$)~times solar metallicity gas, radiative
losses are significant for $\sigma_0\lesssim0.2$ ($0.5$).  
For larger values of
$\sigma_0$ the importance of radiative cooling is generally smaller,
and depends on the neutral fraction at the surface of the evaporating cloud.

We use our conduction models to extend the purely photoionized cloud
models for local high-velocity
metal line absorbers that we presented in Gnat \& Sternberg~(2004; GS04).
In that paper, we modeled the absorbers as warm ($\sim 10^4$~K) clouds 
that are photoionized by the metagalactic radiation field, and
embedded in low-mass
($<2\times 10^9$~M$_\odot$) dark-matter mini-haos. 
Here, we consider the additional 
ionization that occurs in the conductive envelopes,
as these clouds evaporate into a hot corona and/or
intergalactic medium. We consider a wide range of input parameters
(cloud radius, thermal pressure and temperature of the hot medium) as
summarized in Table~\ref{models-table}. This table also lists the 
ionization parameter, global saturation parameter, 
mass-loss rate, maximal Mach number, and evaporation time scale for each model.
We assume that $\phi_s=0.3$,
so that the flow is everywhere subsonic.

In \S3 we compute the ionization states as a function of
distance from the evaporating cloud surface. We plot the resulting ion
fractions in Figure~8.
In \S4 we present the metal absorption column densities arising in
conductively evaporating minihalo clouds.
We list the column densities of the high ions that are produced
in the interface layers of our dwarf-scale models
in Table~\ref{columns-table}.
We then focus on two illustrative models, for which we 
include the contributions to the ion distributions from both a
central photoionized cloud and the conductive interface surrounding it.
In \S4.1 we discuss the ion fractions and projected column densities in 
a low-ionization model embedded in a high-pressure environment.
In \S4.2, we present results for a more highly ionized cloud embedded in
a low-pressure medium.
We find that the contribution of the conductive interface
enhances the formation of high-ions such as \ion{C}{4} and
\ion{O}{6}.
However, in all cases the enhanced columns are still too low to
account for the \ion{O}{6} columns observed in the high-velocity metal-ion
absorbers (e.g.~Table~\ref{collins}). 
The central \ion{O}{6} column density
for evaporating clouds is limited to $<10^{13}$~cm$^{-2}$. 
These limits are reached for small
saturation parameters, just at the point where radiative losses
become significant enough to result in condensation.
While a conductive origin for the \ion{O}{6} absorption observed in the ionized
high-velocity clouds appears unlikely, our models do support the conclusion by
Savage \& Lehner~(2006) that evaporating \ion{O}{6} absorption occurs in the
local ISM, with characteristic columns of $\sim10^{13}$~cm$^{-2}$.

We use the predictions for the high-ion column densities produced in
evaporating clouds (see Table~\ref{columns-table}) to construct ion-ratio
diagnostic diagrams. In Section 4.3, we discuss one example, 
$N_{\rm C~IV} / N_{\rm O~VI}$ versus $N_{\rm N~V} / N_{\rm O~VI}$,
and show how these ratios vary with impact parameter in the evaporating
layers, and how they may be used to probe the ionization processes
affecting metal-line absorbers.

\section*{Acknowledgments}

We thank Blair Savage and Mike Shull for useful comments.
This research was supported by the US-Israel Binational
Science Foundation (BSF) grant 2002317, by the 
Deutsche Forschungsgemeinschaft (DFG)
via German-Israeli Project Cooperation grant STE1869/1-1.GE625/15-1,
and by the National Science Foundation
through grant AST0908553 (CFM).
O.G. acknowledges support provided
by NASA through Chandra Postdoctoral Fellowship grant number PF8-90053
awarded by the Chandra X-ray Center, which is operated by the
Smithsonian Astrophysical Observatory for NASA under contract NAS8-03060.

\appendix
\section{On the Sensitivity to the Spectral Slope Near the Lyman Limit}
Our standard fit for the metagalactic radiation field is given by 
Equation~\ref{radn}, and displayed in Figure~\ref{radfield} (solid dark curve). 
Near the Lyman limit, this representation varies as $\nu^{-3.13}$.
However, the spectral slope may be significantly shallower
at the Lyman limit, with $J_\nu$ varying as $\nu^{-1.8}$, and steepen gradually
with increasing frequency (e.g.~Shull et al.~1999; Telfer et al.~2002;
Scott et al.~2004; Shull et al.~2004; Zheng et al.~2004; 
Faucher-Giguere et al.~2009). In this Appendix we present results for
the ionization properties in evaporating conduction fronts for a 
"maximal field" (Gnat \& Sternberg 2004) that is represented by,
\begin{equation}
J_{\nu} = \left\{ \begin{array}{lcl}
   J_{\nu0}(\frac{\nu}{\nu_0})^{-1.725}&,& 1~<\frac{\nu}{\nu_0}~<~18.4\\
   2.512\times10^{-2}J_{\nu0}(\frac{\nu}{\nu_0})^{-0.46}&,& 18.4~<~\frac{\nu}{\nu_0} \\
                  \end{array}\right.
\end{equation}
and displayed by the dashed curve in Figure~\ref{radfield}. 
As for our standard field,  
$J_{\nu0} = 2\times10^{-23}$~erg~s$^{-1}$~cm$^{-2}$~Hz$^{-1}$~sr$^{-1}$, 
and $\nu_0$ is the Lyman limit frequency. Our standard and maximal
fields span the range of possible spectral slopes suggested by observations
and theory.

We repeat the calculations presented in Sections~\ref{physics}, \ref{results},
and~\ref{columns}
assuming irradiation by the maximal field. The increased intensity above the Lyman
limit yields higher photoionization rates, which affect the column
densities of intermediate ions. In Table~\ref{columns-formax}, we list 
the (two-sided) central ($b=0$) and maximum column densities for the 
different models considered in \S3, assuming photoionization by
the maximal field. A comparison of Tables~\ref{columns-table} 
and~\ref{columns-formax} demonstrates that the high-ion column densities
are generally insensitive to the exact value of the spectral slope near
the Lyman limit.

Table~\ref{columns-formax} confirms that low pressure/density models are
more affected by the increased UV intensity than higher pressure models,
and that intermediate ions are more affected than the high ions. In all
cases, we find that the columns computed with the standard 
field agree with those computed with the maximal field to better than
$0.5$~dex. As we describe below, in most cases the agreement is better.

\begin{figure}[!h]
\epsscale{0.5}
\plotone{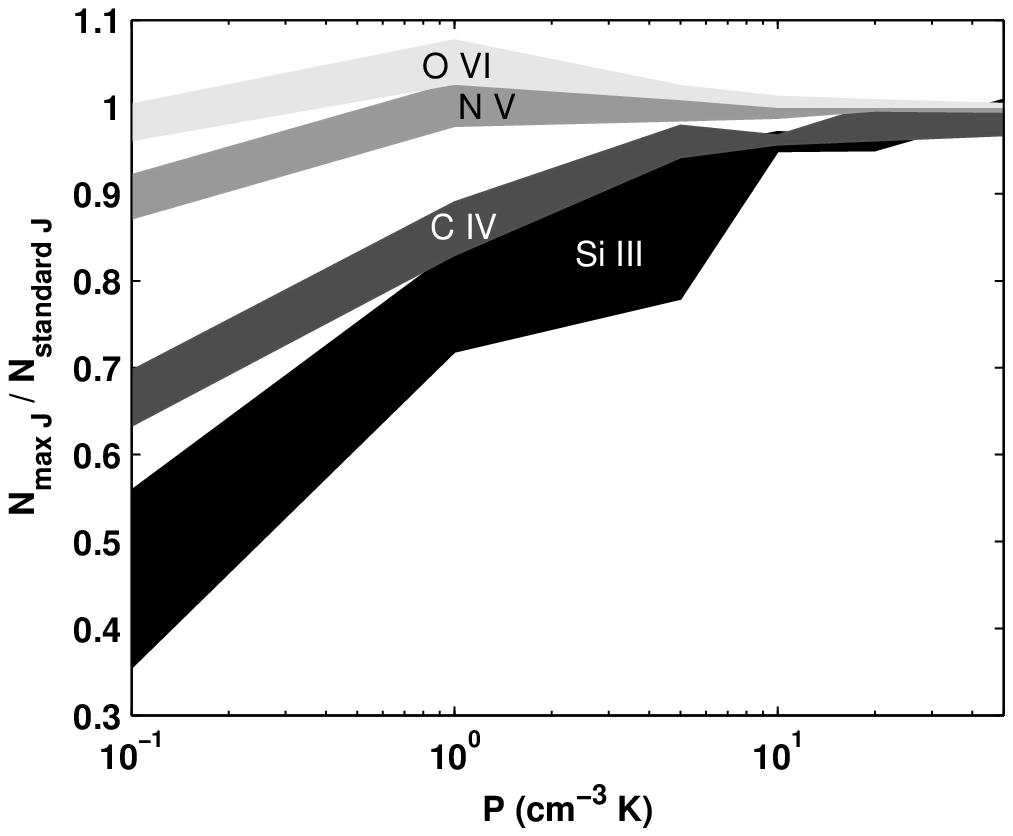}
\caption{The range of $N_{\rm max}/N_{\rm standard}$ ratios as a function of
the external bounding pressure, for \ion{Si}{3}, \ion{C}{4}, \ion{N}{5},
and \ion{O}{6} produced in dwarf-scale-halos with $T=1-2\times10^6$~K.
Here $N_{\rm standard}$ are the columns computed with the standard
field described in Section~\ref{ionization}, and $N_{\rm max}$ are the columns
computed with the maximal field, which is shallower above the Lyman limit.
The high-ion columns are insensitive to spectral slope near the Lyman limit,
and the intermediate-ions are only moderately affected, mostly for low 
bounding pressures.}
\label{Ndiff}
\end{figure}

Figure~\ref{Ndiff} compares the columns computed with the maximal
field, $N_{\rm max}$, to the column computed with the standard 
field, $N_{\rm standard}$. We display the range of $N_{\rm max}/N_{\rm standard}$ 
ratios as a function of the external bounding pressure, for \ion{Si}{3}, \ion{C}{4},
\ion{N}{5}, and \ion{O}{6} produced in dwarf scale halos with $T=1-2\times10^6$~K.

Figure~\ref{Ndiff} shows that for $P\gtrsim1$~cm$^{-3}$~K, the \ion{O}{6}
and \ion{N}{5} columns remain similar, the \ion{C}{4} columns agree to 
within $20\%$, and the \ion{Si}{3} columns agree to within $30\%$.
For the lowest pressure that we consider, $P=0.1$~cm$^{-3}$~K, the
differences are larger, but are still within a factor $2.5$.
We conclude that column densities of the high ions are only weakly
dependent on the exact spectral slope near the Lyman limit.

\begin{deluxetable*}{lllllllllll}
\tablewidth{0pt}
\tablecaption{Column Densities for the Maximal Radiation Field}
\tablehead{
\colhead{T$_{\rm HIM}$}&
\colhead{R}&
\colhead{P/k$_{\rm B}$}&
\colhead{$\sigma_0$}&
\colhead{}&
\colhead{C~IV}&
\colhead{N~V}&
\colhead{O~VI} &
\colhead{Si~III} &
\colhead{Si~IV}&
\colhead{S~III}}
\startdata
$10^{6}$        &  $3$    &  $ 0.1$ & 85.62 & central &   $6.2\times10^{9} $ &	$6.2\times10^{9} $ &	$8.3\times10^{10}$ &	$1.5\times10^{5} $ &	$6.4\times10^{6} $ &	$8.0\times10^{4} $ \\
                &         &         &       & maximal &   $2.6\times10^{10}$ &	$1.8\times10^{10}$ &	$1.9\times10^{11}$ &	$1.0\times10^{6} $ &	$4.1\times10^{7} $ &	$3.6\times10^{5} $ \\
$10^{6}$        &  $3$    &  $ 1$   &  8.83 & central &   $1.4\times10^{11}$ &	$4.9\times10^{10}$ &	$3.9\times10^{11}$ &	$4.1\times10^{8} $ &	$3.3\times10^{9} $ &	$2.0\times10^{8} $ \\
                &         &         &       & maximal &   $5.3\times10^{11}$ &	$8.5\times10^{10}$ &	$5.6\times10^{11}$ &	$5.4\times10^{9} $ &	$2.7\times10^{10}$ &	$2.3\times10^{9} $ \\
$10^{6}$        &  $3$    &  $ 5$   &  1.81 & central &   $3.4\times10^{11}$ &	$1.3\times10^{11}$ &	$1.1\times10^{12}$ &	$5.0\times10^{9} $ &	$1.9\times10^{10}$ &	$2.0\times10^{9} $ \\
                &         &         &       & maximal &   $1.2\times10^{12}$ &	$2.8\times10^{11}$ &	$2.0\times10^{12}$ &	$9.6\times10^{10}$ &	$1.5\times10^{11}$ &	$3.4\times10^{10}$ \\
$10^{6}$        &  $3$    &  $10$   &  0.91 & central &   $4.3\times10^{11}$ &	$1.7\times10^{11}$ &	$1.7\times10^{12}$ &	$1.1\times10^{10}$ &	$2.7\times10^{10}$ &	$5.1\times10^{9} $ \\
                &         &         &       & maximal &   $2.1\times10^{12}$ &	$5.0\times10^{11}$ &	$3.6\times10^{12}$ &	$2.8\times10^{11}$ &	$2.5\times10^{11}$ &	$1.2\times10^{11}$ \\
$10^{6}$        &  $3$    &  $20$   &  0.46 & central &   $5.2\times10^{11}$ &	$2.2\times10^{11}$ &	$2.5\times10^{12}$ &	$1.2\times10^{10}$ &	$3.3\times10^{10}$ &	$6.5\times10^{9} $ \\
                &         &         &       & maximal &   $3.6\times10^{12}$ &	$8.5\times10^{11}$ &	$6.2\times10^{12}$ &	$3.6\times10^{11}$ &	$4.4\times10^{11}$ &	$1.7\times10^{11}$ \\
$10^{6}$        &  $3$    &  $50$   &  0.19 & central &   $7.1\times10^{11}$ &	$3.4\times10^{11}$ &	$4.5\times10^{12}$ &	$5.3\times10^{10}$ &	$4.6\times10^{10}$ &	$3.0\times10^{10}$ \\
                &         &         &       & maximal &   $6.7\times10^{12}$ &	$1.6\times10^{12}$ &	$1.2\times10^{13}$ &	$1.8\times10^{12}$ &	$7.9\times10^{11}$ &	$1.1\times10^{12}$ \\
$2\times10^{6}$ &  $1.3$  &  $50$   &  3.34 & central &   $4.1\times10^{11}$ &	$1.6\times10^{11}$ &	$1.3\times10^{12}$ &	$7.5\times10^{9} $ &	$2.5\times10^{10}$ &	$3.6\times10^{9} $ \\
                &         &         &       & maximal &   $2.5\times10^{12}$ &	$5.4\times10^{11}$ &	$3.2\times10^{12}$ &	$2.3\times10^{11}$ &	$3.3\times10^{11}$ &	$1.0\times10^{11}$ \\
$2\times10^{6}$ &  $3$    &  $ 0.1$ & 666.8 & central &   $5.4\times10^{9} $ &	$5.6\times10^{9} $ &	$6.9\times10^{10}$ &	$1.3\times10^{5} $ &	$4.9\times10^{6} $ &	$7.1\times10^{4} $ \\
                &         &         &       & maximal &   $2.6\times10^{10}$ &	$1.9\times10^{10}$ &	$1.9\times10^{11}$ &	$9.8\times10^{5} $ &	$3.5\times10^{7} $ &	$5.3\times10^{5} $ \\
$2\times10^{6}$ &  $3$    &  $ 1$   & 68.70 & central &   $1.3\times10^{11}$ &	$3.7\times10^{10}$ &	$2.4\times10^{11}$ &	$3.8\times10^{8} $ &	$3.3\times10^{9} $ &	$1.6\times10^{8} $ \\
                &         &         &       & maximal &   $4.4\times10^{11}$ &	$6.5\times10^{10}$ &	$3.2\times10^{11}$ &	$4.3\times10^{9} $ &	$2.4\times10^{10}$ &	$1.5\times10^{9} $ \\
$2\times10^{6}$ &  $3$    &  $ 5$   & 14.04 & central &   $3.0\times10^{11}$ &	$9.0\times10^{10}$ &	$6.2\times10^{11}$ &	$4.6\times10^{9} $ &	$1.7\times10^{10}$ &	$1.7\times10^{9} $ \\
                &         &         &       & maximal &   $9.6\times10^{11}$ &	$1.8\times10^{11}$ &	$9.5\times10^{11}$ &	$9.0\times10^{10}$ &	$1.4\times10^{11}$ &	$3.1\times10^{10}$ \\
$2\times10^{6}$ &  $3$    &  $10$   &  7.08 & central &   $3.6\times10^{11}$ &	$1.2\times10^{11}$ &	$9.0\times10^{11}$ &	$6.1\times10^{9} $ &	$2.1\times10^{10}$ &	$2.7\times10^{9} $ \\
                &         &         &       & maximal &   $1.5\times10^{12}$ &	$3.0\times10^{11}$ &	$1.6\times10^{12}$ &	$1.5\times10^{11}$ &	$1.9\times10^{11}$ &	$6.4\times10^{10}$ \\
$2\times10^{6}$ &  $3$    &  $20$   &  3.58 & central &   $4.1\times10^{11}$ &	$1.5\times10^{11}$ &	$1.3\times10^{12}$ &	$1.6\times10^{10}$ &	$2.6\times10^{10}$ &	$7.7\times10^{9} $ \\
                &         &         &       & maximal &   $2.4\times10^{12}$ &	$5.1\times10^{11}$ &	$2.9\times10^{12}$ &	$6.6\times10^{11}$ &	$3.2\times10^{11}$ &	$3.2\times10^{11}$ \\
$2\times10^{6}$ &  $3$    &  $50$   &  1.45 & central &   $4.7\times10^{11}$ &	$1.9\times10^{11}$ &	$2.0\times10^{12}$ &	$8.2\times10^{9} $ &	$2.9\times10^{10}$ &	$3.9\times10^{9} $ \\
                &         &         &       & maximal &   $4.7\times10^{12}$ &	$1.1\times10^{12}$ &	$6.8\times10^{12}$ &	$3.6\times10^{11}$ &	$6.0\times10^{11}$ &	$1.6\times10^{11}$ \\
$2\times10^{6}$ &  $7.5$  &  $ 0.1$ & 226.7 & central &   $7.7\times10^{9} $ &	$8.6\times10^{9} $ &	$1.1\times10^{11}$ &	$1.8\times10^{5} $ &	$6.7\times10^{6} $ &	$1.0\times10^{5} $ \\
                &         &         &       & maximal &   $4.5\times10^{10}$ &	$3.2\times10^{10}$ &	$3.2\times10^{11}$ &	$1.8\times10^{6} $ &	$6.7\times10^{7} $ &	$6.6\times10^{5} $ \\
$2\times10^{6}$ &  $7.5$  &  $ 1$   & 27.48 & central &   $1.8\times10^{11}$ &	$6.3\times10^{10}$ &	$4.4\times10^{11}$ &	$4.5\times10^{8} $ &	$3.4\times10^{9} $ &	$2.4\times10^{8} $ \\
                &         &         &       & maximal &   $9.5\times10^{11}$ &	$1.5\times10^{11}$ &	$6.2\times10^{11}$ &	$8.8\times10^{9} $ &	$4.2\times10^{10}$ &	$4.1\times10^{9} $ \\
$2\times10^{6}$ &  $7.5$  &  $ 5$   &  5.62 & central &   $3.7\times10^{11}$ &	$1.3\times10^{11}$ &	$1.0\times10^{12}$ &	$6.3\times10^{9} $ &	$1.9\times10^{10}$ &	$2.8\times10^{9} $ \\
                &         &         &       & maximal &   $2.4\times10^{12}$ &	$3.7\times10^{11}$ &	$2.0\times10^{12}$ &	$2.4\times10^{11}$ &	$2.9\times10^{11}$ &	$1.1\times10^{11}$ \\
$2\times10^{6}$ &  $7.5$  &  $10$   &  2.83 & central &   $4.0\times10^{11}$ &	$1.6\times10^{11}$ &	$1.4\times10^{12}$ &	$4.1\times10^{9} $ &	$2.1\times10^{10}$ &	$1.9\times10^{9} $ \\
                &         &         &       & maximal &   $2.8\times10^{12}$ &	$6.0\times10^{11}$ &	$3.6\times10^{12}$ &	$1.6\times10^{11}$ &	$3.2\times10^{11}$ &	$6.9\times10^{10}$ \\
$2\times10^{6}$ &  $7.5$  &  $20$   &  1.43 & central &   $4.7\times10^{11}$ &	$1.9\times10^{11}$ &	$1.9\times10^{12}$ &	$1.9\times10^{10}$ &	$3.1\times10^{10}$ &	$9.2\times10^{9} $ \\
                &         &         &       & maximal &   $4.7\times10^{12}$ &	$1.1\times10^{12}$ &	$6.8\times10^{12}$ &	$1.1\times10^{12}$ &	$5.9\times10^{11}$ &	$5.5\times10^{11}$ \\
$2\times10^{6}$ &  $7.5$  &  $50$   &  0.58 & central &   $5.6\times10^{11}$ &	$2.4\times10^{11}$ &	$2.9\times10^{12}$ &	$3.6\times10^{10}$ &	$4.0\times10^{10}$ &	$1.8\times10^{10}$ \\
                &         &         &       & maximal &   $9.5\times10^{12}$ &	$2.1\times10^{12}$ &	$1.5\times10^{13}$ &	$2.4\times10^{12}$ &	$1.2\times10^{12}$ &	$1.3\times10^{12}$ \\
$10^7$          &  $3$    &  $10$   &  831.1& central &   $3.7\times10^{11}$ &	$1.3\times10^{11}$ &	$9.8\times10^{11}$ &	$7.0\times10^{9} $ &	$1.7\times10^{10}$ &	$2.5\times10^{9} $ \\
                &  	  &         &       & maximal &   $2.1\times10^{12}$ &	$3.5\times10^{11}$ &	$1.8\times10^{12}$ &	$2.2\times10^{11}$ &	$1.9\times10^{11}$ &	$7.5\times10^{10}$ \\
$10^7$          &  $3$    &  $20$   &  419.3& central &   $4.0\times10^{11}$ &	$1.5\times10^{11}$ &	$1.2\times10^{12}$ &	$1.2\times10^{10}$ &	$2.4\times10^{10}$ &	$6.0\times10^{9} $ \\
                &  	  &         &       & maximal &   $2.4\times10^{12}$ &	$4.9\times10^{11}$ &	$2.6\times10^{12}$ &	$6.1\times10^{11}$ &	$3.1\times10^{11}$ &	$3.0\times10^{11}$ \\
$10^7$          &  $3$    &  $50$   & 169.7 & central &   $4.3\times10^{11}$ &	$1.7\times10^{11}$ &	$1.5\times10^{12}$ &	$7.5\times10^{9} $ &	$2.5\times10^{10}$ &	$3.4\times10^{9} $ \\
                &  	  &         &       & maximal &   $4.1\times10^{12}$ &	$8.7\times10^{11}$ &	$5.0\times10^{12}$ &	$3.2\times10^{11}$ &	$5.1\times10^{11}$ &	$1.4\times10^{11}$ \\
$10^7$          &  $7.5$  &  $10$   & 332.4 & central &   $4.2\times10^{11}$ &	$1.5\times10^{11}$ &	$1.3\times10^{12}$ &	$9.7\times10^{9} $ &	$2.3\times10^{10}$ &	$4.1\times10^{9} $ \\
                &  	  &         &       & maximal &   $3.2\times10^{12}$ &	$5.7\times10^{11}$ &	$3.1\times10^{12}$ &	$4.2\times10^{11}$ &	$3.7\times10^{11}$ &	$1.7\times10^{11}$ \\
$10^7$          &  $7.5$  &  $20$   & 167.7 & central &   $4.4\times10^{11}$ &	$1.7\times10^{11}$ &	$1.5\times10^{12}$ &	$1.5\times10^{10}$ &	$2.7\times10^{10}$ &	$6.7\times10^{9} $ \\
                &  	  &         &       & maximal &   $4.1\times10^{12}$ &	$8.8\times10^{11}$ &	$5.0\times10^{12}$ &	$8.4\times10^{11}$ &	$5.2\times10^{11}$ &	$3.7\times10^{11}$ \\
$10^7$          &  $7.5$  &  $50$   & 67.88 & central &   $4.6\times10^{11}$ &	$1.8\times10^{11}$ &	$1.7\times10^{12}$ &	$2.3\times10^{10}$ &	$3.0\times10^{10}$ &	$1.0\times10^{10}$ \\
                &  	  &         &       & maximal &   $7.0\times10^{12}$ &	$1.5\times10^{12}$ &	$9.1\times10^{12}$ &	$1.7\times10^{12}$ &	$9.2\times10^{11}$ &	$7.4\times10^{11}$ \\
\enddata 
\tablecomments{Assuming $\phi_s=0.3$. Columns were computed
for $Z=0.1$ times solar, and are proportional to $Z$.}
\label{columns-formax}
\end{deluxetable*}


\clearpage
\section*{}
{\bf {\sc ~~~~~~~~~~~~~~~~~~~~~~~~~~~~~~~~~~~~~~~Electronic (only) Figure Set 8}}

\begin{figure}[!h]
\figurenum{8.1,~8.2}
\epsscale{1.0}
\plottwo{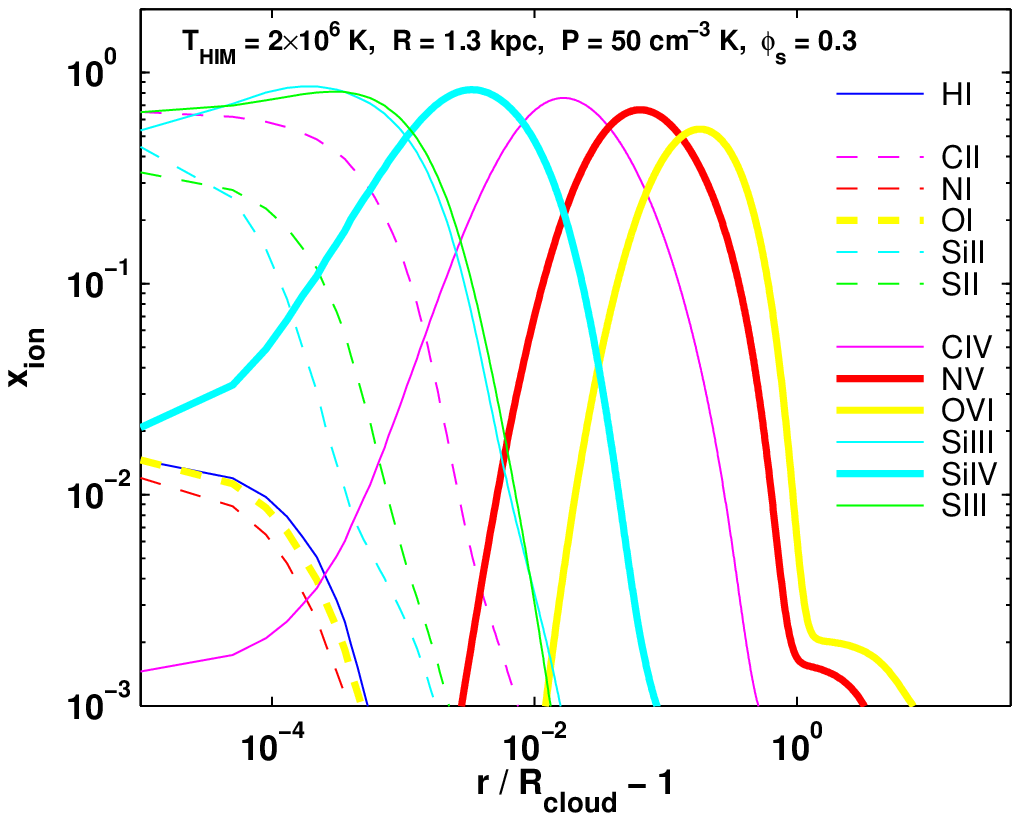}{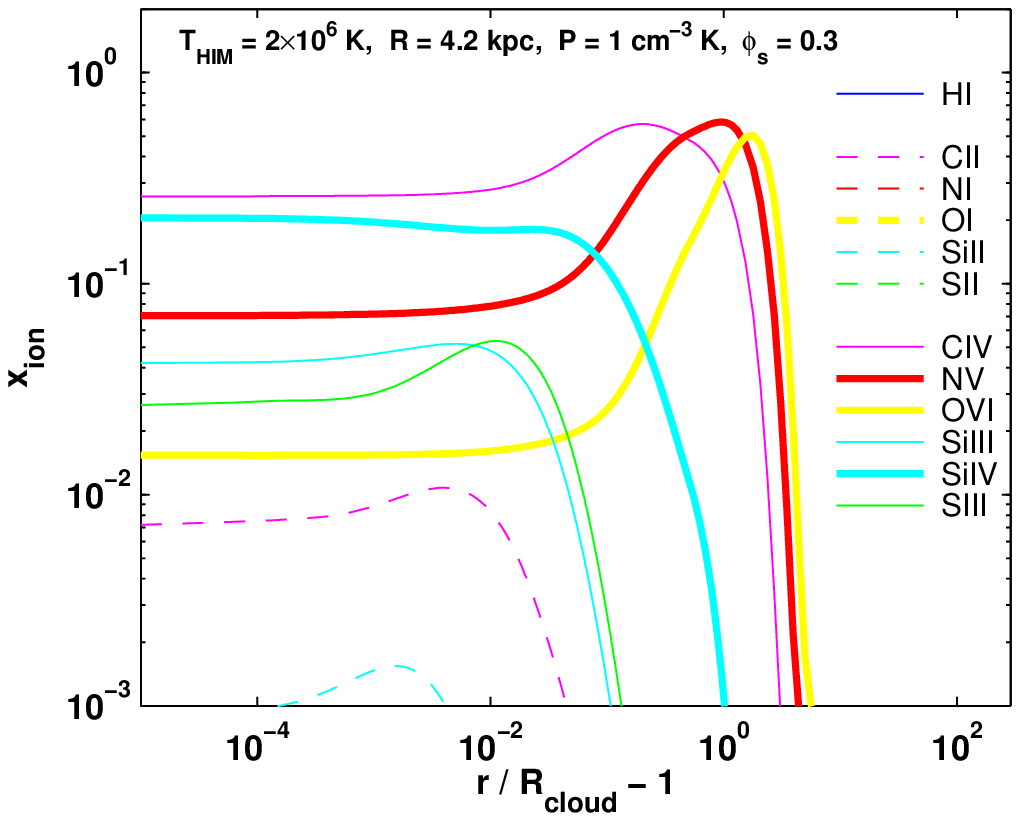}
\caption{Ion fractions as a function of scaled radius
$r/R-1$, in conductive interfaces surrounding dwarf galaxy-scale
halos. The ambient temperature, $T_{\rm HIM}$, cloud radius, $R$, 
and bounding pressure, $P$ are indicated in each panel.}
\end{figure}

\begin{figure}[!h]
\figurenum{8.3,~8.4}
\epsscale{1.0}
\plottwo{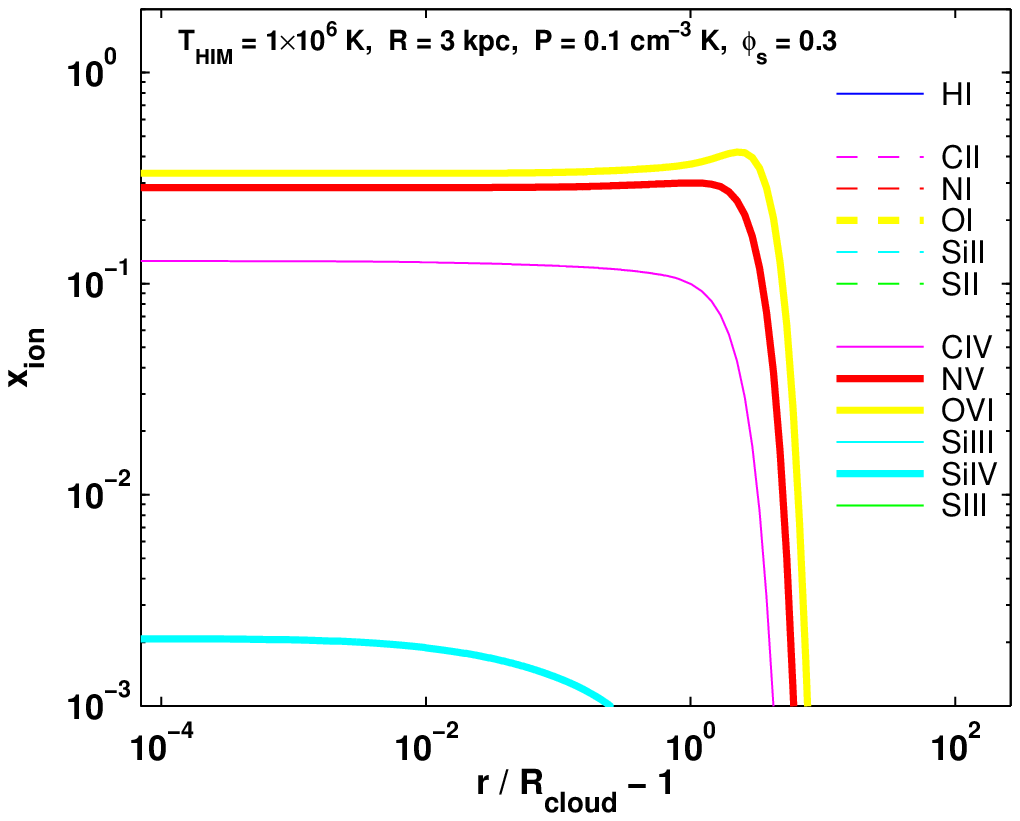}{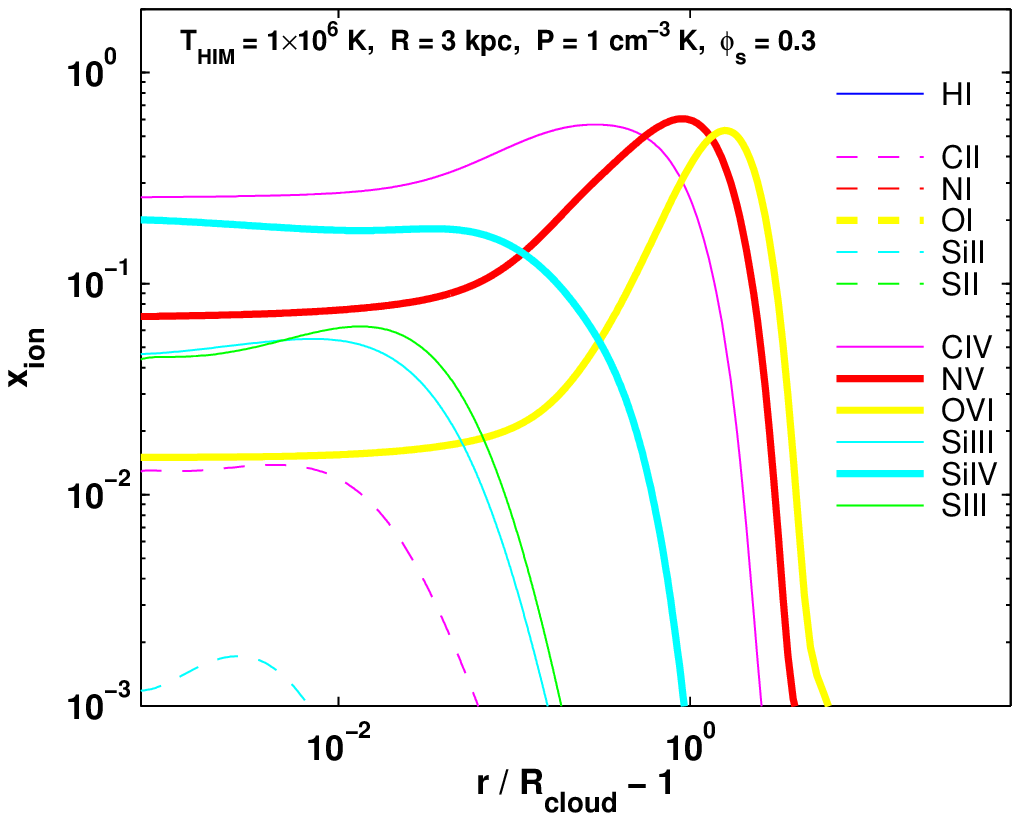}
\caption{Ion fractions versus scaled radius in dwarf galaxy-scale models -- continued.}
\end{figure}

\begin{figure}[!h]
\figurenum{8.5,~8.6}
\epsscale{1.0}
\plottwo{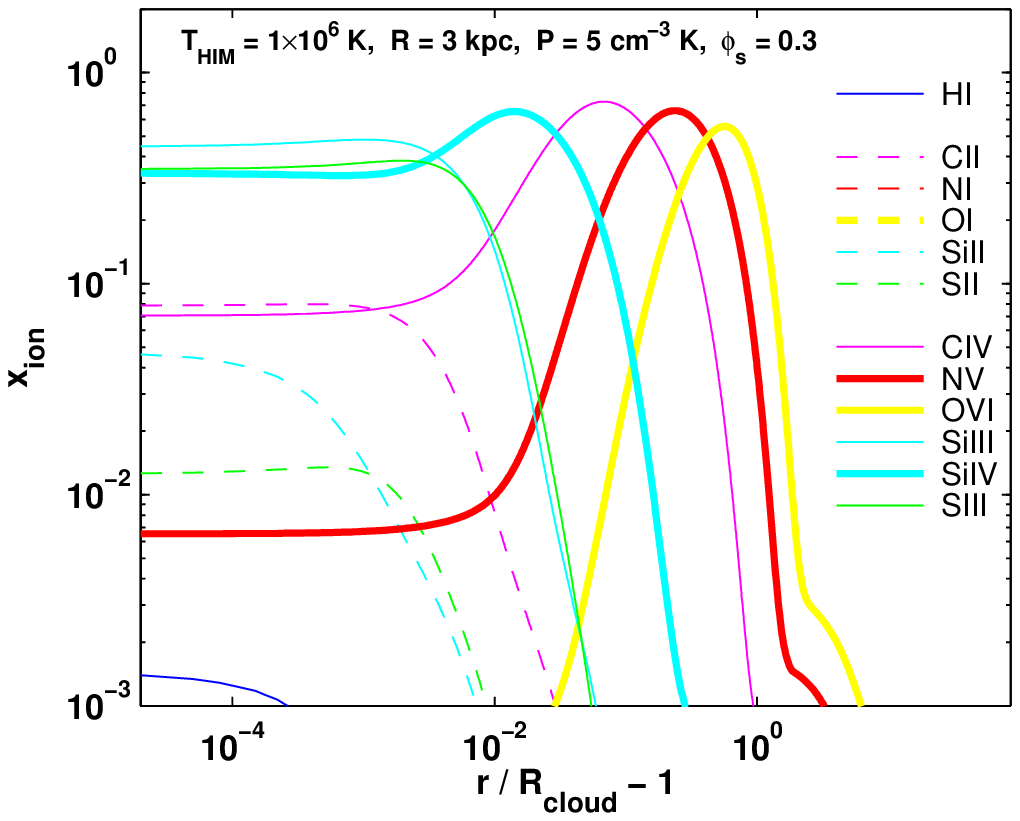}{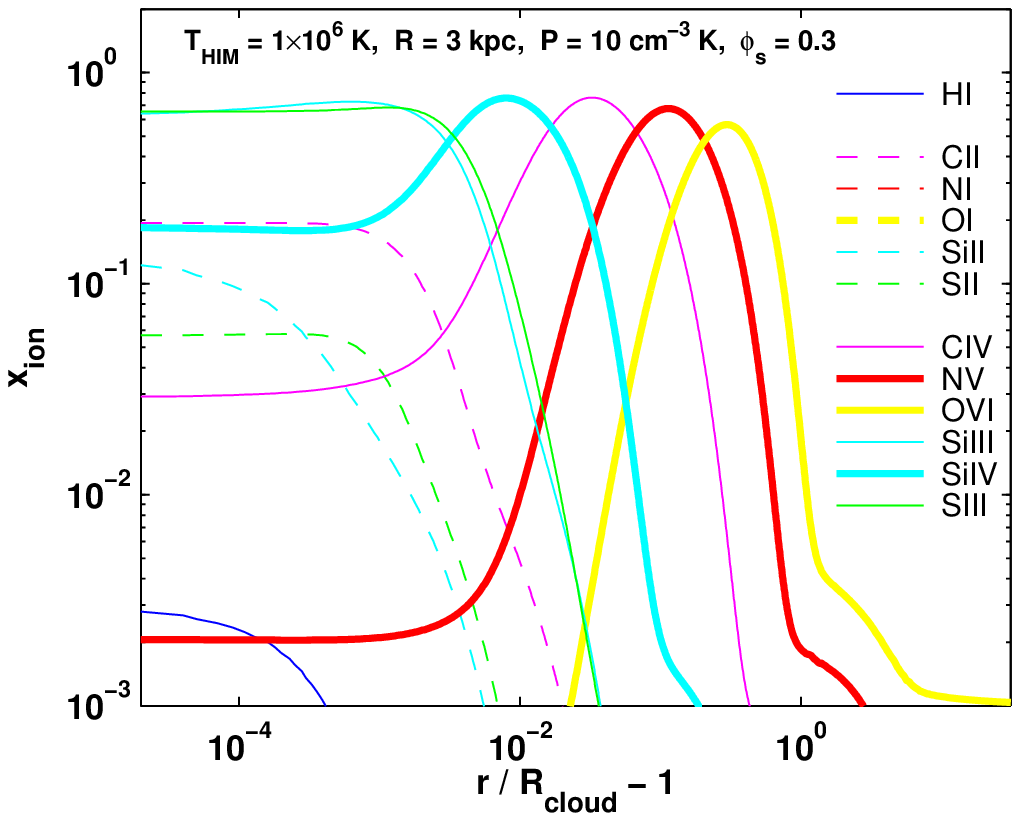}
\caption{Ion fractions versus scaled radius in dwarf galaxy-scale models -- continued.}
\end{figure}

\begin{figure}[!h]
\figurenum{8.7,~8.8}
\epsscale{1.0}
\plottwo{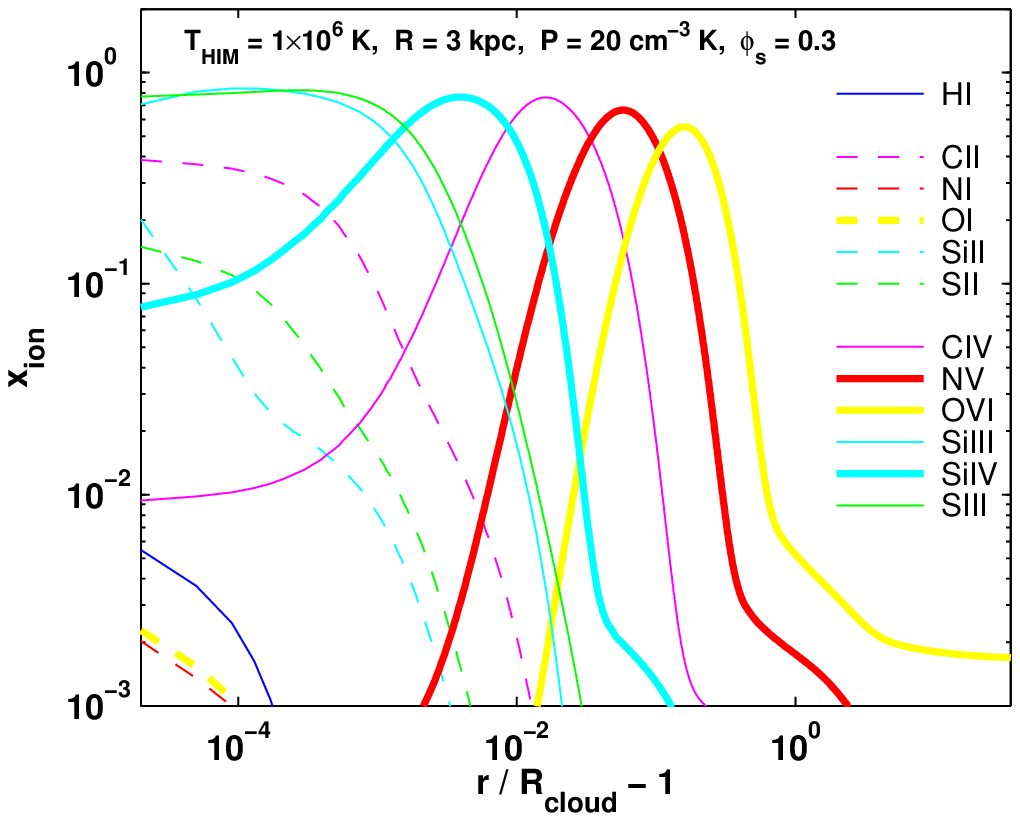}{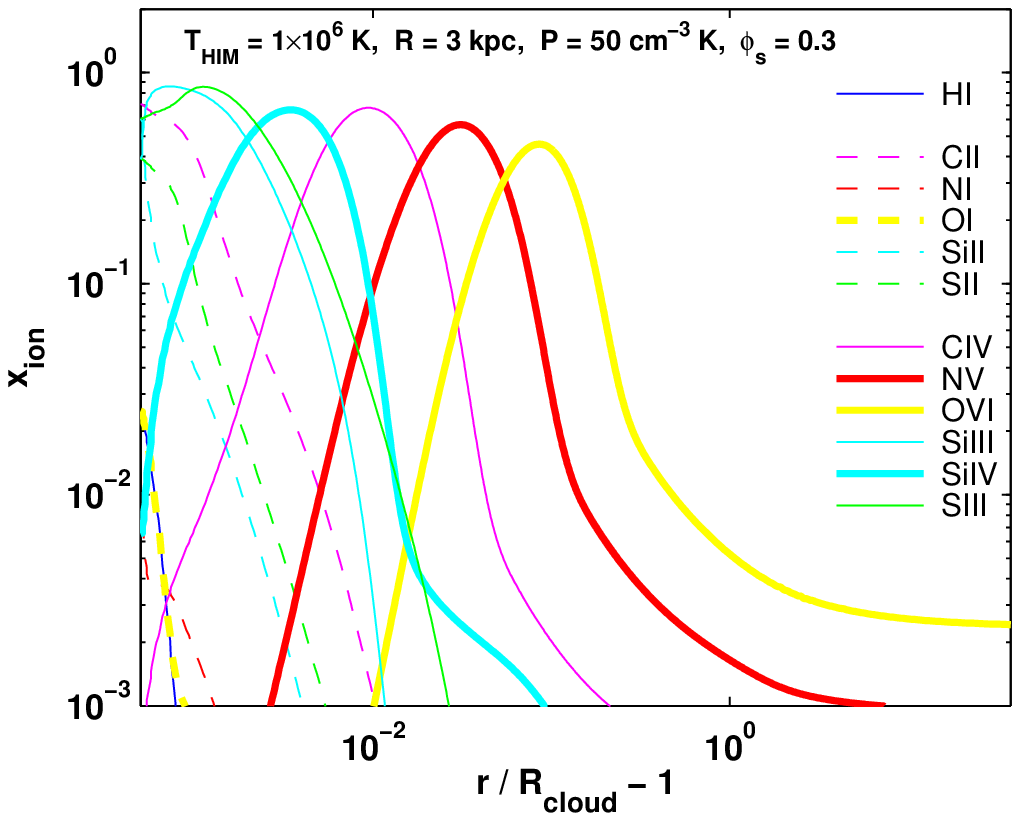}
\caption{Ion fractions versus scaled radius in dwarf galaxy-scale models -- continued.}
\end{figure}

\begin{figure}[!h]
\figurenum{8.9,~8.10}
\epsscale{1.0}
\plottwo{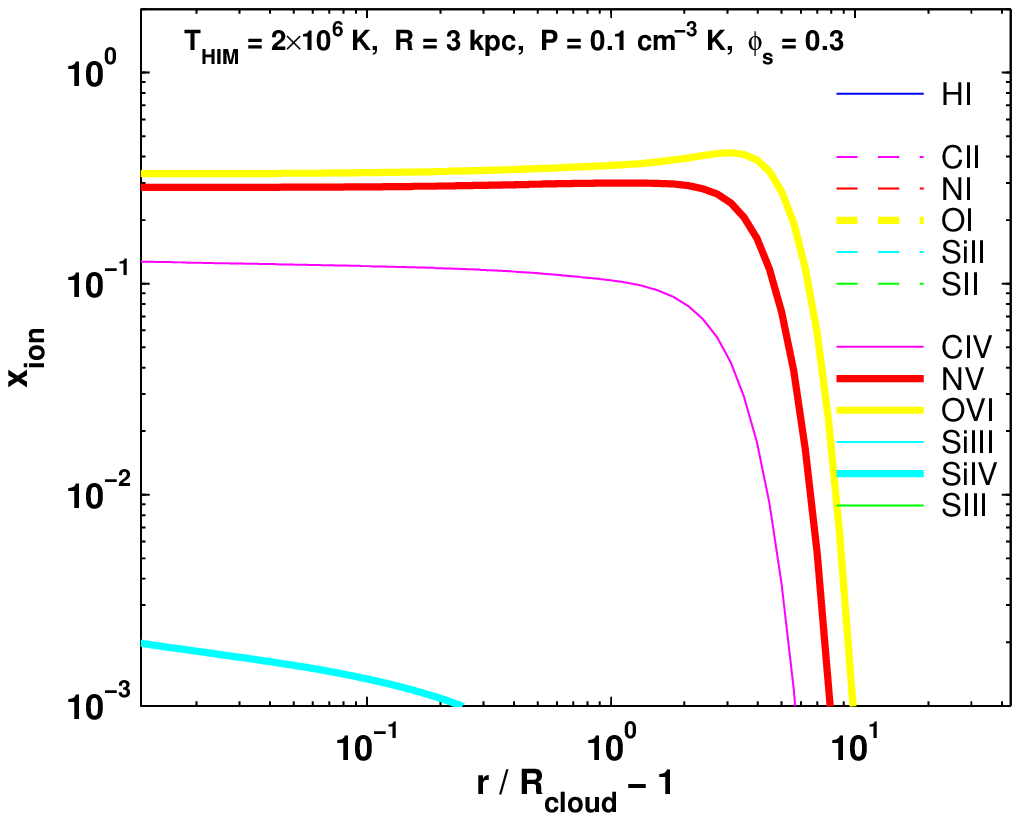}{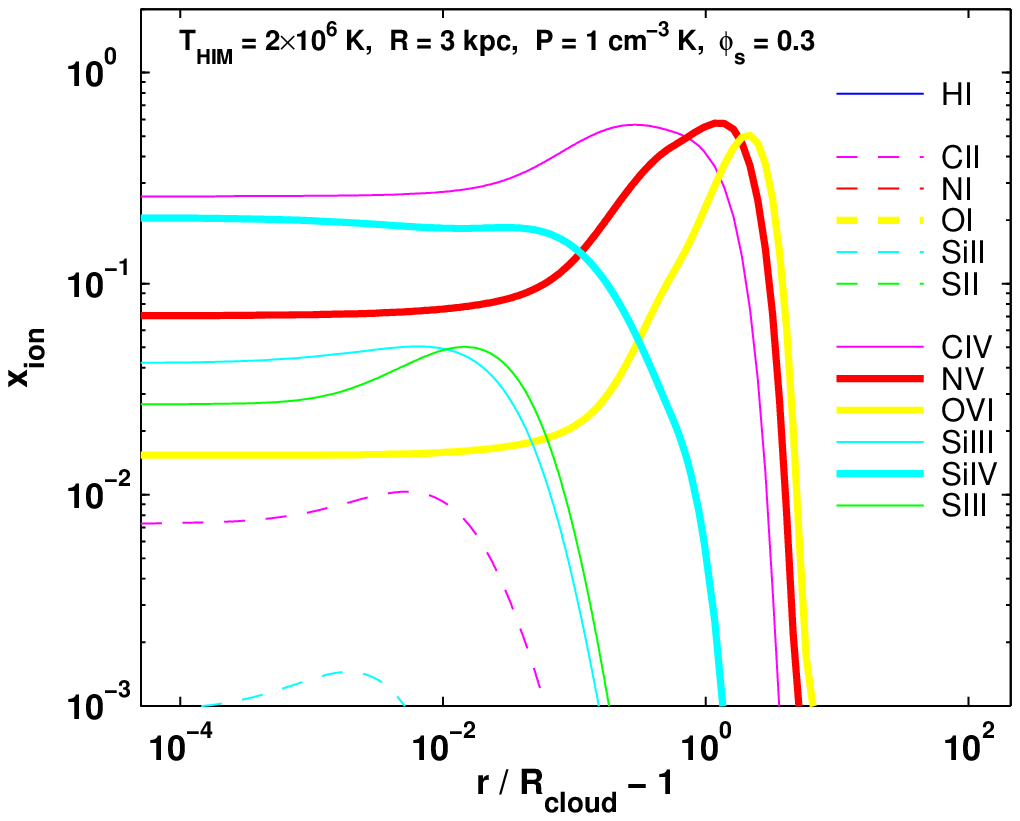}
\caption{Ion fractions versus scaled radius in dwarf galaxy-scale models -- continued.}
\end{figure}

\begin{figure}[!h]
\figurenum{8.11,~8.12}
\epsscale{1.0}
\plottwo{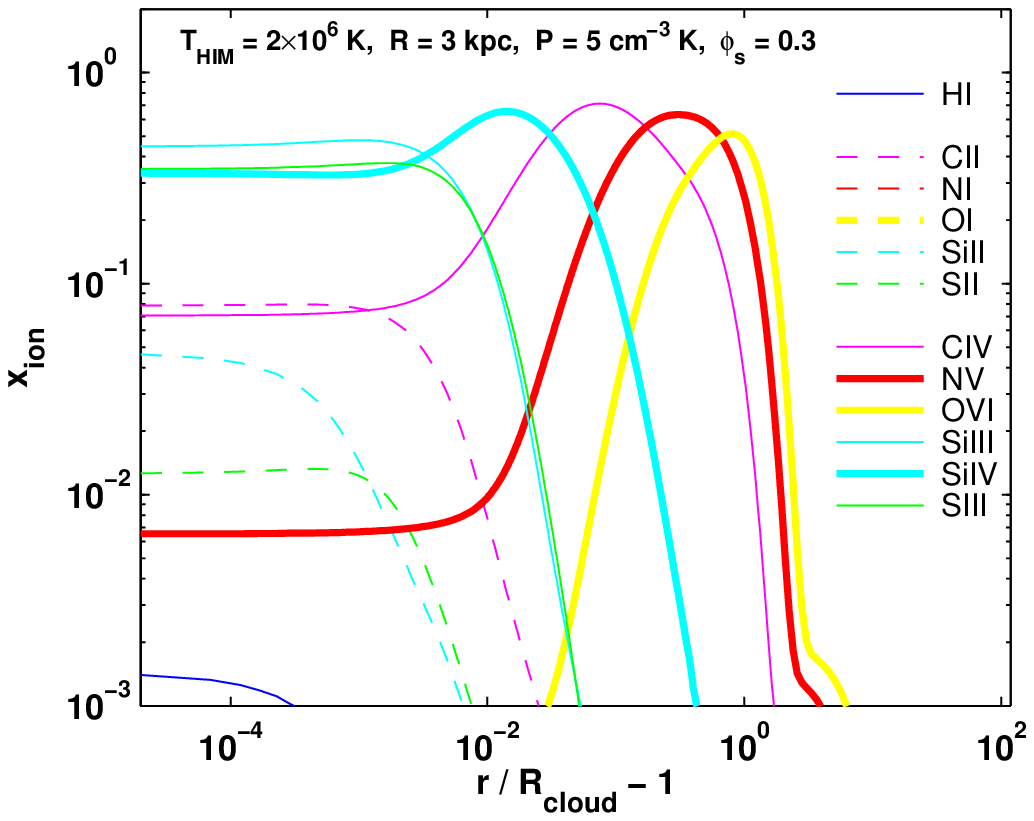}{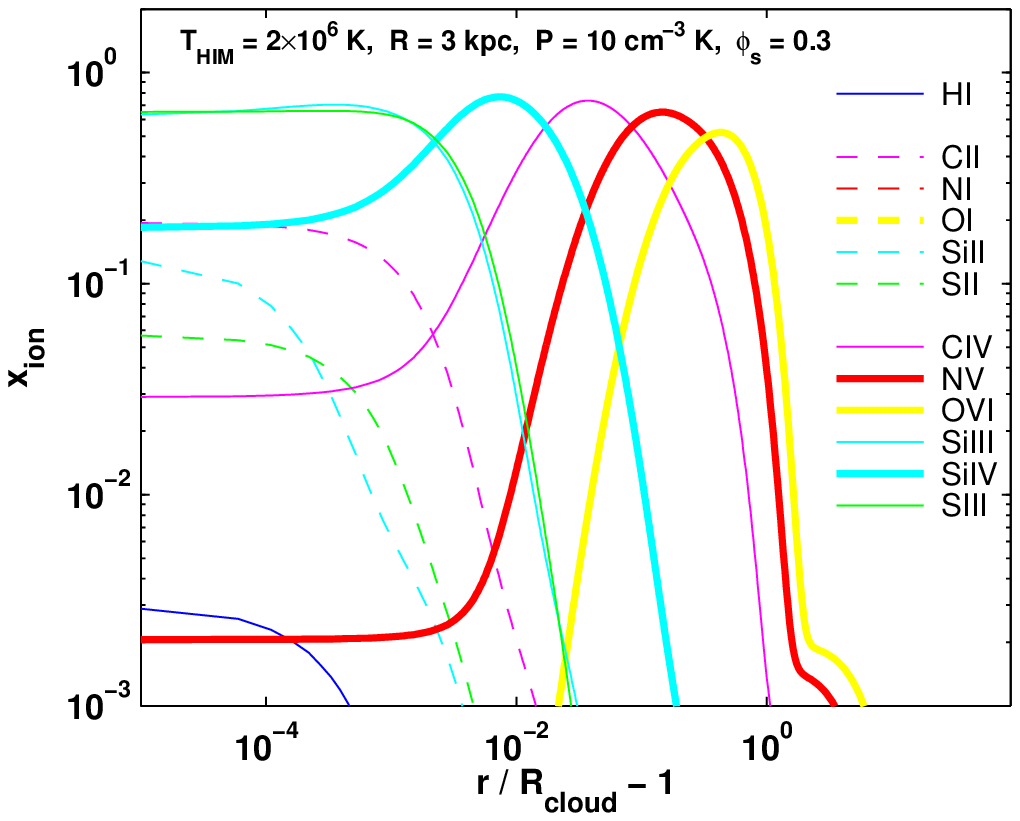}
\caption{Ion fractions versus scaled radius in dwarf galaxy-scale models -- continued.}
\end{figure}

\begin{figure}[!h]
\figurenum{8.13,~8.14}
\epsscale{1.0}
\plottwo{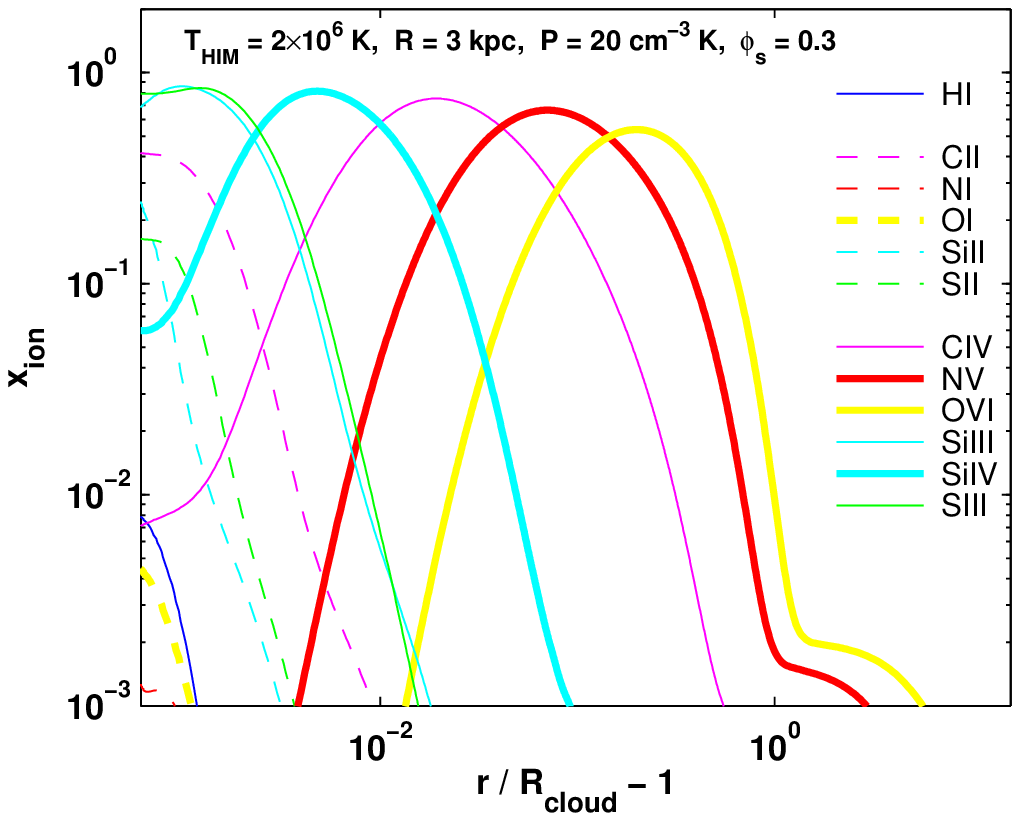}{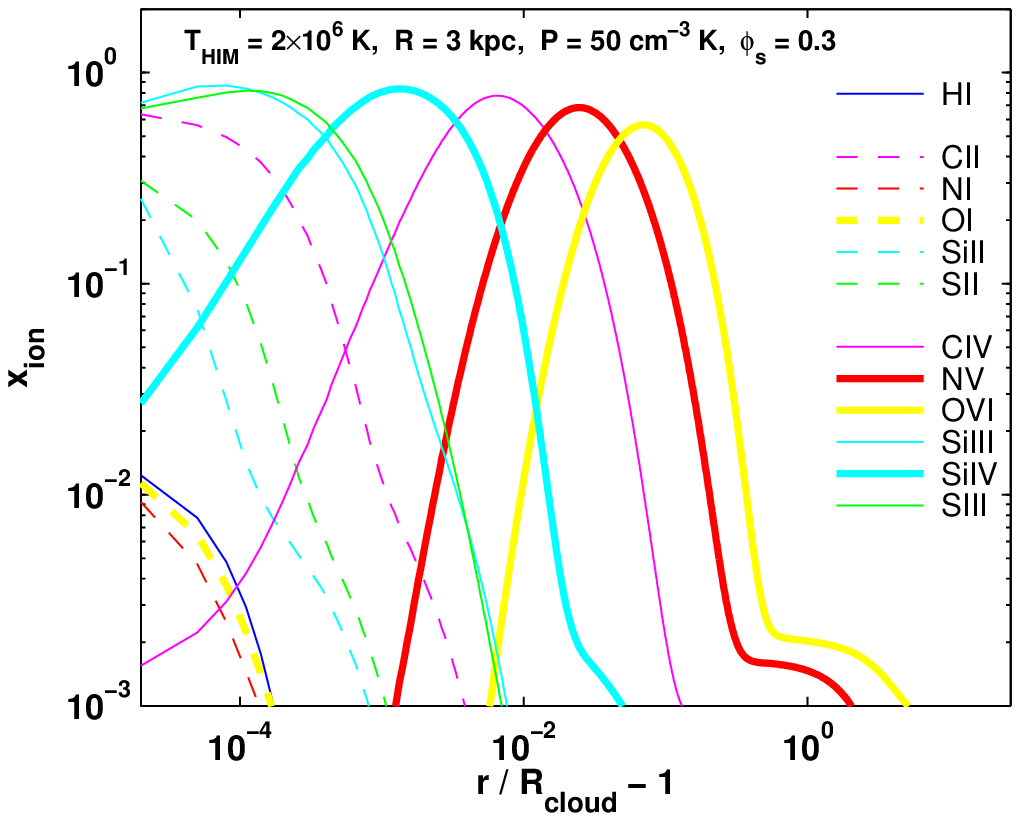}
\caption{Ion fractions versus scaled radius in dwarf galaxy-scale models -- continued.}
\end{figure}

\begin{figure}[!h]
\figurenum{8.15,~8.16}
\epsscale{1.0}
\plottwo{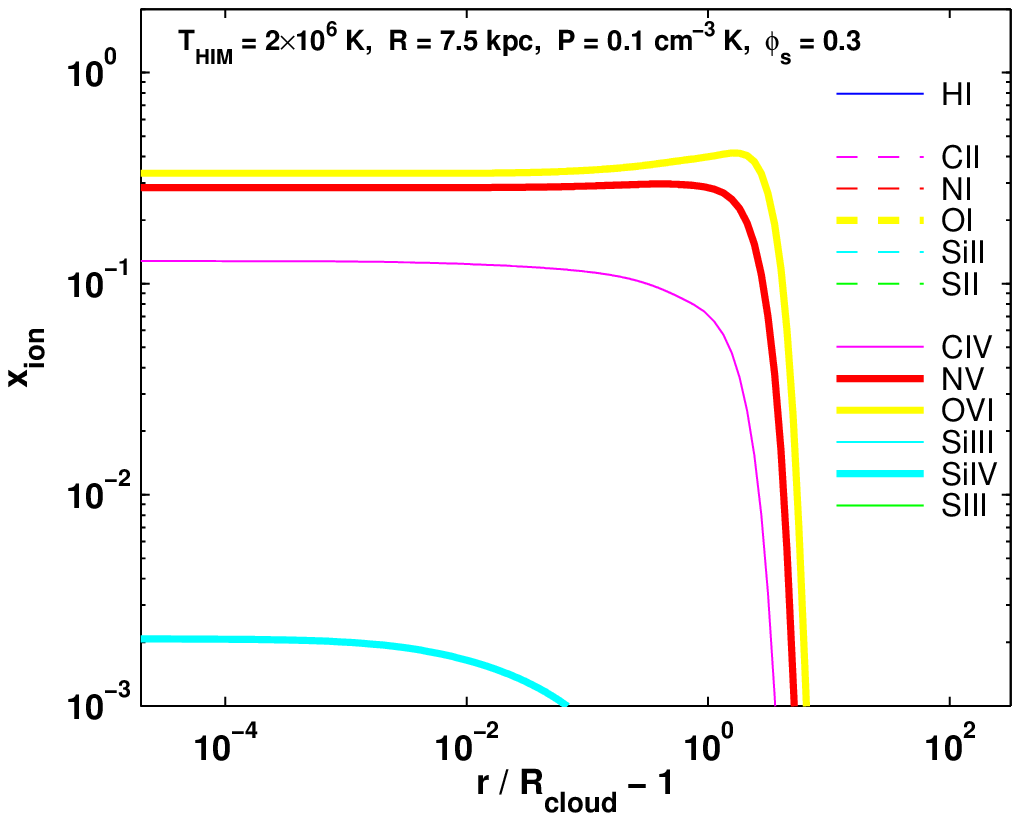}{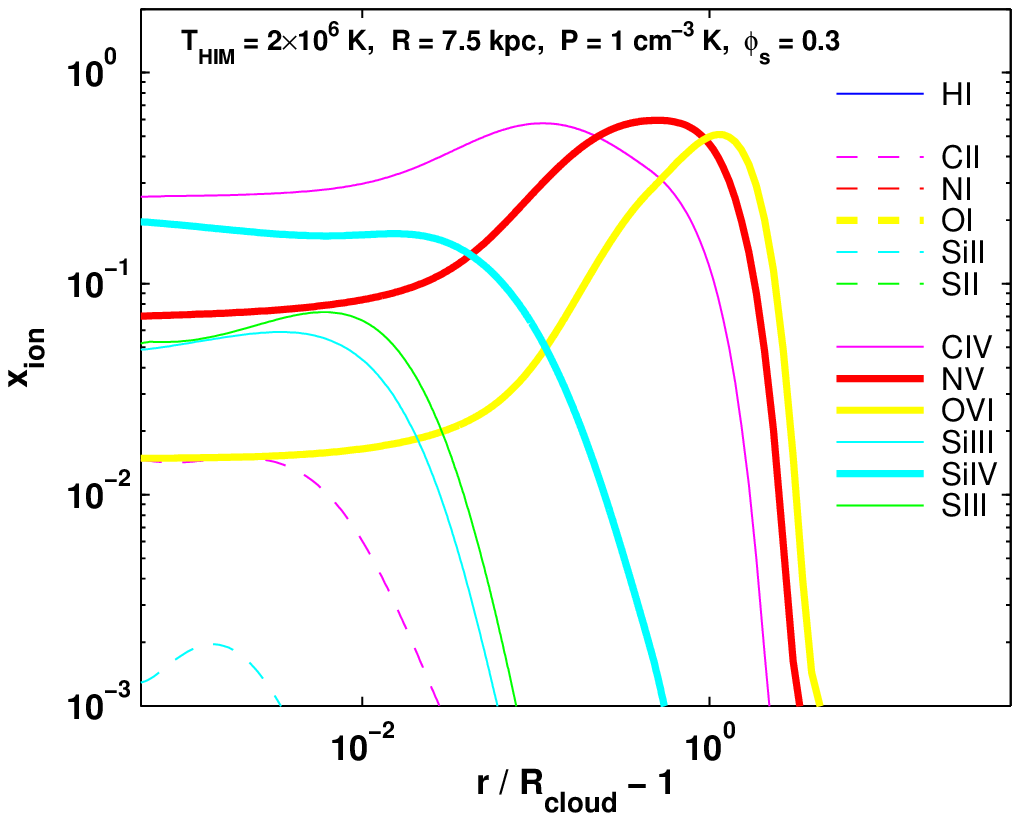}
\caption{Ion fractions versus scaled radius in dwarf galaxy-scale models -- continued.}
\end{figure}

\begin{figure}[!h]
\figurenum{8.17,~8.18}
\epsscale{1.0}
\plottwo{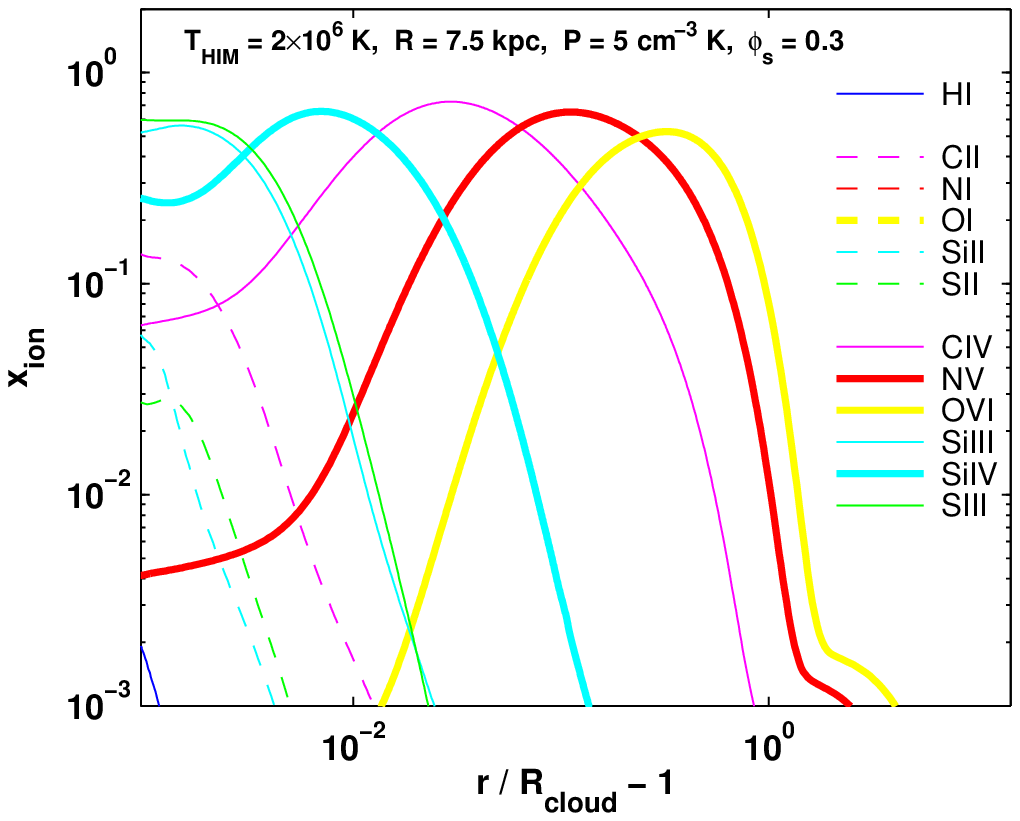}{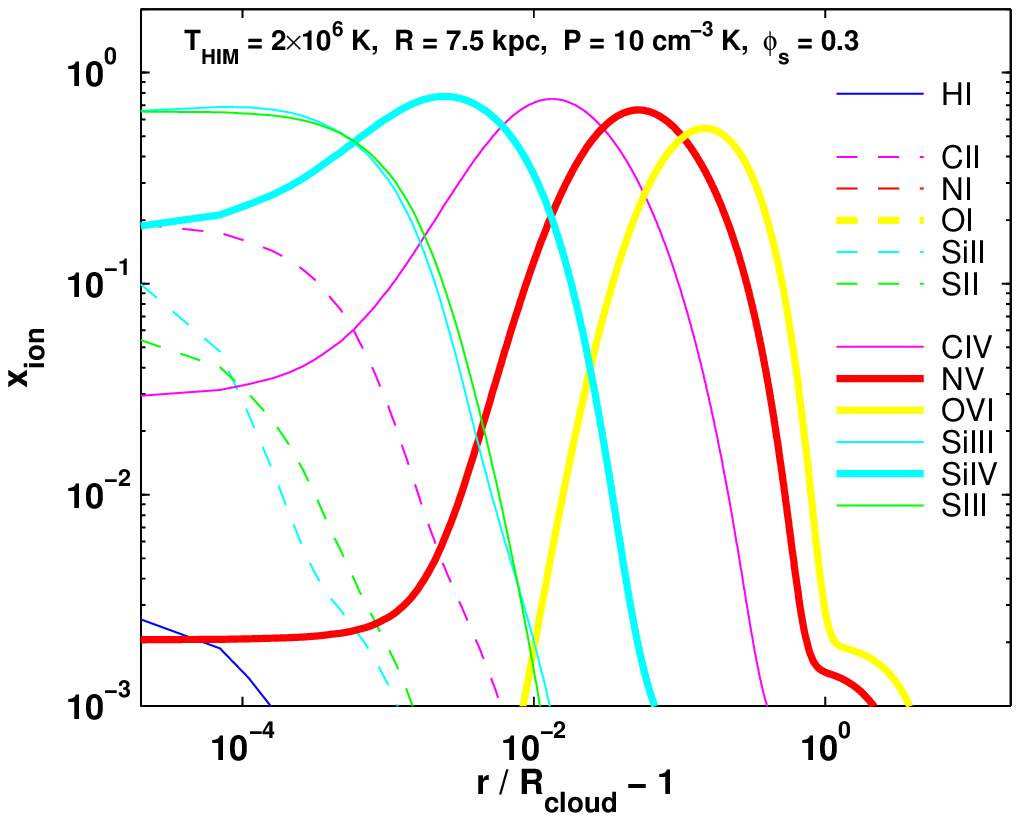}
\caption{Ion fractions versus scaled radius in dwarf galaxy-scale models -- continued.}
\end{figure}

\begin{figure}[!h]
\figurenum{8.19,~8.20}
\epsscale{1.0}
\plottwo{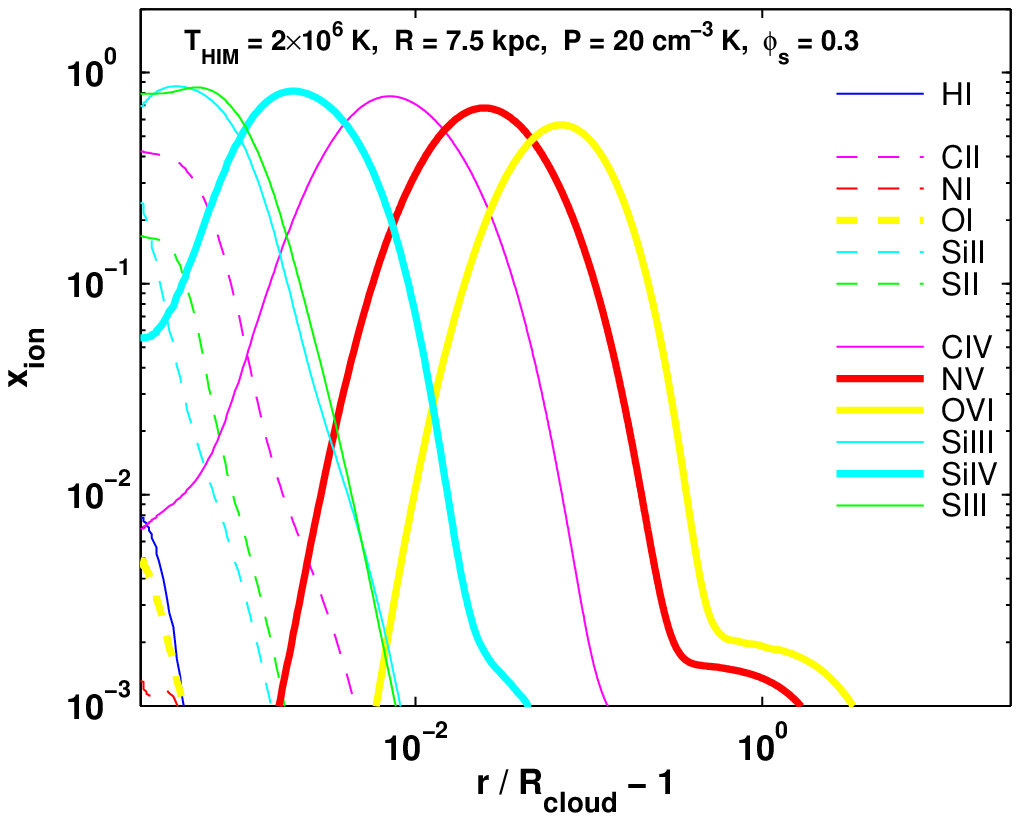}{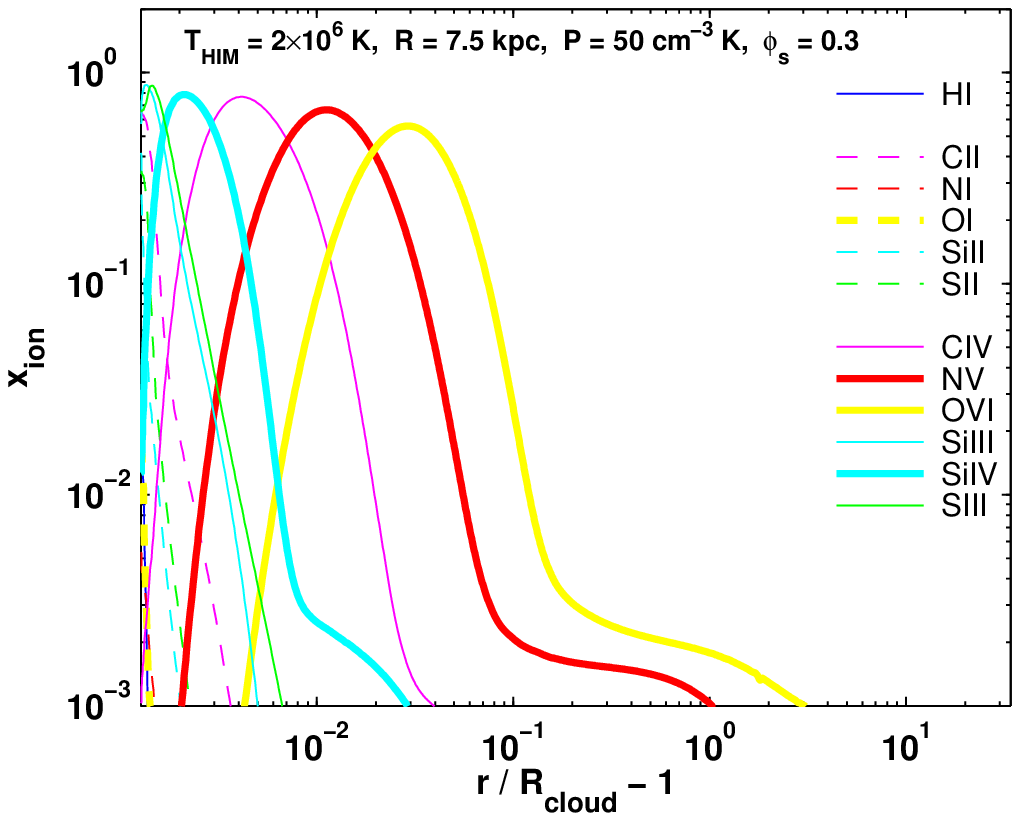}
\caption{Ion fractions versus scaled radius in dwarf galaxy-scale models -- continued.}
\end{figure}

\begin{figure}[!h]
\figurenum{8.21,~8.22}
\epsscale{1.0}
\plottwo{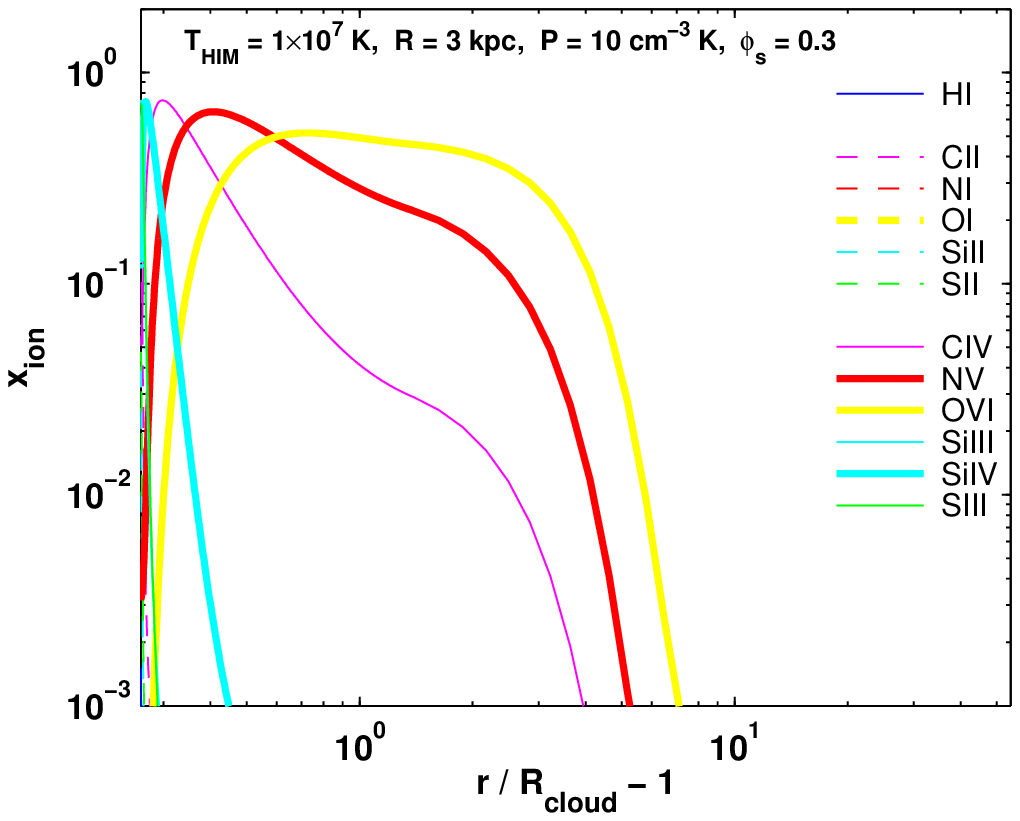}{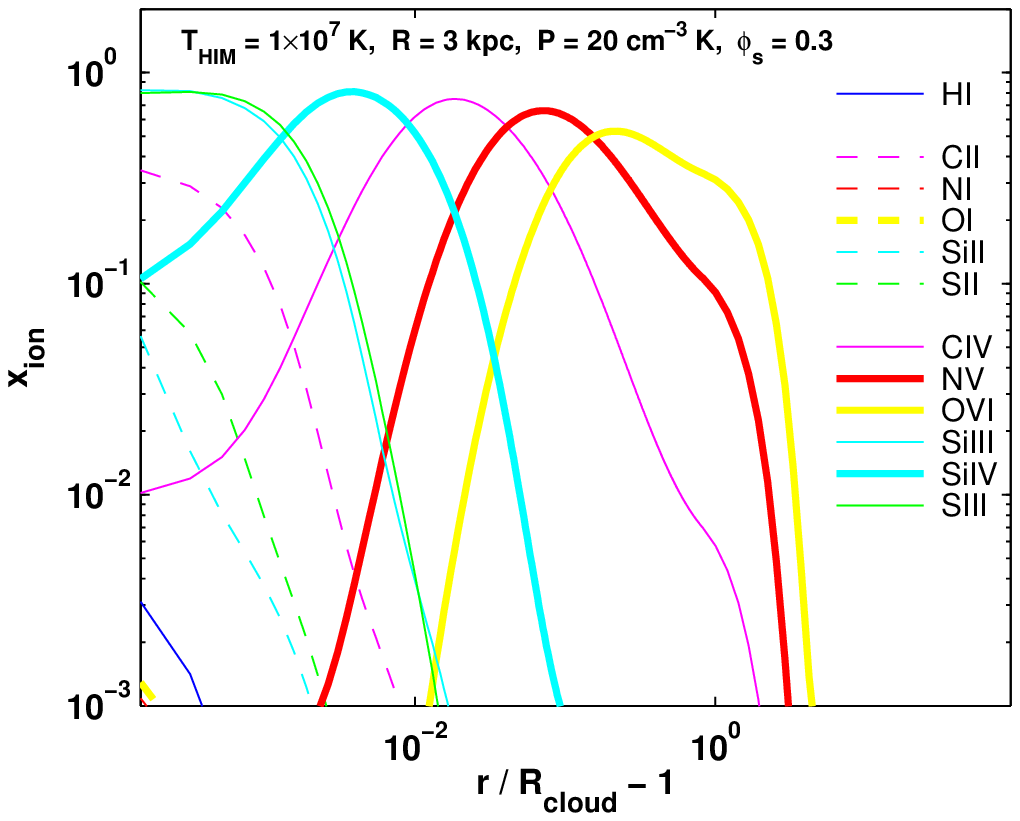}
\caption{Ion fractions versus scaled radius in dwarf galaxy-scale models -- continued.}
\end{figure}

\begin{figure}[!h]
\figurenum{8.23,~8.24}
\epsscale{1.0}
\plottwo{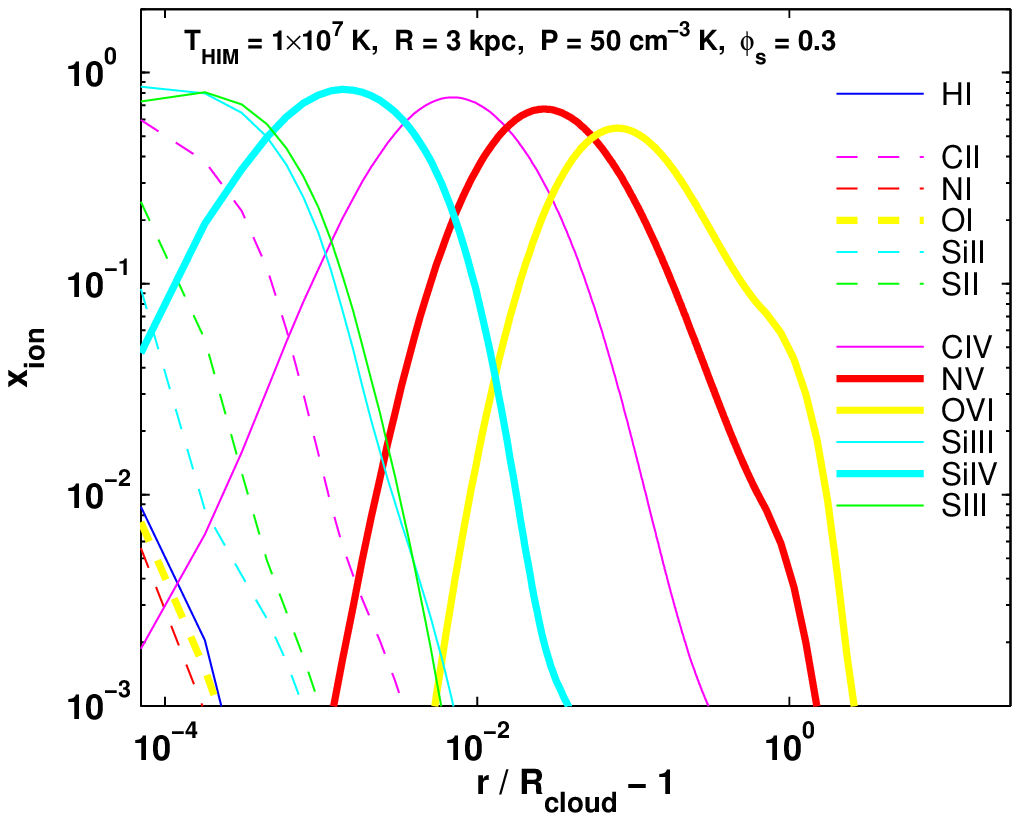}{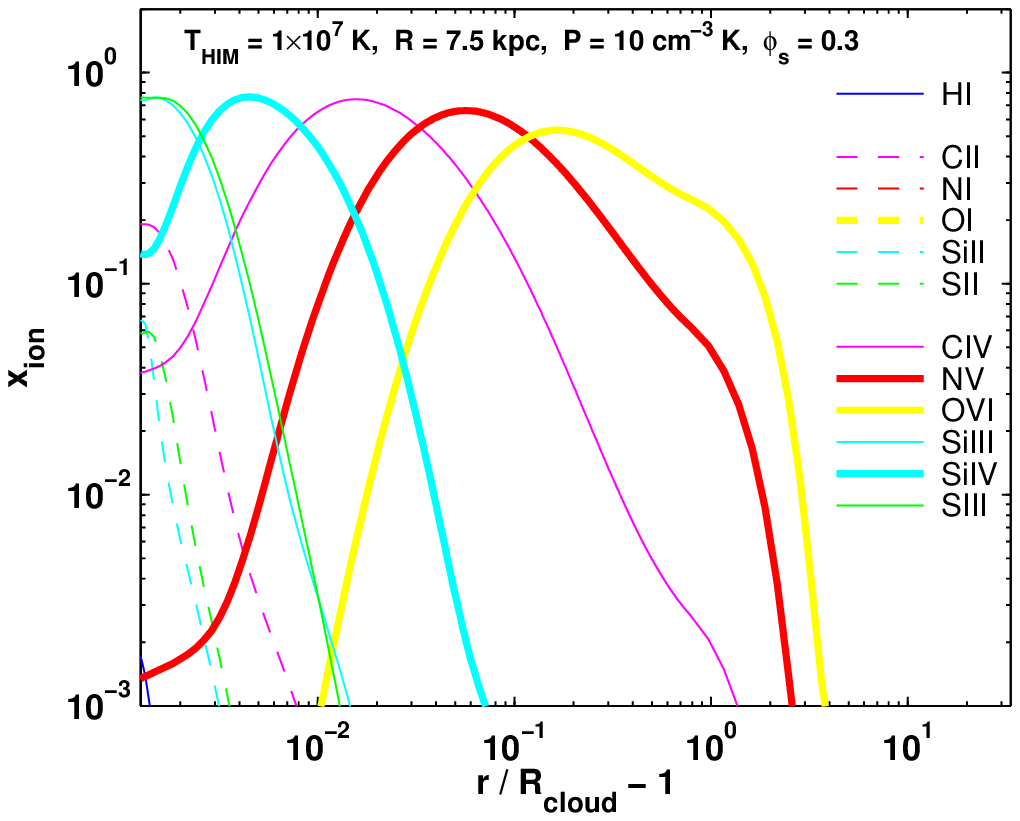}
\caption{Ion fractions versus scaled radius in dwarf galaxy-scale models -- continued.}
\end{figure}

\begin{figure}[!h]
\figurenum{8.25,~8.26}
\epsscale{1.0}
\plottwo{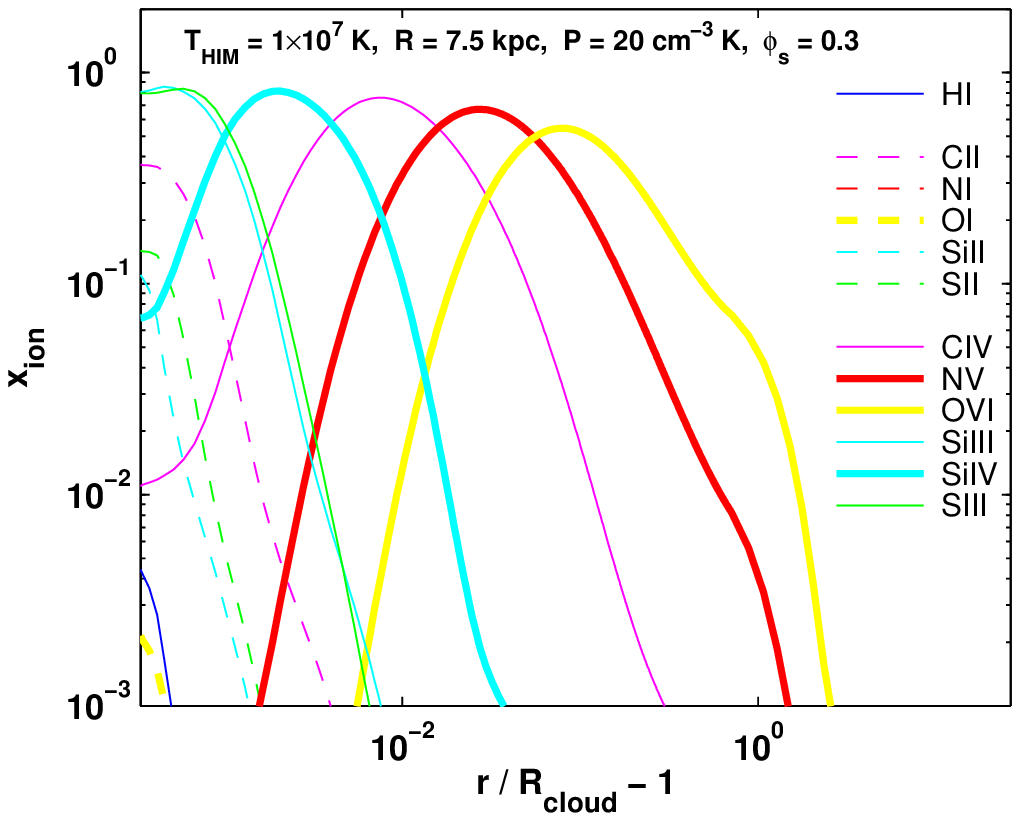}{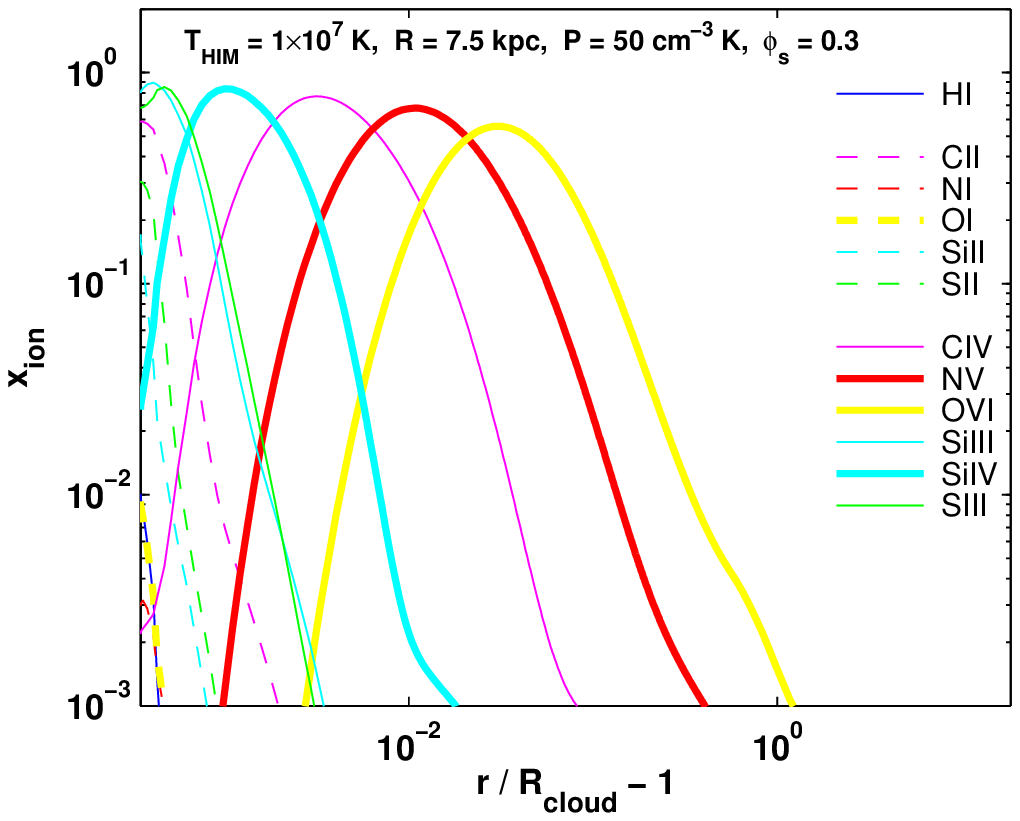}
\caption{Ion fractions versus scaled radius in dwarf galaxy-scale models -- continued.}
\end{figure}


\begin{thebibliography}{}

\bibitem[Aldrovandi \& Pequignot(1973)]{1973A&A....25..137A} Aldrovandi,
S.~M.~V., \& Pequignot, D.\ 1973, \aap, 25, 137

\bibitem[Allen et al.(2008)]{2008ApJS..178...20A} Allen, M.~G., Groves, 
B.~A., Dopita, M.~A., Sutherland, R.~S., 
\& Kewley, L.~J.\ 2008, \apjs, 178, 20 

\bibitem[Altun et al.(2006)]{2006A&A...447.1165A} Altun, Z., Yumak, A.,
Badnell, N.~R., Loch, S.~D., \& Pindzola, M.~S.\ 2006, \aap, 447, 1165

\bibitem[Altun et al.(2005)]{2005A&A...433..395A} Altun, Z., Yumak, A.,
Badnell, N.~R., Colgan, J., \& Pindzola, M.~S.\ 2005, \aap, 433, 395

\bibitem[Altun et al.(2004)]{2004A&A...420..775A} Altun, Z., Yumak, A.,
Badnell, N.~R., Colgan, J., \& Pindzola, M.~S.\ 2004, \aap, 420, 775

\bibitem[Arnaud \& Raymond(1992)]{1992ApJ...398..394A} Arnaud, M., \&
Raymond, J.\ 1992, \apj, 398, 394

\bibitem[Arnaud \& Rothenflug(1985)]{1985A&AS...60..425A} Arnaud, M., \& 
Rothenflug, R.\ 1985, \aaps, 60, 425 

\bibitem[Badnell(2006)]{2006A&A...447..389B} Badnell, N.~R.\ 2006, \aap,
447, 389

\bibitem[Badnell et al.(2003)]{2003A&A...406.1151B} Badnell, N.~R., et al.\ 2003, \aap, 406, 1151

\bibitem[Balbus \& McKee(1982)]{1982ApJ...252..529B} 
Balbus, S.~A., \& McKee, C.~F.\ 1982, \apj, 252, 529 

\bibitem[Ballet et al.(1986)]{1986A&A...161...12B} 
Ballet, J., Arnaud, M., \& Rothenflug, R.\ 1986, \aap, 161, 12 

\bibitem[Balsara et al.(2008)]{2008MNRAS.386..642B} Balsara, D.~S., 
Bendinelli, A.~J., Tilley, D.~A., Massari, A.~R., 
\& Howk, J.~C.\ 2008, \mnras, 386, 642 

\bibitem[Bernstein, Freedman, \& Madore(2002)]{2002ApJ...571...56B} 
Bernstein, R.~A., Freedman, W.~L., \& Madore, B.~F.\ 2002, \apj, 571, 56 

\bibitem[Boehringer \& Hartquist(1987)]{1987MNRAS.228..915B} 
Boehringer, H., \& Hartquist, T.~W.\ 1987, \mnras, 228, 915 

\bibitem[Borkowski et al.(1990)]{1990ApJ...355..501B} Borkowski, K.~J., 
Balbus, S.~A., \& Fristrom, C.~C.\ 1990, \apj, 355, 501 

\bibitem[Chen, Fabian, \& Gendreau(1997)]{1997MNRAS.285..449C} Chen, L.-W., 
Fabian, A.~C., \& Gendreau, K.~C.\ 1997, \mnras, 285, 449

\bibitem[Clarke et al. (1998)]{03}Clarke, N.J., et al.\ 1998, J. Phys. B: At. Mol. Opt. Phys. 33, 533

\bibitem[Colgan et al.(2005)]{2005A&A...429..369C} Colgan, J., Pindzola,
M.~S., \& Badnell, N.~R.\ 2005, \aap, 429, 369

\bibitem[Colgan et al.(2004)]{2004A&A...417.1183C} Colgan, J., Pindzola,
M.~S., \& Badnell, N.~R.\ 2004, \aap, 417, 1183

\bibitem[Colgan et al.(2003)]{2003A&A...412..597C} Colgan, J., Pindzola,
M.~S., Whiteford, A.~D., \& Badnell, N.~R.\ 2003, \aap, 412, 597

\bibitem[Collins et al.(2009)]{2009ApJ...705..962C} Collins, J.~A., Shull, 
J.~M., \& Giroux, M.~L.\ 2009, \apj, 705, 962 

\bibitem[Collins et al.(2007)]{2007ApJ...657..271C} Collins, J.~A., Shull, 
J.~M., \& Giroux, M.~L.\ 2007, \apj, 657, 271 

\bibitem[Collins et al.(2005)]{2005ApJ...623..196C} Collins, J.~A., Shull, 
J.~M., \& Giroux, M.~L.\ 2005, \apj, 623, 196 

\bibitem[Collins et al.(2004)]{2004ApJ...605..216C} Collins, J.~A., Shull, 
J.~M., \& Giroux, M.~L.\ 2004, \apj, 605, 216 

\bibitem[Collins et al.(2003)]{2003ApJ...585..336C} Collins, J.~A., Shull, 
J.~M., \& Giroux, M.~L.\ 2003, \apj, 585, 336 

\bibitem[Cowie \& McKee(1977)]{1977ApJ...211..135C} 
Cowie, L.~L., \& McKee, C.~F.\ 1977, \apj, 211, 135 

\bibitem[Dalton \& Balbus(1993)]{1993ApJ...404..625D} 
Dalton, W.~W., \& Balbus, S.~A.\ 1993, \apj, 404, 625 

\bibitem[Edgar \& Chevalier(1986)]{1986ApJ...310L..27E} 
Edgar, R.~J., \& Chevalier, R.~A.\ 1986, \apjl, 310, L27 

\bibitem[Fang et al.(2006)]{2006ApJ...644..174F} Fang, T., Mckee, C.~F., 
Canizares, C.~R., \& Wolfire, M.\ 2006, \apj, 644, 174 

\bibitem[Faucher-Gigu{\`e}re et al.(2009)]{2009ApJ...703.1416F} 
Faucher-Gigu{\`e}re, C.-A., Lidz, A., Zaldarriaga, M., 
\& Hernquist, L.\ 2009, \apj, 703, 1416 

\bibitem[Ferland et al.(1998)]{1998PASP..110..761F} Ferland, G.~J., 
Korista, K.~T., Verner, D.~A., Ferguson, J.~W., Kingdon, J.~B., \& Verner, 
E.~M.\ 1998, \pasp, 110, 761

\bibitem[Ferland et al.(1997)]{1997ApJ...481L.115F} Ferland, G.~J., 
Korista, K.~T., Verner, D.~A., \& Dalgarno, A.\ 1997, \apjl, 481, L115 

\bibitem[Fox et al.(2006)]{2006ApJS..165..229F} Fox, A.~J., Savage, B.~D., 
\& Wakker, B.~P.\ 2006, \apjs, 165, 229 

\bibitem[Fox et al.(2005)]{2005ApJ...630..332F} Fox, A.~J., Wakker, B.~P., 
Savage, B.~D., Tripp, T.~M., Sembach, K.~R., \& Bland-Hawthorn, J.\ 2005, 
\apj, 630, 332 

\bibitem[Fox et al.(2004)]{2004ApJ...602..738F} Fox, A.~J., Savage, B.~D., 
Wakker, B.~P., Richter, P., Sembach, K.~R., \& Tripp, T.~M.\ 2004, \apj, 
602, 738 

\bibitem[Giovanelli et al.(2010)]{2010ApJ...708L..22G} Giovanelli, R., 
Haynes, M.~P., Kent, B.~R., \& Adams, E.~A.~K.\ 2010, \apjl, 708, L22 

\bibitem[Gnat \& Sternberg(2009)]{2009ApJ...693.1514G} Gnat, O., 
\& Sternberg, A.\ 2009, \apj, 693, 1514 

\bibitem[Gnat \& Sternberg(2007)]{2007ApJS..168..213G} Gnat, O., \& 
Sternberg, A.\ 2007, \apjs, 168, 213 

\bibitem[Gnat \& Sternberg(2004)]{2004ApJ...608..229G} Gnat, O., \& 
Sternberg, A.\ 2004, \apj, 608, 229 

\bibitem[Haardt \& Madau(1996)]{1996ApJ...461...20H} Haardt, F.~\& Madau, 
P.\ 1996, \apj, 461, 20 

\bibitem[Heckman et al.(2002)]{2002ApJ...577..691H} Heckman, T.~M., Norman, 
C.~A., Strickland, D.~K., \& Sembach, K.~R.\ 2002, \apj, 577, 691 

\bibitem[Hindmarsh (1983)]{05}Hindmarsh, A.~C., \ 1983, ODEPACK, 
A Systematized Collection of 
ODE Solvers, in Scientific Computing, R. S. Stepleman et al. (eds.), North-Holland, 
Amsterdam, 1983 (vol. 1 of IMACS Transactions on Scientific Computation), pp. 55-64.

\bibitem[Jenkins(2009)]{2009SSRv..143..205J} Jenkins, E.~B.\ 2009, Space 
Science Reviews, 143, 205 

\bibitem[Kepner et al.(1999)]{1999AJ....117.2063K} Kepner, J., Tripp, 
T.~M., Abel, T., \& Spergel, D.\ 1999, \aj, 117, 2063 

\bibitem[Kingdon \& Ferland(1996)]{1996ApJS..106..205K} Kingdon, J.~B., \& 
Ferland, G.~J.\ 1996, \apjs, 106, 205 

\bibitem[Landini \& Fossi(1991)]{1991A&AS...91..183L} Landini, M., \&
Fossi, B.~C.~M.\ 1991, \aaps, 91, 183

\bibitem[Landini \& Monsignori Fossi(1990)]{1990A&AS...82..229L}
Landini, M., \& Monsignori Fossi, B.~C.\ 1990, \aaps, 82, 229

\bibitem[Martin, Hurwitz, \& Bowyer(1991)]{1991ApJ...379..549M} Martin, C., 
Hurwitz, M., \& Bowyer, S.\ 1991, \apj, 379, 549 

\bibitem[McKee \& Begelman(1990)]{1990ApJ...358..392M} 
McKee, C.~F., \& Begelman, M.~C.\ 1990, \apj, 358, 392 

\bibitem[McKee \& Cowie(1977)]{1977ApJ...215..213M} 
McKee, C.~F., \& Cowie, L.~L.\ 1977, \apj, 215, 213 

\bibitem[Mitnik \& Badnell(2004)]{2004A&A...425.1153M} Mitnik, D.~M., \&
Badnell, N.~R.\ 2004, \aap, 425, 1153

\bibitem[Murphy et al.(2000)]{2000ApJ...538L..35M} Murphy, E.~M.~et al.\ 
2000, \apjl, 538, L35 

\bibitem[McCray(1987)]{1987sap..book..255M}McCray, R. 1987, 
in Spectroscopy of Astrophysical Plasmas,
Eds.~A.~Dalgarno \& D.~Layzer (Cambridge University Press), p. 255

\bibitem[Nagashima et al.(2007)]{2007ASPC..365..121N} Nagashima, M., 
Inutsuka, S., 
\& Koyama, H.\ 2007, SINS - Small Ionized and Neutral Structures in the Diffuse Interstellar Medium, 365, 121 

\bibitem[Narayanan et al.(2010)]{2010AAS...21546005N} Narayanan, A., et 
al.\ 2010, American Astronomical Society Meeting Abstracts, 215, \#460.05 


\bibitem[Nipoti \& Binney(2007)]{2007MNRAS.382.1481N} 
Nipoti, C., \& Binney, J.\ 2007, \mnras, 382, 1481 

\bibitem[Pequignot et al.(1991)]{1991A&A...251..680P} Pequignot, D.,
Petitjean, P., \& Boisson, C.\ 1991, \aap, 251, 680


\bibitem[Savage \& Lehner(2006)]{2006ApJS..162..134S} 
Savage, B.~D., \& Lehner, N.\ 2006, \apjs, 162, 134 

\bibitem[Savage et al.(2005)]{2005ApJ...626..776S} Savage, B.~D., Lehner, 
N., Wakker, B.~P., Sembach, K.~R., \& Tripp, T.~M.\ 2005, \apj, 626, 776 

\bibitem[Scott et al.(2004)]{2004ApJ...615..135S} Scott, J.~E., Kriss, 
G.~A., Brotherton, M., Green, R.~F., Hutchings, J., Shull, J.~M., 
\& Zheng, W.\ 2004, \apj, 615, 135 

\bibitem[Sembach et al.(2003)]{2003ApJS..146..165S} Sembach, K.~R.~et al.\ 
2003, \apjs, 146, 165 

\bibitem[Sembach, Gibson, Fenner, \& Putman(2002)]{2002ApJ...572..178S} 
Sembach, K.~R., Gibson, B.~K., Fenner, Y., \& Putman, M.~E.\ 2002, ApJ, 
572, 178 

\bibitem[Sembach et al.(2000)]{2000ApJ...538L..31S} Sembach, K.~R.~et al.\ 
2000, ApJL, 538, L31

\bibitem[Sembach, Savage, Lu, \& Murphy(1999)]{1999ApJ...515..108S} 
Sembach, K.~R., Savage, B.~D., Lu, L., \& Murphy, E.~M.\ 1999, ApJ, 515, 
108

\bibitem[Shelton(1998)]{1998ApJ...504..785S} Shelton, R.~L.\ 1998, \apj, 
504, 785 

\bibitem[Shull et al.(2009)]{2009ApJ...699..754S} Shull, J.~M., Jones, 
J.~R., Danforth, C.~W., \& Collins, J.~A.\ 2009, \apj, 699, 754 

\bibitem[Shull et al.(2004)]{2004ApJ...600..570S} Shull, J.~M., Tumlinson, 
J., Giroux, M.~L., Kriss, G.~A., \& Reimers, D.\ 2004, \apj, 600, 570 

\bibitem[Shull et al.(1999)]{1999AJ....118.1450S} Shull, J.~M., Roberts, 
D., Giroux, M.~L., Penton, S.~V., \& Fardal, M.~A.\ 1999, \aj, 118, 1450 

\bibitem[Shull \& van Steenberg(1982)]{1982ApJS...48...95S} Shull, J.~M., 
\& van Steenberg, M.\ 1982, \apjs, 48, 95 

\bibitem[Slavin(2007)]{2007ASPC..365..113S} Slavin, J.~D.\ 2007, SINS - 
Small Ionized and Neutral Structures in the Diffuse Interstellar Medium, 
365, 113 

\bibitem[Slavin(1989)]{1989ApJ...346..718S} Slavin, J.~D.\ 1989, \apj, 346, 
718 

\bibitem[Slavin \& Cox(1992)]{1992ApJ...392..131S} 
Slavin, J.~D., \& Cox, D.~P.\ 1992, \apj, 392, 131 

\bibitem[Slavin \& Frisch(2002)]{2002ApJ...565..364S} Slavin, J.~D.~\& 
Frisch, P.~C.\ 2002, \apj, 565, 364 

\bibitem[Smith \& Cox(2001)]{2001ApJS..134..283S} Smith, R.~K.~\& Cox, 
D.~P.\ 2001, \apjs, 134, 283 

\bibitem[Spitzer(1962)]{1962pfig.book.....S} Spitzer, L.\ 1962, Physics of 
Fully Ionized Gases, New York: Interscience (2nd edition), 1962,  

\bibitem[Stancil et al.(1998)]{1998ApJ...502.1006S} Stancil, P.~C., et al.\ 
1998, \apj, 502, 1006 

\bibitem[Sternberg et al.(2002)]{2002ApJS..143..419S} Sternberg, A., McKee, 
C.~F., \& Wolfire, M.~G.\ 2002, \apjs, 143, 419 

\bibitem[Stocke et al.(2006)]{2006ApJ...641..217S} Stocke, J.~T., Penton, 
S.~V., Danforth, C.~W., Shull, J.~M., Tumlinson, J., 
\& McLin, K.~M.\ 2006, \apj, 641, 217 

\bibitem[Sutherland \& Dopita(1993)]{1993ApJS...88..253S} Sutherland, 
R.~S., \& Dopita, M.~A.\ 1993, \apjs, 88, 253 

\bibitem[Telfer et al.(2002)]{2002ApJ...565..773T} Telfer, R.~C., Zheng, 
W., Kriss, G.~A., \& Davidsen, A.~F.\ 2002, \apj, 565, 773 

\bibitem[Tufte et al.(2002)]{2002ApJ...572L.153T} Tufte, S.~L., Wilson, 
J.~D., Madsen, G.~J., Haffner, L.~M., 
\& Reynolds, R.~J.\ 2002, \apjl, 572, L153 

\bibitem[Tumlinson et al.(2005)]{2005ApJ...620...95T} Tumlinson, J., Shull, 
J.~M., Giroux, M.~L., \& Stocke, J.~T.\ 2005, \apj, 620, 95 

\bibitem[Verner et al.(1996)]{1996ApJ...465..487V} Verner, D.~A.,
Ferland, G.~J., Korista, K.~T., \& Yakovlev, D.~G.\ 1996, \apj, 465, 487

\bibitem[Vieser \& Hensler(2007)]{2007A&A...475..251V} 
Vieser, W., \& Hensler, G.\ 2007, \aap, 475, 251 

\bibitem[Voronov(1997)]{1997ADNDT..65....1V} Voronov, G.~S.\ 1997,
Atomic Data and Nuclear Data Tables, 65, 1

\bibitem[Wakker et al.(2003)]{2003ApJS..146....1W} Wakker, B.~P.~et al.\ 
2003, \apjs, 146, 1


\bibitem[Zatsarinny et al.(2006)]{2006A&A...447..379Z} Zatsarinny, O.,
Gorczyca, T.~W., Fu, J., Korista, K.~T., Badnell, N.~R., \& Savin, D.~W.\ 2006, \aap, 447, 379

\bibitem[Zatsarinny et al.(2005)]{2005A&A...438..743Z} Zatsarinny, O.,
Gorczyca, T.~W., Korista, K.~T., Fu, J., Badnell, N.~R., Mitthumsiri,
W., \& Savin, D.~W.\ 2005a, \aap, 438, 743

\bibitem[Zatsarinny et al.(2005)]{2005A&A...440.1203Z} Zatsarinny, O.,
Gorczyca, T.~W., Korista, K.~T., Fu, J., Badnell, N.~R., Mitthumsiri,
W., \& Savin, D.~W.\ 2005b, \aap, 440, 1203

\bibitem[Zatsarinny et al.(2004)]{2004A&A...426..699Z} Zatsarinny, O.,
Gorczyca, T.~W., Korista, K., Badnell, N.~R., \& Savin, D.~W.\ 2004b,
\aap, 426, 699

\bibitem[Zatsarinny et al.(2004)]{2004A&A...417.1173Z} Zatsarinny, O.,
Gorczyca, T.~W., Korista, K.~T., Badnell, N.~R., \& Savin, D.~W.\ 2004a,
\aap, 417, 1173

\bibitem[Zatsarinny et al.(2003)]{2003A&A...412..587Z} Zatsarinny, O.,
Gorczyca, T.~W., Korista, K.~T., Badnell, N.~R., \& Savin, D.~W.\ 2003,
\aap, 412, 587

\bibitem[Zheng et al.(2004)]{2004ApJ...605..631Z} Zheng, W., et al.\ 2004, 
\apj, 605, 631 


\end{thebibliography}
\end{document}